\documentclass[11pt]{article}

\usepackage{authblk}
\usepackage{fancyhdr}
\usepackage{amsmath}
\usepackage{amsthm}
\usepackage{amssymb,amsfonts, enumitem}
\usepackage{threeparttable}
\usepackage[mathscr]{eucal}
\usepackage{graphicx}
\usepackage{geometry}
\usepackage{times}
\usepackage{enumerate}
\usepackage{courier}
\usepackage[round]{natbib}
\usepackage{float}
\usepackage{mathtools}
\usepackage{color}
\usepackage{authblk}
\usepackage{subcaption}
\usepackage{bbm}
\usepackage{array}
\usepackage{booktabs}
\usepackage{multirow}
\usepackage{siunitx}

\newcommand{\RR}{\mathbb{R}} 
\newcommand{\EE}{\mathbb{E}}
\newcommand{\PP}{\mathbb{P}}
\newcommand{\Var}{\text{Var}} 
\newcommand{\calS}{\mathcal{S}}
\newcommand{\bX}{\mathbf{X}}
\newcommand{\bA}{\mathbf{A}}
\newcommand{\bB}{\mathbf{B}}

\newcommand{\pari}{{(i)}}
\newcommand{\bQ}{\mathbf{Q}}

\newcommand{\bI}{\mathbf{I}}

\newcommand{\grad}{{\nabla}}

\newcommand{\calN}{\mathcal{N}}
\newcommand{\one}{\mathbbm{1}}
\newcommand{\Ehat}{\widehat{E}}

\newtheorem{proposition}{Proposition}[section]

\newtheorem{lemma}[proposition]{Lemma}
\newtheorem{corollary}[proposition]{Corollary}
\newtheorem{theorem}[proposition]{Theorem}
\newtheorem{remark}[proposition]{Remark} 
\newtheorem{example}{Example}

\renewcommand{\vec}[1]{\mathrm{#1}}
\newcommand*{\Scale}[2][4]{\scalebox{#1}{$#2$}}

\newcommand\numberthis{\addtocounter{equation}{1}\tag{\theequation}}

\newcommand{\indep}{\rotatebox[origin=c]{90}{$\models$}}

\DeclareMathOperator*{\argmin}{argmin}

\newcount\Comments  
\definecolor{darkgreen}{rgb}{0,0.5,0}
\definecolor{purple}{rgb}{1,0,1}
\newcommand{\kibitz}[2]{\ifnum\Comments=1\textcolor{#1}{#2}\fi}

\begin{document}

\title{Integrative Methods for Post-Selection Inference Under Convex Constraints}
\author[$\dagger$]{Snigdha Panigrahi}
\author[$\star$]{Jonathan Taylor}
\author[$\ddag$]{Asaf Weinstein}
\affil[$\dagger$]{Department of Statistics\\
University of Michigan}
\affil[$\star$]{Department of Statistics\\
Stanford University}
\affil[$\ddag$]{School of Computer Science and Engineering\\
Hebrew University of Jerusalem}
\date{}                     

\maketitle
\thispagestyle{empty}

\begin{abstract}
Inference after model selection has been an active research topic in the past few years, with numerous works offering different approaches to addressing the perils of the reuse of data. 
In particular, major progress has been made recently on large and useful classes of problems by harnessing general theory of hypothesis testing in exponential families, but these methods have their limitations. 
Perhaps most immediate is the gap between theory and practice: implementing the exact theoretical prescription in realistic situations---for example, when new data arrives and inference needs to be adjusted accordingly---may be a prohibitive task. 

In this paper we propose a Bayesian framework for carrying out inference after model selection in the linear model. 
Our framework is very flexible in the sense that it naturally accommodates different models for the data, instead of requiring a case-by-case treatment. 
At the core of our methods is a new approximation to the exact likelihood conditional on selection, the latter being generally intractable. 
We prove that, under appropriate conditions, our approximation is asymptotically consistent with the exact truncated likelihood. 
The advantages of our methods in practical data analysis are demonstrated in simulations and in application to HIV drug-resistance data. 
\end{abstract}

\section{Introduction}
\subsection{Background} \label{subsec:background}
Any meaningful statistical problem consists of a model and a set of matching parameters for which decisions are required. In the classical, ``textbook" paradigm, the model and this set of target parameters are assumed to be chosen independently of the data subsequently used for statistical inference (e.g., estimating the parameters). In practice, however, more often than not, data analysts examine some aspect of the data before deciding on a model and/or the target parameters. 
For example, in fitting a simple regression model, a plot of the data might help the analyst decide whether to model the relationship between the response and the explanatory variable as linear or nonlinear; in fitting  linear regression, one might decide to discard variables with large p-values and re-fit the smaller model before reporting any findings; and in multiple hypothesis testing, the analyst might be tempted to report confidence intervals only for rejected nulls.
\smallskip

Of course, ignoring such form of adaptivity in choosing the model and/or the target parameters, may result in the loss of inferential guarantees and lead to flawed conclusions. Still, most would agree that instructing the analyst to avoid such exploration of the data altogether, is not only impractical, but also not recommended. This realization on the one hand, and a serious concern about a replication problem in science on the other hand, have elicited an effort in the statistical community to develop tools for {\it selective inference}. 
In general, such tools allow to take into account the fact that the same data used to select target parameters, is used when providing inference, thus attempting to restore validity of inference following selection. 
This includes bias-reducing methods based on an extended definition of Uniform Minimum-Variance Unbiased estimation \citep{robbins1988uv}, Bayesian approaches \citep{efron2011tweedie} and bootstrapping \citep{simon2013estimating}; 
as well as methods that work by performing simultaneous inference \citep{berk2013valid}. 
Recently, tools from information theory \citep{russo2015controlling} and Differential Privacy \citep{dwork2014preserving, dwork2015generalization} have also been proposed to quantify and control the effect of selection. 
\smallskip

Another common approach is to condition on selection, namely, base inference on the likelihood of the observed data when truncated to the set of all realizations that would result in the analyst posing the same question. A conditional approach has been pursued by several authors that addressed selective inference in so-called large-scale inference problems \citep{efron2012large}. 
Underlying such problems is, classically, a sequence model, where each observation corresponds to a single parameter, and the parameters are typically not assumed to have any relationship with one another. 
Conditional inference for the effects corresponding to the $K$ largest statistics in the sequence model was proposed in \citet{reid2014post}. 
\citet{zollner2007overcoming} and \citet{zhong2008bias} suggested point and interval estimates for the parameter of a univariate gaussian distribution conditional on exceeding a fixed threshold, in the context of genome-wide association studies; 
\citet{weinstein2013selection} constructed confidence intervals, and \citet{benjamini2014selective} explored point estimators, for the univariate truncated gaussian problem with a more flexible (random) choice of cutoff; 
\citet{yekutieli2012adjusted} proposed to base inference on the truncated likelihood while incorporating a prior on the parameters; and \citet{simonsohn2014p} proposed a frequentist method to assess effect sizes from the distribution of the p-values corresponding to only the selected. 
\smallskip

More recently, the practicability of the conditional approach has been extended considerably from the sequence model to the realm of linear models and GLMs \citep[][among others]{lee2016exact, taylor2013tests, taylor2014exact, lee2014exact, fithian2014optimal, tian2018selective}. 
In these works the selection protocol is assumed to partition the sample space into polyhedral (or at least convex) sets, fitting many popular `automatic' model selection procedures such as marginal screening, Lasso, forward-selection etc. 
The techniques developed in that line of work have practical importance as they allow to carry out exact inference after model selection, which is one of the most popular situations where the problem of selective inference arises. 
At the core of these new contributions is the realization, first made in \citet{lee2016exact}, that when inference for the projection of the mean vector of $Y$ (in a fixed-$X$ homoscedastic, gaussian linear model with known $\sigma$) onto a one-dimensional subspace is desired, then conditioning further on the projection of $Y$ onto the orthogonal complement of this subspace, reduces the problem to inference for a univariate truncated normal variable, the distribution of which depending only on the (scalar) parameter of interest. 
\smallskip

The polyhedral lemma of \citet{lee2016exact} was indeed a significant step forward, because it made feasible exact inference after variable selection in the gaussian regression case, for many popular variable selection rules. 
However, this method has its limitations. 
If attention is restricted to hypothesis testing, then the polyhedral lemma yields a Uniformly Most Powerful Unbiased (UMPU) test in the fixed-$X$ setting and under a specific model for the data, entailing that the expectation of the response is an arbitrary vector in $\RR^n$; \citet{fithian2014optimal} call this the {\it saturated} model. 
Furthermore, the selection rule is assumed to involve the same set of observations used for subsequent inference. 
Lastly, the target of inference in \citet{lee2016exact} is restricted to a real-valued linear function of the model parameters. 
If these assumptions are violated, the polyhedral lemma may not be applicable, or lead to suboptimal procedures.

\subsection{Our contribution} \label{subsec:contrib}

In the current paper we explore the problem of selective inference when one moves away from the aforementioned assumptions. 
To circumvent the limitations related to the polyhedral lemma, we offer an approach that works {\it directly} with a full selective likelihood function while implementing the principles of {\it data carving} \citep{fithian2014optimal}. Carving is a form of randomization that resembles data-splitting in that selection operates on a subset of the data, but is different from data-splitting in that inference uses the entire data set instead of relying on the held-out portion alone. 
While the flavor is frequentist (we use a {\it conditional} likelihood), in our framework we incorporate a prior on the parameters of the (adaptively) chosen model. This allows us to give inference for general functions of the parameter vector by integrating out nuisance parameters. 
Our approach contrasts with the strictly frequentist approach of \citet{fithian2014optimal}, which deals with nuisance parameters by conditioning them out. 
By adopting formally a Bayesian approach, we are able to exploit the usual advantages of the Bayesian machinery. 
For example, because we have a mechanism for generating samples from the posterior distribution, credible intervals can be estimated easily by finding appropriate quantiles in the posterior sample of the parameter. 
{\it Because our methods are amenable to randomization, we handle data carving naturally, and in that sense offer considerably more flexibility as compared to the methods of \citet{lee2016exact}.}
\smallskip

Our main contribution is a tractable approximation to the full {\it truncated} likelihood function, which allows to overcome serious computational objections. 
On the theoretical side, much of our effort is focused on motivating the approximation and especially to proving that it enjoys various desirable properties. 
Specifically, our main technical results establish that in a fixed-$p$, growing-$n$ regime, and under our randomization framework, the proposed approximation:
\begin{itemize}[leftmargin=*]
\item is consistent for the exact selection probability, see Theorem \ref{thm:approx}.
\item leads to consistency of the approximate posterior distribution in the sense of Theorem \ref{posterior:consistency:gen}. 
\item from a computational perspective, it presents a convex optimization problem when solving for the approximate posterior mode, see Theorem \ref{lem:map-convex}. 
\end{itemize}

To emphasize the utility of the proposed methods, in our framework we divide the data into two parts, where selection operates on the first part only. 
The second, held-out portion, is reserved for inference after selection. 
By preventing the statistician from using the held-out portion of the data for selection, this scheme ensures that there is enough leftover information \citep{fithian2014optimal} after selection, thereby allowing to give more ``powerful" inference, for example, to construct shorter confidence intervals; see also \citet{kivaranovic2018expected}. 
This paradigm arises naturally in many realistic situations: in almost any field of science, the researcher begins with an initial data set which she might use for selection and (post-selection) inference. But, usually further observations are made available at a future point in time. This is either because the researcher decided to collect more data after seeing the outcome of the initial analysis, or simply because another data set comes in later on. 
At this point the researcher is faced with the question of how to combine the two data sets to provide inference for parameters selected based only on the first data set. 
As was already pointed out in \citet{fithian2014optimal}, this situation can be cast into the selective inference framework, where the data is the augmented set of observations, and selection ``so happens" to operate on only part of the data \citep[see the simple Example 4 in][]{fithian2014optimal}. 
In other words, the optimal thing to do is base inference on both the initial and the follow-up parts of the data, while taking into account the fact that selection affects the distribution in the first part---what we referred to as ``data carving" earlier.  
Our Bayesian methods naturally implement the principles of data carving, through the updating of the {prior distribution} by the truncated likelihood. 
\smallskip

Another difference between our work and, e.g., \citet{lee2016exact}, is that we allow the researcher to refine the choice of the model after seeing the output of the automatic algorithm. 
Indeed, in many realistic situations this kind of flexibility is essential. 
For example, suppose that we run Lasso to select among $p$ main effects, but after observing the output, we want to add interactions between some of the selected variables. 
Including such interactions may account for some unexplained variance \citep{gunter2011variable, yu2019reluctant}. 
In other cases, domain-specific information might be available a priori for the predictors. 
For instance, in statistical genomics it is common to use existing biological annotations to inform selection of variables \citep{gao2017incorporating}. 
Such information may suggest to include a group of genes known to interact with each other, even if some of them were not selected by the automatic protocol \citep{stingo2011incorporating}.

\begin{figure}
  \centering
    \includegraphics[width=\textwidth]{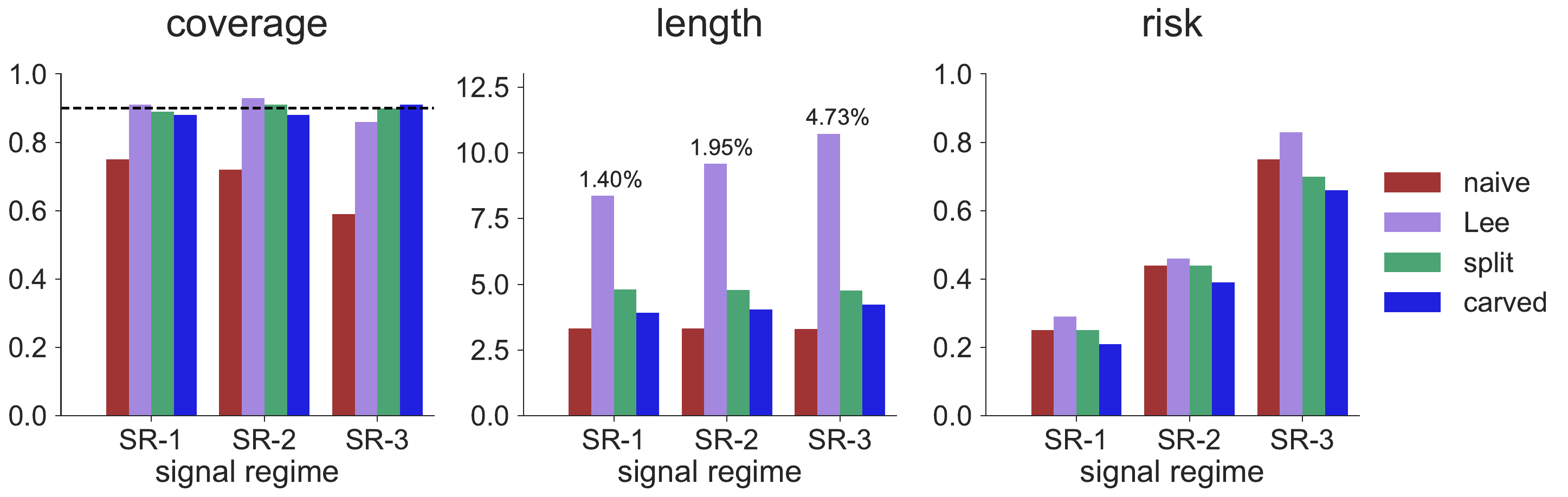}
     \vspace{-0.3cm}
  \caption{}
  \label{fig:carv}
  \vspace{-0.7cm}
\end{figure}

\smallskip
Figure \ref{fig:carv} showcases the advantages of our methods on simulated data. 
In this experiment, the data obeys a linear model with a $500$-by-$100$ design matrix $\bX$, such that the rows of $\bX$ are i.i.d.~copies of a correlated multivariate normal vector. 
The model coefficients $\beta_j$ are i.i.d.~from a mixture of two zero-mean gaussians, one with small variance $0.01$ and the other with large variance $V$. 
We select variables with the Lasso, using a fixed value for the tuning parameter. 
Four different methods are compared on three criteria, for three different values of $V$: average coverage of interval estimates for the $\beta_j$ with nominal FCR (false coverage rate) level of $10\%$; length of the interval estimates; and the prediction risk. 
The figure shows that unadjusted inference, labeled as ``naive" is indeed not valid. 
FCR is roughly maintained at the nominal level for the other methods, including our new ``carving" method that constructs (approximate) Bayesian intervals with respect to a vague prior. 
It can be seen that the carving method produces the shortest intervals. 
In the middle panel of the figure we also indicate the percentage of intervals constructed with the methods of \citet{lee2016exact}, that had infinite length. 
Our method also attains smaller prediction risk than the other three estimators. 
The simulation setting and all four methods methods are described precisely in Section \ref{sec:approximation}, where we also provide a table corresponding to Figure \ref{fig:carv}.

\smallskip
The rest of the paper is organized as follows. 
Section \ref{sec:setup} describes the general setup and introduces the central component, namely, a selection-adjusted likelihood. 
In the following two sections we instantiate the general framework by considering two particular cases. 
Section \ref{sec:univar} begins with the simple example of univariate data and a simple selection rule. 
In Section \ref{sec:lasso} we analyze the example of Lasso selection in the linear model, that features all of the complexity of the general framework; further examples appear in the Appendix. 
Section \ref{sec:approximation}, which includes the main technical novelty, develops an approximation to the selection-adjusted likelihood, provides supporting analysis, and demonstrates its usefulness in a simulation. 
We treat point estimation separately in Section \ref{sec:estimation} and prove the consistency guarantees associated with our estimates. 
Computational details for implementing our methods are provided in Section \ref{sec:computation}, and Section \ref{sec:discussion} concludes with a discussion. 
Proofs are deferred to the Appendix.

\section{Basic setup} \label{sec:setup}

Let
\begin{equation*}
(X_{i1},...,X_{ip}, Y_i) \sim P, \ \ \ \ \ \ \ \ i=1,...,n, 
\end{equation*}
be $(p+1)$-dimensional i.i.d.~vectors with $Y_i\in \RR$. 
Denote $\vec{y}=(Y_1,...,Y_n)^T$, and denote by $\bX$ the $n\times p$ matrix with $(X_{i1},...,X_{ip})$ in its $i$-th row. 
Further, for any matrix $\boldsymbol{D}\in \RR^{k\times l}$, any subset $N \subseteq \{1,...,k\}$ and any subset $M\subseteq \{1,...,l\}$, we denote by $\boldsymbol{D}^N$ the $|N|\times l$ matrix obtained from $\boldsymbol{D}$ by keeping only the {\it rows} with indices in $N$, and we denote by $\boldsymbol{D}_M$ the $k\times |M|$ matrix obtained from $\boldsymbol{D}$ by keeping only the {\it columns} with indices in $M$. 
Lastly, $\calN_k(\cdot;\boldsymbol{\eta}, \boldsymbol{\Gamma})$ is the density of an $k$-dimensional multivariate normal vector with mean $\boldsymbol{\eta}$ and covariance $\boldsymbol{\Gamma}$. 


\smallskip
Our framework consists of a model selection stage followed by an inference stage. 
For a subset $\calS \subset \{1,...,n\}$ determined independently of the data, let $(\bX^{\calS},\vec{y}^{\calS})$ be a subsample of size $|\calS|=n_1$. 
We will refer to $(\bX^{\calS},\vec{y}^{\calS})$ as the original data henceforth. 
On observing the original data, the statistician looks for relevant variables by applying some predetermined variable selection rule, a mapping
$$
(\bX^\calS,\vec{y}^\calS) \longmapsto \widehat{E}\subseteq\{1,...,p\}.
$$
The statistician is allowed at this point to refine the selection by dropping variables from (or adding variables to) the output $\widehat{E}$. 
Specifically, let $\widehat{E}' \subseteq\{1,...,p\}$ be a subset obtained by applying an arbitrary but deterministic function to $\widehat{E}$. 
We denote by $E$ and $E'$ the realizations of the variables $\widehat{E}$ and $\widehat{E}'$, respectively. 

\smallskip
In the second stage inference is provided for $P$ assuming that 
\begin{equation}\label{eq:model}
\vec{y} | \bX \ \sim\ \calN_n(y; \bX_{E'}\beta^{E'}, \sigma^2 \bI). 
\end{equation}
If not indicated otherwise, we treat $\sigma^2 = \sigma^2_{E'}$ as known. 
Nevertheless, we would like to emphasize that our method easily accommodates the case where $\sigma^2$ is unknown and itself given a prior, see Sections \ref{sec:approximation} and \ref{sec:computation}. 
Naive inference would now proceed under the model \eqref{eq:model}, ignoring the fact that $E'$ was chosen after seeing (part of) the data. 
Instead, the fact that $E'$ is data-dependent will be accounted for by {\it truncating} the likelihood in \eqref{eq:model} to the event $\{\Ehat=E\}$. 
Formally, this means that, conditionally on $\bX$, \eqref{eq:model} should be replaced by
\begin{equation} \label{E':model}
\vec{y} | \bX,\ \Ehat=E \ \sim\ \frac{ \calN_n(y; \bX_{E'}\beta^{E'}, \sigma^2 \bI) }{ \int \one_{\{\Ehat=E\}}(\vec{y}) \cdot \calN_n(\vec{y}; \bX_{E'}\beta^{E'}, \sigma^2 \bI) d\vec{y} } \one_{\{\Ehat=E\}}(y),
\end{equation}
remembering that $\widehat{E}'$ is determined by $\widehat{E}$, and that $\widehat{E}$ is in turn measurable with respect to $(\bX^{\calS},\vec{y}^{\calS})$. 

\smallskip
In principle, post-selection inference for $\beta^{{E}'}$ could now be based on \eqref{E':model}. 
Unfortunately, in practice, the denominator in \eqref{E':model} is intractable, and we will seek simplifications. 
Taking a step back, note that if no selection had been involved in specifying $E'$, standard inference for $\beta^{E'}$ would rely on the distribution of the least squares estimator, 
\begin{equation}\label{eq:ls}
\widehat{\beta}^{E'}=\bX_{E'}^\dagger \vec{y},
\end{equation}
where $\bX_{E'}^\dagger$ is the pseudo-inverse of $\bX_{E'}$; if $\bX_{E'}$ has full rank, this is $(\bX_{E'}^T \bX_{E'})^{-1}\bX_{E'}^T$. 
If the selection event could be expressed in terms of $\widehat{\beta}^{E'}$ and nothing else, this would give rise to a truncated distribution for $\widehat{\beta}^{E'}$. 
This is, however, not the case for the selection rules considered in this paper for two reasons. First of all, because we implement data carving, the choice of $E$ cannot be a deterministic function of $\widehat{\beta}^{E'}$, the latter being a function of the full data $(\bX,\vec{y})$ rather than $(\bX^{\calS},\vec{y}^{\calS})$. 
Second, even if we allowed selection to be based on the full data $(\bX,\vec{y})$, we cannot restrict it to be a function of $\widehat{\beta}^{E'}$ alone. For example, Lasso selection cannot be expressed in terms of only the least squares statistic even if selection operates on the entire data set. 
Suppose for the moment, then, that the event $\Ehat=E$ could be written in a polyhedral form, 
\begin{equation}\label{constraints:carved}
{\bA}_{E} \sqrt{n}\vec{T}_n + {\bB}_{E}\vec{\Omega}_n < \vec{b}_{E},
\end{equation}
with
$$
\vec{T}_n = 
\begin{pmatrix}
\widehat{\beta}^{E'}\\ \vec{U}
\end{pmatrix},
$$
where $\vec{U}$ is a vector that may itself depend on $E'$ and, importantly, ${\Omega}_n$ has the property that (i) it is independent of $\vec{T}_n$ and (ii) it follows a distribution that does not depend on $\beta^{E'}$. 
Using the terminology of \citet{tian2018selective}, $\vec{\Omega}_n$ is a {\it randomization} term. 
As we will see later on, the selection rules that we will be concerned with can indeed be approximately represented in the form \eqref{constraints:carved} with appropriate choices of $\vec{U}$, $\Omega_n$ and ${\bA}_E, {\bB}_{E}, \vec{b}_E$; see equations \eqref{stat:carved:univar} and \eqref{K.K.T.:selection}. 
By ``approximately represented" we mean that we allow a term that goes to zero in probability in \eqref{constraints:carved} and also that $\Omega_n$ satisfies the properties (i) and (ii) above asymptotically, as in the sense formalized  in Section \ref{sec:lasso}.\footnote{A similar representation of the selection rule as an approximate polyhedral set appears in \citet{tian2018selective}, see their Lemma 14 for example.} 
We would like to emphasize at this point that one of the technical challenges is to find an {\it explicit} representation as such for a given selection rule. This task will be relatively easy in Section \ref{sec:univar}, but considerably more involved in Section \ref{sec:lasso}, when we address selection with the Lasso.

\smallskip
Treating \eqref{constraints:carved} first as if it holds exactly instead of asymptotically, we start by considering the truncated distribution of $(T_n, \Omega_n)$, 
\begin{equation*}
\begin{aligned}
& \frac{L^n(\beta^{E'}; t_n, \omega_n)}{\PP({\bA}_{E} \sqrt{n}\vec{T}_n + {\bB}_{E}\vec{\Omega}_n < \vec{b}_{E}\lvert \beta^{E'})}\one_{\{{\bA}_{E} \sqrt{n}\vec{t}_n + {\bB}_{E}\vec{\omega}_n < \vec{b}_{E}\}}(t_n, \omega_n),
\end{aligned}
\end{equation*}
where $L^n(\beta^{E'}; t_n, \omega_n)$ denotes the (marginal) likelihood function corresponding to $({T}_n, \vec{\Omega}_n)$. 
Note that there is no conditioning on $\bX$. 
Here and elsewhere, $\PP(\cdot)$ denotes probability with respect to $L^n(\beta^{E'}; \cdot, \cdot)$, that is, 
$$
\PP(A) = \int \one_{A}(t_n, \omega_n) \cdot L^n(\beta^{E'}; t_n, \omega_n) \ dt_n d\omega_n
$$
for any event $A$ measurable with respect to $({T}_n, \vec{\Omega}_n)$. 
The dependence on $\beta^{E'}$ is sometimes emphasized by writing $\PP(\cdot|\beta^{E'})$ instead of $\PP(\cdot)$. 
By integrating with respect to $\omega_n$, and using the fact that $\Omega_n$ is a randomization term, we obtain the {\it selection-adjusted likelihood},  
\begin{equation}\label{eq:sel:lik}
\begin{aligned}
L^n_S(\beta^{E'}; t_n) &:= \frac{L^n(\beta^{E'}; t_n)}{\PP({\bA}_{E} \sqrt{n}\vec{T}_n + {\bB}_{E}\vec{\Omega}_n < \vec{b}_{E}\lvert \beta^{E'})} \PP \left({\bA}_{E} \sqrt{n}\vec{t}_n + {\bB}_{E}\vec{\Omega}_n < \vec{b}_{E} \lvert \vec{t}_n \right) \\[7.5pt]
&\propto \frac{L^n(\beta^{E'}; t_n)}{\PP({\bA}_{E} \sqrt{n}\vec{T}_n + {\bB}_{E}\vec{\Omega}_n < \vec{b}_{E}\lvert \beta^{E'})},
\end{aligned}
\end{equation}
where we use $\propto$ to indicate that two terms are proportional as functions of $\beta^{E'}$. 
Note that the probability in the numerator of the second term above is taken over $\Omega_n$ only, hence this term does not depend on $\beta^{E'}$. 
We remark that we borrowed the term ``selection-adjusted likelihood" from \citet{yekutieli2012adjusted}, although the framework here is very different \citep[for example, there is no randomization in][]{yekutieli2012adjusted}.

\smallskip
The point of view has thus far been frequentist. 
Inspired by the basic ideas in \citet{yekutieli2012adjusted}, we now introduce a prior into our model, 
\begin{equation}\label{eq:prior}
\beta^{E'} | \Ehat = E \ \ \sim\ \  \pi. 
\end{equation}
The conditioning on $\Ehat=E$ above is consistent with the same conditioning in the likelihood. Since we pose a generative model for the data conditionally on selection, the prior as well is a conditional prior. 
Combining the selection-adjusted likelihood \eqref{eq:sel:lik} and the prior \eqref{eq:prior} results formally in the {\it selection-adjusted posterior} distribution, 
\begin{equation}\label{eq:sel-adj-gen}
\pi_S(\beta^{E'}|\vec{t}_n) \propto \pi(\beta^{E'})L^n_S(\beta^{E'}; t_n).
\end{equation}
We use again a subscript $S$ in $\pi_S$ to indicate that in the selection-adjusted posterior, $\pi$ is updated with the selection-adjusted likelihood rather than a marginal likelihood. 
The idea is, in general, to provide adjusted inference for $\beta^{E'}$, or functions thereof, based on \eqref{eq:sel-adj-gen}. 

\smallskip
The central component in our Bayesian model is the selection-adjusted likelihood \eqref{eq:sel:lik}, which also presents the main technical challenge. In order to sample from \eqref{eq:sel-adj-gen}, we need to be able to evaluate the {\it adjustment factor},
\begin{equation}\label{eq:adj-factor}
\PP({\bA}_{E} \sqrt{n}\vec{T}_n + {\bB}_{E}\vec{\Omega}_n < \vec{b}_{E}\lvert \beta^{E'}),
\end{equation}
as a function of $\beta^{E'}$, which is still (generally) intractable. 
Hence, most of our effort will be devoted to developing an amenable approximation that can be plugged in instead. 



\section{A Primer}\label{sec:univar}
We find it instructive to first demonstrate our methods in a simple situation, namely, the special case where there are no covariates and the selection rule takes on a simple form. 
Besides priming the sequel, the example below is a case where an exact analysis can be carried out. 
Hence, unlike in the general situation, we can compare our approximate methods against the ground truth when assessing their effectiveness. 

\smallskip
Throughout this section, then, suppose that the observations are
\begin{equation}\label{model:univar}
Y_i\sim \mathcal{N}(\beta, 1),\ \ \ \ \ \ \ i=1,...,n.
\end{equation}
For the theoretical analysis that follows, $\beta=\beta_n$ depends in general on $n$, and we use the parameterization $\sqrt{n}\beta_n \equiv n^{\delta}{\beta^*}$ with ${\beta^*}$ a constant and $\delta \in (0, 1/2]$. 
Let $\mathcal{S} \subseteq \{1,...,n\}$ be a random subset of size $n_1=\rho n$, and denote
$$
\bar{Y}^{\mathcal{S}}:= \frac{1}{n_1}\sum_{i\in \calS} Y^{(i)}. 
$$
We provide inference for $\beta_n$ conditionally on 
\begin{equation}\label{eq:sel:iid}
\sqrt{n_1}\bar{Y}^{\mathcal{S}}>0,
\end{equation}
that is, the selection event entails the (scaled) least squares estimate for $\beta_n$ from the original data, exceeds a fixed threshold (zero). 
We first note that, on defining 
$$
W_n := \bar{Y}^{\mathcal{S}} - \bar{Y}_n,
$$
we have 
$$
\sqrt n W_n \sim \mathcal{N}\left(0, \dfrac{1-\rho}{\rho}\right),\ \ \ \ \ \ \ \ \ \ \ \ \ \ \sqrt n W_n \indep \sqrt{n} \bar{Y}_n.
$$
Therefore, letting 
\begin{equation}\label{stat:carved:univar}
\sqrt n \widehat{\beta}_n = \sqrt n\bar{Y}_n,\ \ \ \ \ \ \ \Omega_n = \sqrt n W_n,
\end{equation}
we see that the selection event \eqref{eq:sel:iid} can be written exactly in the form \eqref{constraints:carved} with
$T_n = \widehat{\beta}_n$, and $\bA_E = -1, \bB_E = -1, b_E = 0$. 

\smallskip
In this simple example, we have an exact form for the probability of the selection event, 
\begin{equation}\label{eq:sel:iid:pr}
\PP\left({\bA}_{E} \sqrt{n}\vec{T}_n + {\bB}_{E}\vec{\Omega}_n < \vec{b}_{E} \lvert \beta_n\right) = \PP \left(\sqrt{n_1}\bar{Y}^{\mathcal{S}}>0 \lvert \beta_n\right) = \bar{\Phi}\left(-\sqrt{\rho}\cdot\sqrt{n}\beta_n\right).
\end{equation}
This will not be the case in the more complicated example of the following section, and approximations to the selection probability will be crucial. 
The following result suggests an approximation for the selection probability \eqref{eq:sel:iid:pr} in the simple univariate example, which we will generalize in the next section. 

\begin{figure}[h]
\centering
  \begin{subfigure}[t]{0.425\textwidth}
    \includegraphics[width=\textwidth]{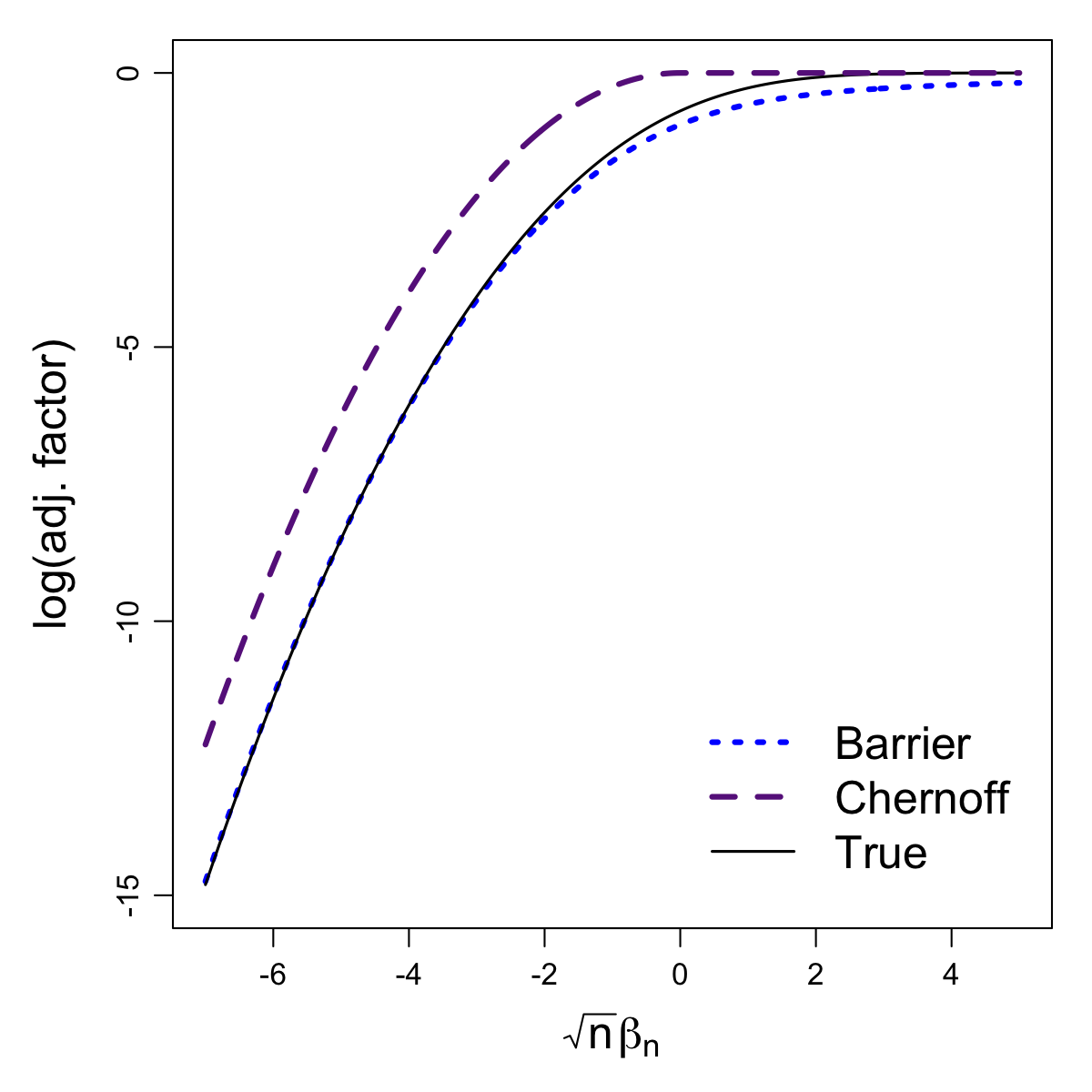}
    \caption{}
    \label{fig:adj-factor-approx}
  \end{subfigure}
  \begin{subfigure}[t]{0.425\textwidth}
    \includegraphics[width=\textwidth]{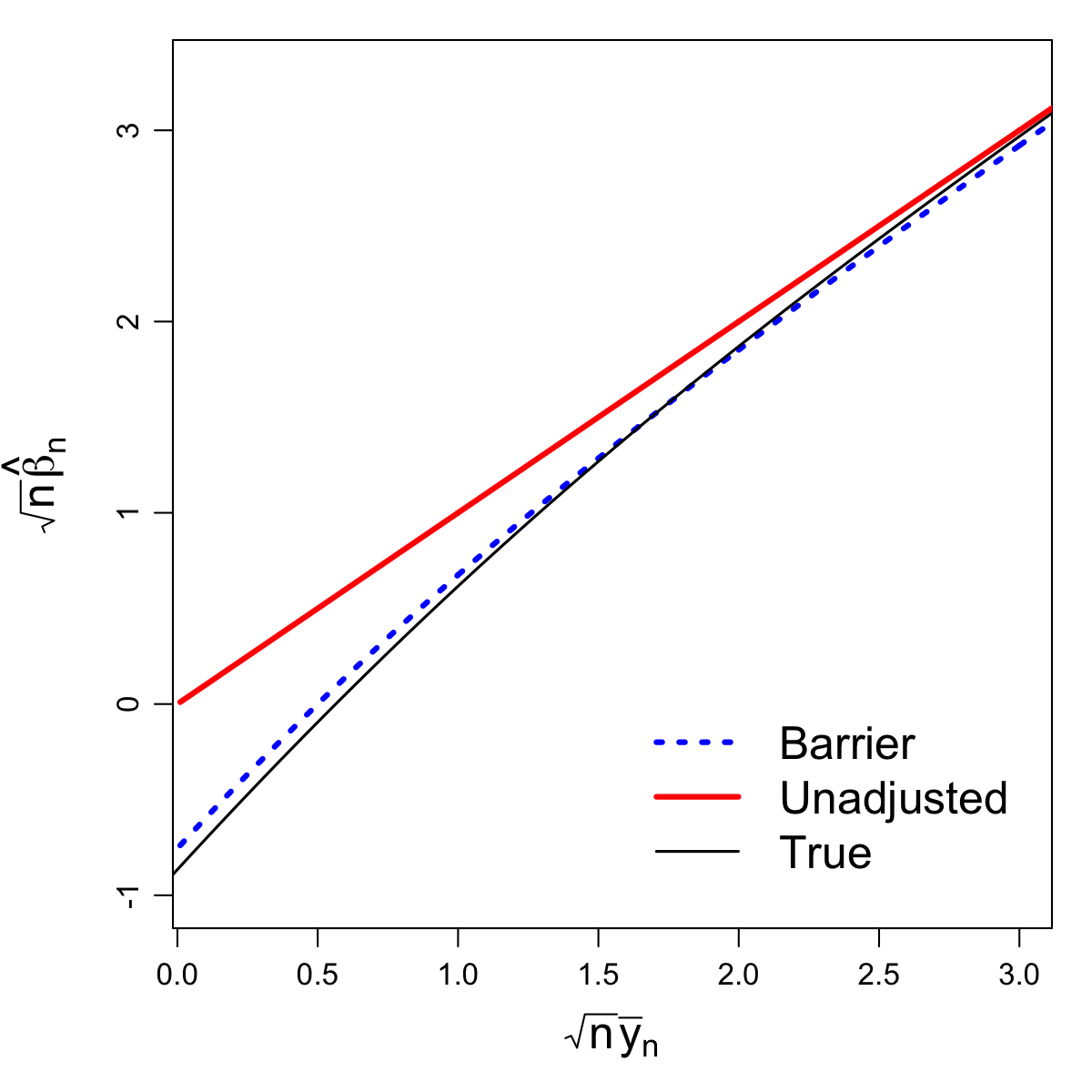}
    \caption{}
    \label{fig:mle-univar}
  \end{subfigure}
  \caption{
(a) Approximations to log-adjustment factor in the univariate normal setting. 
  In the legend, ``Chernoff" refers to the approximation from \eqref{ineq:sel:prob:carved}, and ``Barrier" to the approximation from \eqref{univar-barrier}; ``True" corresponds to the exact expression \eqref{eq:sel:iid:pr}. Note that the curve for the Chernoff approximation lies above the true one, as predicted by our theory. 
(b) Maximum-likelihood estimation for the univariate gaussian setting. Broken line corresponds to the approximate MLE, incorporating the proposed approximation to the selection probability. Solid black curve is the exact (true) MLE, and solid red line is the identity line, representing the unadjusted estimate.
  }
  \label{fig:selective:mle}
\end{figure}

\begin{theorem}
\label{carved:gaussian}
Let $\beta = \beta_n = n^{\delta}{\beta^*}/\sqrt{n}$, where ${\beta^*}$ is a constant. 
Let $\mathcal{K}=(0,\infty)$. Then 
\begin{equation}
\label{ineq:sel:prob:carved}
\log \mathbb{P}\left(\sqrt{n}(\bar{Y}_n + W_n)/n^{\delta}\in \mathcal{K}\big| \beta_n \right) \leq -n^{2\delta} \cdot\inf_{(z,w):z+w \in \mathcal{K}} \dfrac{(z-{\beta^*})^2}{2} + \dfrac{\rho w^2}{2(1-\rho)}\ \ \ \ \ \ \ \text{for any $n\in \mathbb{N}$}
\end{equation}
whenever $0<\delta\leq 1/2$.
Furthermore, the logarithm of the sequence of selection probabilities satisfies
\begin{equation}
\label{lim:sel:prob:carved}
\lim_{n\to \infty} \frac{1}{n^{2\delta}}\log \mathbb{P}\left(\sqrt{n} (\bar{Y}_n + W_n)/n^{\delta}\in \mathcal{K}\big| \beta_n \right) = -\inf_{(z,w):(z+w) \in \mathcal{K}} \dfrac{(z-{\beta^*})^2}{2} + \dfrac{\rho w^2}{2(1-\rho)}.
\end{equation}
\end{theorem}
We call the exponent of the right hand side of \eqref{ineq:sel:prob:carved} the {\it Chernoff approximation} to the selection probability \eqref{eq:sel:iid:pr}. 
Note that the infimum on right hand side of \eqref{ineq:sel:prob:carved} is taken over a constrained set in $\mathbb{R}^2$. 
To obtain a more computationally amenable expression, we propose to replace the right hand side of \eqref{ineq:sel:prob:carved} with the unconstrained optimization problem 
\begin{equation}
\label{univar-barrier}
-n^{2\delta}\cdot \inf_{(z, w)\in \mathbb{R}^{2}}\left\{\dfrac{(z-{\beta^*})^2}{2} + \dfrac{\rho w^2}{2(1-\rho)} + \frac{1}{n^{2\delta}}\psi_{n^{-\delta}} (z + w)\right\},
\end{equation}
where $\psi_{n^{-\delta}}$ is a function satisfying that in the limit as $n\to \infty$,
$$
\frac{1}{n^{2\delta}}\psi_{n^{-\delta}}(x) \longrightarrow I(x) =
\begin{cases} 
      0  & \text{ if } x\in (0,\infty) \\
      \infty  & \text{ otherwise}
\end{cases}.
$$
Specifically, using 
\begin{equation}\label{eq:psi:barrier}
\psi_{n^{-\delta}} (x) = \log(1+ n^{-\delta}/x)
\end{equation}
whenever $x\in (0,\infty)$ and $\infty$ otherwise
in \eqref{univar-barrier}, we obtain our {\it barrier approximation} to the selection probability \eqref{eq:sel:iid:pr}, so called because \eqref{eq:psi:barrier} serves as a ``soft" barrier alternative to the indicator function $I(x)$. 
Figure \ref{fig:adj-factor-approx} compares the approximations in \eqref{ineq:sel:prob:carved} and \eqref{univar-barrier} to the exact expression \eqref{eq:sel:iid:pr}. 
The plots for both approximations follow the exact curve fairly closely. Note that the curve corresponding to the Chernoff approximation lies above that for the true probability, as predicted by Theorem \ref{carved:gaussian}. 

\begin{remark}
\label{theoretical:delta}
In practice, of course, we will not have access to $\delta$, but, by substituting $\sqrt{n}z' = n^{\delta} z, \sqrt{n}w' = n^{\delta} w$, \eqref{univar-barrier} is equivalent to
$$
-n\cdot \inf_{(z', w')\in \mathbb{R}^{2}}\left\{\dfrac{(z'-{\beta_n})^2}{2} + \dfrac{\rho {w'}^2}{2(1-\rho)} + \frac{1}{n}\psi_{n^{-1/2}} (z' + w')\right\},
$$ 
and we can simply work with $\beta_n$ without knowing the underlying parameterization.
\end{remark}

\smallskip
The logarithm of the selection-adjusted posterior is
\begin{equation}\label{eq:log-post:univar}
\log\pi_S(\beta_n\lvert\bar{y}_n) =\log \pi(\beta_n) -n(\bar{y}_n - \beta_n)^2/2 -\log \mathbb{P}(\sqrt{n}(\bar{Y}_n + W_n)/n^{\delta} \in \mathcal{K}\lvert \beta_n),
\end{equation}
and a corresponding {\it approximate} selection-adjusted posterior, denoted as $\log\widetilde{\pi}_S(\beta_n\lvert \bar{y}_n)$ is 
\begin{equation}\label{eq:log-post:approx:univar}
\log \pi(\beta_n) -n(\bar{y}_n - \beta_n)^2/2 + n^{2\delta}\cdot \inf_{(z, w)\in \mathbb{R}^{2}}\left\{\dfrac{(z-{\beta^*})^2}{2} + \dfrac{\rho w^2}{2(1-\rho)} + \frac{1}{n^{2\delta}}\psi_{n^{-\delta}} (z+ w)\right\},
\end{equation}
where $\psi_{n^{-\delta}}(\cdot)$ is a (convex) barrier function associated with $\mathcal{K}=(0,\infty)$. 
The following Corollary, a consequence of Theorem \ref{carved:gaussian}, says that the sequence of selection-adjusted posteriors based on a suitable barrier approximation, indeed converges to the corresponding (true) selection-adjusted posterior.
\begin{corollary}
\label{carved:LDP:approximate:selective:posterior}
For $\beta=\beta_n$ such that $\sqrt{n}\beta_n \equiv n^{\delta}{\beta^*}, \; \delta \in (0,1/2]$, we have
\[\lim_{n\to \infty} \frac{1}{n^{2\delta}} \left\{\log\pi_S(\beta_n\lvert\bar{y}_n)- \log\widetilde{\pi}_S(\beta_n\lvert \bar{y}_n)\right\} \to 0\]
as $n\to \infty$, whenever $n^{-2\delta}\psi_{n^{-\delta}}(x)$ converges pointwise to 
$$
I_{\mathcal{K}}(x) =\begin{cases} 
      0  & \text{ if } x\in \mathcal{K} \\
      \infty  & \text{ otherwise}\\
         \end{cases}.
$$
Here $\mathcal{K}=(0,\infty)$, and $\log\pi_S(\beta_n\lvert\bar{y}_n)$, $\log\widetilde{\pi}_S(\beta_n\lvert \bar{y}_n)$ are given by \eqref{eq:log-post:univar} and \eqref{eq:log-post:approx:univar}, respectively. 
\end{corollary}
In Section \ref{sec:estimation} we discuss estimation with an approximate maximum a-posteriori (MAP) statistic, here an approximation to the maximizer of $\log\pi_S(\beta\lvert\bar{y}_n)$, and focus on establishing theoretical guarantees relating the approximate MAP (using our barrier approximation to the adjustment factor) to the exact MAP. 
Specifically, if we use a constant prior, the MAP is equivalent to the maximum-likelihood estimate (MLE). 
Without delving into the technical details that will follow, Figure \ref{fig:mle-univar} shows how the approximate MLE compares to the exact MLE (and to the unadjusted MLE) in the univariate gaussian setting of this section.

\section{Selection with the Lasso}\label{sec:lasso}
We now turn to the case of primary interest, specializing the general framework of Section \ref{sec:setup} to selection with the Lasso in the linear model. 
It is important to note again that our methodology accommodates a rich class of selection protocols, of which Lasso is no more than a special case: in the Appendix we show how marginal-screening and logistic Lasso for a binary response, can be reduced to fit our framework. 

\smallskip
From now on, unless specified otherwise, we consider
\begin{equation}\label{eq:lasso-sel}
\widehat{E} = \{j: \widehat{\beta}_j^{\lambda}\neq 0\},
\end{equation}
where $\widehat{\beta}^{\lambda}$ is the solution to 
\begin{equation}
\label{carved:lasso}
\underset{\beta\in\RR^p}{\text{minimize}} \;\;\dfrac{1}{2\sqrt{n}\rho} \Big\|\vec{y}^{\mathcal{S}} - \bX^{\mathcal{S}}\beta\Big\|_2^2 +\lambda \|\beta\|_1,
\end{equation}
and $\rho= n_1/n$. 
We emphasize that $\widehat{\beta}^{\lambda} = \widehat{\beta}^{\lambda}_{\calS}$ is obtained from only the original data $(\vec{y}^{\mathcal{S}}, \bX^{\mathcal{S}})$, but we suppress the subscript throughout. 
We also denote by $\widehat{\beta}_E^{\lambda}$ the vector in $\RR^{|E|}$ consisting of only the nonzero coordinates of $\widehat{\beta}^{\lambda}$. 
While the general prescription in Section \ref{sec:setup} would now entail conditioning on
the event $\widehat{E} = E$, as in \citet{lee2016exact} we in fact condition on the more specific event 
\begin{equation}\label{eq:lasso-sel-signs}
\left( \widehat{E},\widehat{S}^{\widehat{E}} \right) = \left(E,s^E \right),
\end{equation}
where $\widehat{S}^{\widehat{E}} = \text{sgn}(\widehat{\beta}_E^{\lambda})$ is the vector of signs of $\widehat{\beta}_E^{\lambda}$. 
This refinement is important for obtaining a convex truncation region, crucial for our methods to be applicable. 
As in the general setting, the object of inference is $\beta^{E'}$, with $E'\subseteq\{1,...,p\}$ chosen upon observing \eqref{eq:lasso-sel-signs}. 

\smallskip
To fit the general framework of Section \ref{sec:setup}, in the previous section we re-expressed the selection event, originally involving $\widehat{\beta}^\calS = \bar{Y}^{\mathcal{S}}$, as a sum involving $\widehat{\beta}_n = \bar{Y}_n$ and an independent variable $W_n$. 
Furthermore, the distribution of $W_n$ was fully specified, in particular it did not depend on the unknown parameter $\beta$. 
We will now work out an analogous asymtotic representation for the Lasso selection event \eqref{eq:lasso-sel-signs}. 
Thus, let
\begin{equation}\label{T_n}
\sqrt{n}\vec{T}_n := 
\begin{pmatrix}
\sqrt{n}\widehat{\beta}^{E'}\\\sqrt{n} \vec{N}_{-E'}
\end{pmatrix} := 
\begin{pmatrix}
\sqrt{n}\widehat{\beta}^{E'} \\ \dfrac{1}{\sqrt{n}} \bX_{-E'}^T\left(\vec{y} - \bX_{E'}\widehat{\beta}^{E'}\right)  
\end{pmatrix} 
\end{equation}
where $\bX_{-E'}$ is the matrix obtained from $\bX$ by deleting the columns with indices in $E'$. 
Define also 
\begin{equation}
\label{carved:randomization}
\begin{aligned}
\vec{\Omega}_n = \sqrt{n} W_n &= \dfrac{\partial}{\partial \beta} \left\{ - \dfrac{1}{2\sqrt{n}\rho} \Big\|\vec{y}^{\mathcal{S}} - \bX^{\mathcal{S}}\beta\Big\|_2^2+ \dfrac{1}{2\sqrt{n}} \Big\|\vec{y}- \bX\beta\Big\|_2^2\right\}\Big\lvert_{\beta = \widehat{\beta}^\lambda}. 
\end{aligned}
\end{equation}
As observed in \cite{markovic2016bootstrap}, $\vec{\Omega}_n$ can be treated asymptotically as a randomization term. 
We verify this in the following proposition.

\begin{proposition}
\label{asymptotic:distribution}

Denote ${\mu} := (\beta^{E'}, {0})^T \in \mathbb{R}^{p}$ where $\beta^{E'} \in \mathbb{R}^{|E' |}$. 
Define also the matrices
\begin{align*}
\mathbf{{Q}} := \mathbb{E}_{P}(\bX_{E'}^T \bX_{E'}/n), \ \ \mathbf{N} := \left\{\mathbb{E}_{P}(\bX_{-E'}^T \bX_{-E'}/n)-  \mathbb{E}_{P}(\bX_{-E'}^T \bX_{E'}/n)\mathbf{Q}^{-1} \mathbb{E}_{P}(\bX_{-E'}^T \bX_{E'}/n)^T\right\}^{-1}.
\end{align*}
\noindent Then, under the modeling assumption \eqref{eq:model},
the distribution of 
\begin{equation*}
\begin{pmatrix}
\sqrt{n}\left({T}_n - {\mu} \right)\\
\Omega_n
\end{pmatrix}
\in \RR^{2p}
\end{equation*}
is asymptotically normal with mean zero and covariance
$$
\boldsymbol{\Sigma} = 
\begin{bmatrix} \boldsymbol{\Sigma}_{P} & 0 \\ 0 & \boldsymbol{\Sigma}_{\mathcal{G}} \end{bmatrix},
$$
where $\boldsymbol{\Sigma}_{P}= \sigma^2 \begin{bmatrix} \mathbf{Q}^{-1} & 0 \\ 0 & \mathbf{N}^{-1}\end{bmatrix}$ and $\boldsymbol{\Sigma}_{\mathcal{G}}$ are $p\times p$ matrices. 
\end{proposition}

To fit the framework of Section \ref{sec:setup}, we now show that the event \eqref{eq:lasso-sel-signs} can be asymptotically written as in \eqref{constraints:carved}. 
This is established in the following proposition. 
\begin{proposition}
\label{K.K.T.:selection}
The event $\left( \widehat{E},\widehat{S}^E \right) = \left( E, s^E \right)$ is equivalent to
\begin{equation}
\mathbf{A}_E \sqrt{n}T_n + \mathbf{B}_E\Omega_n + o_p(1) < b_E,
\end{equation}
where
\begin{equation*}
\mathbf{A}_E= \begin{bmatrix} -\text{diag}(s^E)(\mathbf{Q}^E)^{-1}\mathbf{P}^{E, E'} & -\text{diag}(s^E)(\mathbf{Q}^{E})^{-1}\mathcal{I}^{E, E'} \\  \mathbf{F}^{E, E'} - \mathbf{C}^E(\mathbf{Q}^E)^{-1}\mathbf{P}^{E, E'} & \mathcal{J}^{E, E'} - \mathbf{C}^E(\mathbf{Q}^E)^{-1}\mathcal{I}^{E, E'}\\  
-\mathbf{F}^{E, E'} + \mathbf{C}^E(\mathbf{Q}^E)^{-1}\mathbf{P}^{E, E'} & -\mathcal{J}^{E, E'} + \mathbf{C}^E(\mathbf{Q}^E)^{-1}\mathcal{I}^{E, E'}\end{bmatrix},
\end{equation*} 
\medskip
\begin{equation*}
\mathbf{B}_E= \begin{bmatrix} -\text{diag}(s^E)(\mathbf{Q}^{E})^{-1} & \mathbf{0} \\-\mathbf{C}^{E}(\mathbf{Q}^{E})^{-1} & \mathbf{I} \\\mathbf{C}^{E}(\mathbf{Q}^{E})^{-1} & -\mathbf{I}\end{bmatrix}, \;\; b_E = \lambda \begin{pmatrix}- \text{diag}(s^E) (\bQ^{E})^{-1} s^E\\ \mathbf{1} -\mathbf{C}^{E}(\mathbf{Q}^{E})^{-1} s^E \\ \mathbf{1} + \mathbf{C}^{E}(\mathbf{Q}^{E})^{-1} s^E\end{pmatrix};
\end{equation*}
\medskip
$$\mathbf{P}^{E, E'} = \mathbb{E}_{P}(\bX_{E}^T \bX_{E'}/n), \ \ \ \mathbf{F}^{E, E'}= \mathbb{E}_{P}(\bX_{-E}^T \bX_{E'}/n),$$ 
$$\mathbf{Q}^E = \mathbb{E}_{P}(\bX_{E}^T \bX_{E}/n), \ \ \ \ \mathbf{C}^E = \mathbb{E}_{P}(\bX_{-E}^T \bX_{E}/n);$$ 
$\mathcal{I}^{E, E'} \in \mathbb{R}^{|E|\times (p-|E'|)}$ and $\mathcal{J}^{E, E'} \in \mathbb{R}^{(p-|E|)\times (p-|E'|)}$ are matrices with entries that are all zero except for the $(j,j)$ entries, that are defined as follows:
Suppose that $E = \{k_1,k_2,...,k_{|E|}\}$, and that $E^c = \{l_1,l_2,...,l_{p-|E|}\}$. 
Then 
$$
\mathcal{I}^{E, E'}_{j,j}  = 
\begin{cases}
1 & \text{ if } k_j \notin E'\\
0 & \text{otherwise}
\end{cases}, \ j=1,...,\text{min}(|E|,p-|E'|);$$
$$\mathcal{J}^{E, E'}_{j,j}  = 
\begin{cases}
1 & \text{ if } l_j \notin E'\\
0 & \text{otherwise}
\end{cases}, \ j=1,...,\text{min}(p-|E|,p-|E'|).
$$
\end{proposition}

\begin{remark} 
When $E'=E$, then note that $\mathcal{I}^{E, E}= \mathbf{0}$, $\mathcal{J}^{E, E} = \mathbf{I}$. 
In particular, the polyhedron in Proposition \ref{K.K.T.:selection} has
\begin{equation*}
\mathbf{A}_E= \begin{bmatrix} -\text{diag}(s^E) & \mathbf{0}\\ \mathbf{0} & \mathbf{I} \\ \mathbf{0} & -\mathbf{I}\end{bmatrix}, \; \; \mathbf{B}_E= \begin{bmatrix} -\text{diag}(s^E){\mathbf{Q}^E}^{-1} & \mathbf{0} \\-\mathbf{C}^E{\mathbf{Q}^E}^{-1} & \mathbf{I} \\\mathbf{C}^E{\mathbf{Q}^E}^{-1} & -\mathbf{I}\end{bmatrix}, \;\; b_E = \lambda \begin{pmatrix}- \text{diag}(s^E) {\bQ^E}^{-1} s^E\\ 1 -\mathbf{C}^E{\mathbf{Q}^E}^{-1} s^E \\ 1 + \mathbf{C}^E{\mathbf{Q}^E}^{-1} s^E\end{pmatrix}.
\end{equation*}
Lemma 14 in \cite{tian2018selective} establishes asymptotic normality for $\sqrt{n}(T_n-\mu)$ (compare to Proposition \ref{asymptotic:distribution}), and gives the approximate polyhedron for the specific case $E'=E$, under the Lasso objective with additive heavy-tailed randomization. 
\end{remark}

\section{Approximations to the selection-adjusted likelihood}\label{sec:approximation}

As indicated before, to be able to put the Bayesian machinery to work, an immediate challenge that presents itself is evaluating the selection-adjusted likelihood---more specifically, the adjustment factor \eqref{eq:adj-factor}---as a function of $\beta^{E'}$. 
Since \eqref{eq:adj-factor} is in general unmanageable, we will propose to replace it with a computationally tractable approximation. 
In this section, we use $\beta^{E'}_n$ instead of $\beta^{E'}$ and represent $\mu$ defined in Proposition \ref{asymptotic:distribution} by $\mu_n$ to make explicit the dependence of the parameter vector in \eqref{E':model} on $n$. 
Furthermore, we assume without loss of generality that the variance term $\sigma^2=1$. 
The case where $\sigma^2$ is unknown is discussed after stating Theorem \ref{thm:approx}. 

\smallskip
Theorem \ref{thm:approx} below is a moderate deviations-type of result allowing to obtain the limiting value of the probability of a polyhedral region, and provides motivation for our approximation.  
Before stating the theorem, we recall the following representation from the proof of Proposition \ref{asymptotic:distribution}, 
\begin{equation*}
\begin{aligned}
\begin{pmatrix}\sqrt{n}T_n \\ \Omega_n \end{pmatrix}- \sqrt{n}\begin{pmatrix} \mu_n \\ 0 \end{pmatrix} 
&= \sqrt{n}\bar{Z}_n + E_n,
\end{aligned}
\end{equation*}
where $E_n = o_p(1)$ and $\mu_n = (\beta_n^{E'}, 0)^T \in \mathbb{R}^{p}$, and $\bar{Z}_n$ is a mean statistic based on $n$ i.i.d. centered observations in our framework.

\begin{theorem}
\label{thm:approx}
Suppose that \eqref{eq:model} holds for some vector $\beta_n^{E'}$ satisfying $\sqrt{n}\beta_n^{E'} = n^{\delta}{\beta^*}\in {\mathbb{R}^{|E'|}}$, where $\beta^*$ does not depend on $n$, and where $\delta \in (0, 1/2)$. 
Defining $W_n$ through $\Omega_n := \sqrt{n}W_n$, we assume that
 \begin{equation}
  \label{moment:condition}
\mathbb{E}_P\left[\exp(\lambda_0\| {X}(Y-  {X_{E'}}^T\beta_n^{E'})\|\right],  \mathbb{E}_P\left[\exp(\eta_0\| {X}(Y-  {X_E}^T\mathbb{E}[\widehat\beta_E^{\lambda}])\|)\right] <\infty
 \end{equation}
for some $\lambda_0, \eta_0 >0$, where $(X,Y) = (X_1,...,X_p,Y) \sim P$. 
 We also assume that
 \begin{equation} \label{error:condition}
 n^{-2\delta}\cdot \log \mathbb{P}[n^{-\delta}\| E_n \| > \epsilon ] =- \infty \ \ \ \ \ \ \ \text{ for every } \epsilon >0,
 \end{equation}
 and that
\begin{equation} \label{b:condition}
\Scale[0.88]{\displaystyle\lim_{n\to \infty} \dfrac{1}{n^{2\delta}} \left\{ \log \mathbb{P}(\mathbf{A}_E \sqrt{n}T_n + \mathbf{B}_E\sqrt{n}W_n +o_p(1) < b_E\lvert \beta^{E'}_n) - \log \mathbb{P}(\mathbf{A}_E \sqrt{n}T_n + \mathbf{B}_E\sqrt{n}W_n < 0 \lvert \beta^{E'}_n)\right\} = 0.}
\end{equation}
Then, denoting $\mathcal{H}_n =\{ (b, \eta, w): \mathbf{A}_E(b, \eta)^T+ \mathbf{B}_E w< n^{-\delta}b_E\}$, we have
\begin{equation}
\label{eq:approx}
\begin{aligned}
& \displaystyle\lim_{n\to \infty} \dfrac{1}{n^{2\delta}} \log \mathbb{P}(\mathbf{A}_E \sqrt{n}T_n + \mathbf{B}_E\sqrt{n}W_n < b_E\lvert \beta^{E'}_n) \\
& \;\;\;\;\;\;\;\;\;+ \displaystyle\inf_{(b, \eta, w) \in \mathcal{H}_n} \dfrac{(b- {\beta^*})^T \bQ (b- {\beta^*})}{2} + \dfrac{\eta^T \mathbf{N} \eta}{2} +\dfrac{w^T \boldsymbol{\Sigma}_{\mathcal{G}}^{-1} w}{2} = 0,
\end{aligned}
\end{equation}
where the optimization variables $(b, \eta, w) \in \RR^{E'} \times \RR^{p-|E'|} \times \RR^{p}$, and matrices $\mathbf{Q}, \mathbf{N}$ and $\boldsymbol{\Sigma}_{\mathcal{G}}$ are defined in Proposition \ref{asymptotic:distribution}.
\end{theorem}

Theorem \ref{thm:approx} suggests using the negative of the second term in \eqref{eq:approx}, scaled by $n^{2\delta}$, as an approximation in computing the log of the probability in  \eqref{eq:adj-factor}, whenever $\sqrt{n}(T_n, W_n)$ satisfies a central limit property and has exponential moments. 
In fact, the polyhedron can be more generally replaced with any open and convex subset $\mathcal{K}\subset \mathbb{R}^{2p}$. 
Furthermore, if $\sigma^2$ is unknown, we can put $$\mathcal{K}=\{(t, w): \mathbf{A}_E\sqrt{n}t +  \mathbf{B}_E\sqrt{n}w \leq b_E/\sigma\}$$ and compute the probability with respect to the law of $(\sqrt{n}T_n/\sigma, \sqrt{n}W_n/\sigma)^T$ where $\mu_n:=(\beta^{E'}_n/\sigma, 0)^T$. 
In this case, we note that \eqref{eq:approx} becomes 
\begin{equation}
\label{approx:sigma}
\begin{aligned}
& \displaystyle\lim_{n\to \infty} \dfrac{1}{n^{2\delta}} \log \mathbb{P}(\mathbf{A}_E (\sqrt{n}T_n/\sigma) + \mathbf{B}_E(\sqrt{n}W_n/\sigma) < b_E/\sigma\; \lvert \; \beta^{E'}_n, \sigma) \\
& \;\;\;\;\;\;\;\;\;+ (\sigma^2)^{-1}\cdot \displaystyle\inf_{(b, \eta, w) \in \mathcal{H}_n} \dfrac{(b- {\beta^*})^T \bQ (b- {\beta^*})}{2} + \dfrac{\eta^T \mathbf{N} \eta}{2} +\dfrac{w^T \boldsymbol{\Sigma}_{\mathcal{G}}^{-1} w}{2} = 0,
\end{aligned}
\end{equation}
where $\mathcal{H}_n$ is defined in Theorem \ref{thm:approx}. 
Hence our approximation readily accommodates the case of unknown $\sigma^2$. 
In Section \ref{sec:computation} we use this to outline a scheme for posterior sampling when imposing a joint prior on $(\beta^{E'}, \sigma)^T$.

\begin{remark}
Assumption \eqref{moment:condition} is a condition on the existence of an exponential moment. This is a necessary condition for a mean statistic based on i.i.d. observations, in our case $\bar{Z}_n$,  to satisfy a moderate deviation principle; see \cite{arcones2002moderate, eichelsbacher2003moderate}. 
Assumption \eqref{error:condition} is necessary to allow a moderate deviation principle for the statistic $(\sqrt{n} (T_n-\mu_n), \Omega_n)^T$, which is obtained by adding to the centered mean statistic $\sqrt{n}\bar{Z}_n$ a remainder term that is converging in probability to $0$. This assumption is typically needed to apply moderate deviations to statistics such as M-estimators; see \citet{arcones2002moderate}.
Moderate deviations approximations are used to compute probabilities of the form 
$$
\mathbb{P}(\sqrt{n} (T_n, W_n)^T/ n^{\delta} \in \mathcal{K}).
$$
If we take $\mathcal{K} =\left\{(t, w): \;\begin{bmatrix}\mathbf{A}_E & \mathbf{B}_E\end{bmatrix} (t, w)^T <0\right\}$, then assumption \eqref{b:condition} allows us to approximate 
$$ 
n^{-2\delta}\cdot\log \mathbb{P}(\begin{bmatrix}\mathbf{A}_E & \mathbf{B}_E\end{bmatrix}\sqrt{n} (T_n, W_n)^T/ n^{\delta} + n^{-\delta} o_p(1)   < n^{-\delta} b_E\lvert \beta^{E'}_n)
$$ 
by 
$$
n^{-2\delta}\cdot \log \mathbb{P}(\begin{bmatrix}\mathbf{A}_E & \mathbf{B}_E\end{bmatrix}\sqrt{n} (T_n, W_n)^T/ n^{\delta}   <0\lvert \beta^{E'}_n).
$$
\end{remark}

\medskip
From now on, we write the selection event as $\{\mathbf{A}_E \sqrt{n}T_n + \mathbf{B}_E\sqrt{n}W_n < b_E\}$, ignoring the $o_p(1)$ term. 
In practice, to obtain an approximation for $\log \mathbb{P}(\mathbf{A}_E \sqrt{n}T_n + \mathbf{B}_E\sqrt{n}W_n < b_E\lvert \beta^{E'}_n)$, we solve an {\it unconstrained} version of the optimization problem in \eqref{eq:approx}, as we describe next. 

First we introduce a change of variable for the optimization arguments, which simplifies the constraints in the left hand side of \eqref{eq:approx} considerably. 

\begin{proposition}
\label{eq:change-variables-optimization}
For $0<\delta<1/2$ and for a sequence $\beta^{E'}_n$ such that $\sqrt{n}\vec{\beta}^{E'}_n = n^{\delta}\beta^*$, 
define a change of variable $\omega\mapsto o$ through
\begin{equation}
\label{change:variables}
 n^{\delta}w = n^{\delta} P_E \begin{pmatrix} b \\ \eta \end{pmatrix} + n^{\delta} Q_E o + r_E,
\end{equation}
where 
\[
P_E = -\begin{bmatrix} \mathbf{P}^{E, E'} & \mathcal{I}^{E, E'} \\ \mathbf{F}^{E, E'} & \mathcal{J}^{E, E'}\end{bmatrix},\ \ \ \ Q_E = \begin{bmatrix} \mathbf{Q}^{E} & 0 \\ \mathbf{C}^{E} & I\end{bmatrix},\ \ \ \ r_E =  \begin{pmatrix}\lambda s^E\\ 0\end{pmatrix}.
\]
Then, minimizing $\; n^{2\delta}\cdot \displaystyle\inf_{(b, \eta, w) \in \mathcal{H}_n} \left\{(b- \beta^*)^T \bQ (b- \beta^*)/2 + \eta^T \mathbf{N} \eta/2 +w^T \boldsymbol{\Sigma}_{\mathcal{G}}^{-1} w/2\right\}$
is equivalent to minimizing
\begin{align*}
& n^{2\delta}\cdot\inf_{\{(b, \eta, o) \in \RR^{2p}: o \in \mathcal{O}_n\}} \Big\{(b- \beta^*)^T \bQ (b- \beta^*)/2 + \eta^T \mathbf{N} \eta/2\\ 
&\Scale[0.9]{\;\;\;\;\;\;\;\;\;\;\;\;\;\;\;\;\;\;\;\;\;\;\;+ \left(P_E \begin{pmatrix} b \\ \eta \end{pmatrix} + Q_E o + r_E/n^{\delta}\right)^T \mathbf{\Sigma}_{\mathcal{G}}^{-1} \left(P_E \begin{pmatrix} b \\ \eta \end{pmatrix} + Q_E o + r_E/n^{\delta}\right)\Big/2\Bigg\}}, \numberthis
\label{opt:variable:change}
\end{align*}
where the constraints in the two objectives are given, respectively, by
\[
\Scale[0.9]{\mathcal{H}_n=\left\{(b, \eta, w)\in \mathbb{R}^{2p}: \mathbf{A}_E \begin{pmatrix} b\\ \eta \end{pmatrix} + \mathbf{B}_E w \leq  n^{-\delta} b_E\right\}; \mathcal{O}_n=\{o\in \mathbb{R}^p: \textnormal{sgn}(n^{\delta}o_{E}) = s^E,\;\|n^{\delta}o_{-E}\|_{\infty} \leq \lambda\}.}
\]
\end{proposition}

Note that in the new form of the optimization problem, the variables $b$ and $\eta$ are unconstrained, while $o$ has the simple constraint given above. 
For the parametrization of $\beta^{E'}_n$ considered, we obtain a more flexible form of the optimization problem as
\begin{equation}
\label{eq:optimization-general}
\begin{aligned}
&\Scale[0.95]{\log \widetilde{\mathbb{P}}(\mathbf{A}_E \sqrt{n}T_n + \mathbf{B}_E\sqrt{n}W_n < b_E\lvert \beta^{E'}_n) = -n^{2\delta}\cdot\inf_{(b, \eta, o) \in \RR^{2p}} \Big\{(b-{\beta^*})^T \bQ (b- {\beta^*})/2 + \eta^T \mathbf{N} \eta/2} \\
&\Scale[0.95]{\;\;\;\;\;\;+ \left(P_E \begin{pmatrix} b & \eta \end{pmatrix}^T  + Q_E o + r_E/n^{\delta}\right)^T  \mathbf{\Sigma}_{\mathcal{G}}^{-1} \left(P_E \begin{pmatrix} b & \eta \end{pmatrix}^T  + Q_E o + r_E/n^{\delta}\right)\Big/2+{n^{-2\delta}} \psi_{n^{-\delta}}(o_E, o_{-E})\Big\}},
\end{aligned}
\end{equation}
where $\psi_s(o)=\psi_s(o_E, o_{-E})$ is some penalty function corresponding to the set $\mathcal{O}_n$, with a scaling factor $s$. 
This specializes to \eqref{opt:variable:change} by taking $\psi_s(o_E, o_{-E})$ to be the characteristic function
\[I_\mathcal{O}(o_E, o_{-E}) =\begin{cases} 
      0  & \text{ if } o\in \mathcal{O} \\
      \infty  & \text{ otherwise}\\
         \end{cases}.\]
In the next and crucial step, instead of the characteristic function that restricts the optimizing variables to the set $\mathcal{O}_n$, we use a smoother nonnegative penalty function: 
we replace $I_{\mathcal{O}_n}(o_E, o_{-E})$ with a suitable ``barrier" penalty function $\psi_s$  that reflects preference for values of $o$ farther away from the boundary and inside the constraint region $\mathcal{O}_n$, by taking on smaller values for such $o$. 
Specifically, we use $\psi_{n^{-\delta}}$ defined by
\begin{equation}
\label{eq:barrier}
\begin{aligned}
&\psi_{n^{-\delta}}(o)\equiv\psi_{n^{-\delta}}(o_E, o_{-E})=\\
&\Scale[0.9]{=\left(\displaystyle\sum_{i=1}^{E} \log\left(1+ \dfrac{1}{s_{i,E} n^{\delta}o_{i,E}}\right) + \sum_{i=1}^{p-|E|} \log\left(1+ \dfrac{1}{\lambda_{i,-E} - n^{\delta}o_{i,-E}}\right)+ \log\left(1+ \dfrac{1}{\lambda_{i,-E} + n^{\delta}o_{i,-E}}\right)\right)}.
\end{aligned}
\end{equation}
In line with Remark \ref{theoretical:delta} for the univariate thresholding example, we remark that $\delta$ in the parameterization of $\beta_n^{E'}$ is only a theoretical construct. 

\smallskip
This ultimately leads to an approximation to the log of the selection-adjusted posterior \eqref{eq:sel-adj-gen}, as
\begin{equation}
\label{eq:approx-posterior-barrier}
\begin{aligned}
\log {\widetilde{\pi}}_{S}(\beta^{E'}_n\lvert\widehat{\beta}^{E'}) = \Scale[1.]{ \log \pi(\beta^{E'}_n)} &\Scale[1.]{-n(\widehat{\beta}^{E'}- \beta^{E'}_n)^T \bQ ( \widehat{\beta}^{E'}- \beta^{E'}_n)/2} \\
&\Scale[1.]{- \log \widetilde{\mathbb{P}}(\mathbf{A}_E \sqrt{n}T_n + \mathbf{B}_E\sqrt{n}W_n < b_E\lvert \beta^{E'}_n)},
\end{aligned}
\end{equation}
where we use \eqref{eq:barrier} in \eqref{eq:optimization-general}. 
To be more explicit, the last term in \eqref{eq:approx-posterior-barrier} can be obtained from \eqref{eq:optimization-general} as
\begin{equation*}
\small{
\begin{aligned}
&-n\cdot\inf_{(b', \eta', o') \in \RR^{2p}} \Big\{(b'-{\beta^{E'}_n})^T \bQ (b'- {\beta^{E'}_n})/2 + \eta^{'T} \mathbf{N} \eta'/2 \\
&\Scale[0.95]{\;\;\;\;\;\;+ \left(P_E \begin{pmatrix} b' & \eta' \end{pmatrix}^T  + Q_E o' + r_E/\sqrt{n}\right)^T  \mathbf{\Sigma}_{\mathcal{G}}^{-1} \left(P_E \begin{pmatrix} b' & \eta' \end{pmatrix}^T  + Q_E o' + r_E/\sqrt{n}\right)\Big/2+ \psi_{n^{-1/2}}(o'_E, o'_{-E})/n\Big\}}
\end{aligned}
}
\end{equation*}
by substituting $\sqrt{n}b' = n^{\delta}b, \sqrt{n}\eta' = n^{\delta}\eta, \sqrt{n}o' = n^{\delta}o$.

\medskip
We now present some simulation results which demonstrate the effectiveness of our methods. 
\begin{example}\label{ex:regression}
In each of $50$ rounds, we draw a $n\times p$ matrix $\bX$ with $n=500, p=100$ such that the rows $\vec{x}^\pari\sim N_p(0, \mathbf{\Sigma}), i=1,2,...,500,$ where the $(j,k)$-th entry of $\mathbf{\Sigma}$ equals $0.20^{|j-k|}$. 
We then draw a pair $(\beta, \vec{y})$, where the components of $\beta \in \real^{100\times 1}$ are i.i.d.~from
\begin{equation}
 \label{eq:true-prior}
0.9\cdot \calN(0,0.1) + 0.1 \cdot \calN(0, V),
\end{equation}
a mixture of two zero-mean normal distributions, one with small variance $0.1$ and the other with larger variance $V$, and 
$\vec{y}| \bX, \beta \sim N_n(\bX\beta, \mathbf{I})$. 
The variance $V\in \{5, 3, 2\}$, roughly corresponding to signal-to-noise ratio $0.50, 0.25, 0.10$, respectively. 
In the selection stage, we select at random a subset $\mathcal{S}\subset \{1,...,n\}$ of size $|\mathcal{S}|=n/2=250$, and denote by $(\vec{y}^\mathcal{S},\bX^\mathcal{S})$ the corresponding data. 
For a theoretical value of the tuning parameter, $\lambda = \EE[\|\bX^T \Psi\|_{\infty}], \;\Psi \sim N_n(0, \mathbf{I})$ \citep[as proposed in][]{negahban2009unified}, denote by $E \subset \{1,...,100\}$ the set corresponding to the nonzero Lasso estimates, obtained by solving
\[
\argmin_{\beta} \frac{1}{2\rho} \|\vec{y}^{\mathcal{S}}- \bX^{\mathcal{S}}\beta\|^2 + \lambda \|\beta\|_1.
\]
In the inference stage we have access to the entire data $(X^\pari,Y^\pari),\ i=1,...,n$. 
We give inference assuming that the model is 
$$
\vec{y}|\bX\sim \calN(\bX_{E}\beta^{E},\sigma^2 \bI),
$$
in other words, we take $E' = E$ in \eqref{E':model}; of course, this model is (usually) misspecified, in the sense that $E\neq \{j:\beta_j\neq 0\}$. 
Ancillarity now entails conditioning on $\bX$  when giving inference for $\beta^{E'}$. 
Four different methods for inference are compared:

\medskip
\noindent {\bf Unadjusted (Naive)}. 
Bayesian inference for $\beta^{E}$ using a noninformative prior $\pi(\beta^{E})\propto 1$ and the unadjusted likelihood $\vec{y}\lvert \bX \sim \calN_n(\bX_E\beta^{E}, \sigma^2 \mathbf{I})$

\smallskip
\noindent {\bf Split}. 
Bayesian inference using only the confirmatory (held-out) data: a noninformative prior $\pi(\beta^{E})\propto 1$ is prepended to the unadjusted likelihood $\vec{y}^{\mathcal{S}^c}\lvert \bX^{\mathcal{S}^c} \sim \calN_n(\bX^{\mathcal{S}^c}_E\beta^{E}, \sigma^2 \mathbf{I})$

\smallskip
\noindent {\bf Carving}. 
Bayesian inference for $\beta^{E}$ using a noninformative prior $\pi(\beta^{E})\propto 1$ and the approximate selection-adjusted likelihood, 
\begin{equation}\label{eq:approx-lik}
\widetilde{\pi}_S(\beta^{E}|\widehat{\beta}^{E}) \propto \dfrac{\exp(-n(\widehat{\beta}^{E} -\beta^{E})^T \bQ(\widehat{\beta}^{E} -\beta^{E})/2\sigma^2)}{\widetilde{\mathbb{P}}(\mathbf{A}_E \sqrt{n}T_n + \mathbf{B}_E\sqrt{n}W_n < b_E\lvert \beta^{E})}
\end{equation}
where the denominator on the right-hand side is given by \eqref{eq:optimization-general} with the penalty defined in \eqref{eq:barrier}.

\smallskip
\noindent {\bf Lee et al.}. 
We use the Selective Inference package in R to obtain estimates with the methods of \citet{lee2016exact}. 
Because there is no implementation for carving, the entire data is used for selection, that is, $\widehat{E}$ is the subset of indices corresponding to the nonzero elements of
\[
\argmin_{\beta} \frac{1}{2} \|\vec{y}- \bX\beta\|^2 + \lambda \|\beta\|_1.
\]
Exact inference conditional on $\widehat{E}$ and the corresponding signs is given coordinate-wise using the methods of \citet{lee2016exact}, which correct for selection.

\medskip
We compare the methods above on the following criteria: (i) we use each method to construct (marginal) $90\%$ interval estimates, and calculate the average (over simulation rounds) proportion of covering intervals (this is reported as $1-FCR$ in the tables below); (ii) lengths of constructed CIs; and (iii) relative prediction risk, 
\[
\dfrac{(\widehat{\beta}- \beta)^T (\bX^T \bX) (\widehat{\beta}- \beta)}{\beta^T (\bX^T \bX) \beta},
\]
where the `inactive' coordinates $\{j\notin E\}$ of $\beta$ and $\widehat{\beta}$ are set to zero, and the estimates $\widehat{\beta}$ are the posterior means for the first three methods, and the plain Lasso estimate for ``Lee et al.".

\begin{table}
  \begin{tabular}{lSSSSSSSSS}
    \toprule
    \multirow{2}{*}{} &
      \multicolumn{3}{c}{$\bf{V= 5\ (SNR=0.5)}$} &
      \multicolumn{3}{c}{$\bf{V= 3.5\ (SNR=0.25)}$} &
      \multicolumn{3}{c}{$\bf{V= 2\ (SNR=0.10)}$} \\
      & {1-FCR} & {Length} & {Rel. Risk} & {1-FCR} & {Length} & {Rel. Risk} & {1-FCR} & {Length} & {Rel. Risk} \\
      \midrule
    Carving & .884 & 3.92 & .21 & .887 & 4.04 & .39 & .905 & 4.22 & .66 \\
    Naive & .753 & 3.33 & .25 & .728 & 3.33 & .44 & .598 & 3.31 & .75 \\
    Split & .9 & 4.80 & .25 & .915 & 4.79 & .44 & .902 & 4.77 & .70 \\
    Lee et al & .913 & 8.38 & .29 & .939 & 9.58 & .46 & .863 & 10.73 & .83 \\
    \bottomrule
  \end{tabular}
  \caption{Summary of Example 1. Four methods are compared in three SNR regimes. 
  For Lee et al, the numbers displayed for length of CI are averages of constructed intervals that had finite length; for $V=5,3.5,2$, Lee et al produced infinitely long intervals for 1.4\%, 1.95\% and 4.73\% of the selected parameters, on average. }
\end{table}

\smallskip
We see that for all methods except the unadjusted, the coverage, as measured by one minus the false coverage rate (FCR), is roughly the nominal level $0.9$. 
In particular, the CIs constructed based on the proposed approximation to the selection-adjusted posterior have good coverage; we emphasize again that this is in spite of the fact the the assumed model is (usually) misspecified. 
Meanwhile, the length of the intervals for the proposed method (this is ``carving" in the tables), is much smaller than that for Lee et al intervals, which are sometimes infinitely long. More importantly, the carved intervals based on our method are smaller in length than the intervals for sample splitting. 
This matches our expectations, because left-over information is not utilized at all in sample-splitting. 
\end{example}

\section{Point estimates}\label{sec:estimation}
This section elaborates on point estimation after selection. 
In our Bayesian selection-adjusted framework, a natural estimate for the model parameters $\beta_n^{E'}$ is the selection-adjusted maximum a-posteriori (MAP) statistic, the maximizer in $\beta_n^{E'}$ of \eqref{eq:sel-adj-gen}. 
The selection-adjusted MLE obtains as a special case when we use a constant prior. 
In any case, this would again require the ability to evaluate the adjustment factor as a function of $\beta_n^{E'}$, hence we cannot implement the exact MAP. 
However, we may consider an {\it approximate} MAP by replacing the adjustment factor in the selection-adjusted likelihood with our workable approximation. 
Specifically, we will show that using the approximation given by \eqref{eq:optimization-general} with the choice of penalty \eqref{eq:barrier}, has various desirable and nontrivial features. 
Before we proceed, it is worth mentioning that point estimates can be obtained with the methods of \citet{lee2016exact} which rely on an exact truncated {univariate} normal distribution. Unsurprisingly, such estimates might be suboptimal because, stated informally, they do not utilize all of the information about the parameters in the sample and are tied exclusively to the saturated model. 
By contrast, the methods suggested below are based on the full (conditional) likelihood, defined for a fairly flexible class of models. 

\smallskip
We study first the approximate MAP in the univariate gaussian example of Section \ref{sec:univar}. 
Again, in that case we can implement the exact MAP. 
As before, we assume \eqref{model:univar} with $\beta = \beta_n$ such that $\sqrt{n} \beta_n = n^{\delta}{\beta^*}, 0<\delta\leq 1/2$. 
The selection event $\{\sqrt{n_1}\bar{Y}^{\mathcal{S}}>0\} \equiv \{\sqrt{n}(\bar{Y}_n + W_n)>0\}$ is of the form
$$
\sqrt{n}(\bar{Y}_n + W_n)/n^{\delta} \in \mathcal{K},
$$ 
where $\mathcal{K}$ is an interval on the real line. 
The selection-adjusted likelihood is
\begin{equation}
\label{true:carved:sel:likelihood}
\Scale[1]{L_S^{n}(\beta_n)  =  -n(\bar{y}_n -\beta_n)^2/2 - \log \mathbb{P}(\sqrt{n}(\bar{Y}_n + W_n)/n^{\delta} \in \mathcal{K}\lvert \beta_n)},
\end{equation} 
and our approximate version for the selection-adjusted (log-) likelihood is 
\begin{equation}
\label{carved:sel:likelihood}
 \Scale[1]{\widetilde{L}_S^{n}(\beta_n) = -{n(\bar{y}_n -\beta_n)^2}/{2} + n^{2\delta}\cdot \inf_{(z, w)\in \mathbb{R}^{2}}\left\{\dfrac{(z-{\beta^*})^2}{2} + \dfrac{\rho w^2}{2(1-\rho)} + \dfrac{1}{n^{2\delta}}\psi_{n^{-\delta}} (z+w)\right\}}.
 \end{equation}
First we prove consistency of the approximate selective MLE. 
As in Theorem \ref{carved:gaussian}, the statement in Theorem \ref{consistency:approximate:MLE} holds also for $\delta=1/2$; compare to Theorem \ref{consistency:gen}. 
We will note that consistency of the approximate selective MLE hinges on strong convexity of the negative logarithms of the approximate selection-adjusted likelihood, which leads to a contraction identity as in Lemma \ref{identity:convexity}. 
In Theorem \ref{consistency:approximate:MLE} we additionally use the fact that the variance of a gaussian random variable is smaller when restricted to a convex set \citep{kanter1977reduction}. 
Before stating the main theorem, in the following two lemmas we make crucial observations about the sequence of approximate likelihoods. 

\begin{lemma}[Strong convexity] \label{strong:convexity}
Let $\sqrt{n}\beta_n = n^{\delta}{\beta^*}, 0<\delta\leq 1/2$. The approximate selection-adjusted log-likelihood in \eqref{carved:sel:likelihood} equals
\[\widetilde{L}_S^{n}(\beta_n) = n\bar{y}_n\beta_n - n^{2\delta}\cdot \widetilde{C}_n({\beta^*}) - n\bar{y}_n^2/2,\]
where 
\begin{equation}
\label{carved:partition:function}
\widetilde{C}_n(\beta^*) =  (1-\rho)\cdot {\beta^*}^2/2 + \bar{H}_n^*(\rho\beta^*),
\end{equation}
and with $\bar{H}_n^*(.)$ denoting the convex conjugate of 
$\bar{H}_n(\bar{z}) = \rho\cdot \bar{z}^2/2+ {n^{-2\delta}}\cdot \psi_{n^{-\delta}}(\bar{z}).$
Moreover, $\widetilde{C}_n(\cdot)$ is strongly convex with index of convexity lower bounded by $(1-\rho)$.
\end{lemma}

\begin{lemma}
\label{identity:convexity} 
For a sequence $\beta_n$ as in Lemma \ref{strong:convexity}, the maximizer $\widehat{\beta}_S$ of \eqref{carved:sel:likelihood} satisfies
\[
n^{1-2\delta}(\widehat{\beta}_S -\beta_n)^2\leq \cfrac{1}{(1-\rho)^2}(n^{1/2-\delta}\bar{y}_n - \grad \widetilde{C}_n(\beta^*))^2, 
\]
where $\widetilde{C}_n$ is given by \eqref{carved:partition:function}. 
\end{lemma}

We are now ready to prove the consistency guarantees associated with our approximate selective MLE.

\begin{theorem}
\label{consistency:approximate:MLE}
Let $\mathcal{K}\subset \mathbb{R}$ be a convex set, and denote by $\widehat{\beta}_S$ the maximizer of \eqref{carved:sel:likelihood}. 
Then, for $0<\delta\leq 1/2$, 
\[
\mathbb{P}(n^{1/2-\delta}|\widehat{\beta}_S-\beta_n|>\epsilon \big|\; \sqrt{n}(\bar{Y}_n + W_n)/n^{\delta} \in \mathcal{K}) \longrightarrow 0
\]
as $n\to \infty$. 
\end{theorem}

Figure \ref{fig:mle-univar} shows the approximate MLE against the exact MLE. 
To complete the picture, we show that without randomization, the maximizer of the approximate truncated likelihood is not consistent. 
This further highlights the importance of holding out some samples at the selection stage, exclusively for inference. If selection was based on the entire data, then we could still
employ our approximation for the selection probability $\mathbb{P}(\sqrt{n}\bar{Y}_n/n^{\delta} \in \mathcal{K}\lvert \beta_n)$. In particular, when $\mathcal{K}=(0, \infty)$ 
(as considered before), the approximate log-selection probability takes the form 
$$-n \cdot  \inf_{z \in \mathbb{R}}\left\{\frac{(z-\beta_n)^2}{2} + \frac{1}{n}\log\left(1+\frac{1}{\sqrt{n}z}\right)\right\}.$$

\begin{theorem} 
\label{thm:inconsistency}
Let $\beta_n \equiv{\beta^*}<0$. 
Consider the approximate (non-randomized) selection-adjusted log likelihood,
\begin{equation} \label{soft:max:MLE}
\widetilde{L}_S^{n}(\beta_n) = -n(\bar{y}_n -\beta_n)^2/2 + n \cdot\inf_{z \in \mathbb{R}}\left\{\frac{(z-\beta_n)^2}{2} + \frac{1}{n}\log\left(1+\frac{1}{\sqrt{n}z}\right)\right\}.
\end{equation}
Then the maximizer $\widehat{\beta}_S$ of \eqref{soft:max:MLE} does not converge in probability to ${\beta^*}$ as $n \to \infty$. 
\end{theorem}

\begin{remark}
Theorem \ref{thm:inconsistency} is stated for the barrier approximation in \eqref{eq:psi:barrier} for clarity of exposition. 
More generally, the approximate non-randomized selective MLE is not consistent as long as the barrier function $\psi(\cdot)$ satisfies
\[
x \nabla \psi(x) \to \textit{constant} \ \ \ \ \text{ as } x \downarrow 0.
\]
\end{remark}

As a consequence of Theorem \ref{consistency:approximate:MLE}, we show next a form of consistency of the selection-adjusted posterior law with respect to a fixed prior.  This guarantees that the approximate selection-adjusted posterior concentrates around the true parameter $\beta_n$ in an asymptotic sense.
\begin{theorem}
\label{posterior:consistency:uni}
Consider a ball of radius $\delta$ around $\beta_n$, 
\[
\mathcal{B}(\beta_n, \delta) := \{b_n: |b_n - \beta_n| \leq \delta\},
\]
and suppose that $\pi$ is a prior which assigns nonzero probability to $\mathcal{B}(\beta_n, \delta)$ for any $\delta>0$.
Under the conditions in Theorem \ref{consistency:approximate:MLE}, for any $\epsilon>0$ we have
\[
\mathbb{P}(\Pi_S(\mathcal{B}^c( \beta_n, \delta)\lvert \bar{Y}_n)>\epsilon \lvert\;  \sqrt{n}(\bar{Y}_n + W_n)/n^{\delta} \in \mathcal{K}) \to 0
\]
as $n \to \infty$, where
\[
\Pi_S\left(\mathcal{B}^c( \beta_n, \delta)\lvert \bar{y}_n \right) := \dfrac{\underset{\mathcal{B}^c( \beta_n, \delta)}{\int}\pi(b_n)\cdot \exp(\widetilde{L}_S^{n}(b_n))db_n}{\int \pi(b_n)\cdot\exp(\widetilde{L}_S^{n}(b_n))db_n}
\]
is the posterior probability under the approximate truncated likelihood .\end{theorem}

\smallskip
Having established consistency for the univariate example, we now move on to the general case. 
Thus, consider the MAP estimator, given as the maximizer of \eqref{eq:approx-posterior-barrier}. 
For a constant prior this reduces to the approximate maximum-likelihood estimate,
\begin{equation}
\label{eq:approx-sel-mle}
\begin{aligned}
\widehat{\beta}^{E'}_S &= \argmin_{\beta^{E'}_n} \Big\{ 
n(\widehat{\beta}^{E'} -\beta^{E'}_n)' \bQ(\widehat{\beta}^{E'} -\beta^{E'}_n)/2 \\
&\;\;\;\;\;\;\;\;\;\;\;\;\;\;\;\;\;\;\;\;\;\;\;\;\;\;\;+ \log \widetilde{\mathbb{P}}(\mathbf{A}_E \sqrt{n}T_n + \mathbf{B}_E\sqrt{n}W_n < b_E\lvert \beta^{E'}_n)
\Big\},
\end{aligned}
\end{equation}
where in \eqref{eq:optimization-general} we use \eqref{eq:barrier}. 
Before presenting the main result for this section, we state two key lemmas. 
Recall the quantitites $P_E, Q_E, r_E$ defined in Proposition \ref{eq:change-variables-optimization}. 
Let $P_E^{E'}$ denote the matrix that consists of the columns in $P_E$ corresponding to the coordinates in $E'$. Similarly, $P_E^{-E'}$ denotes the matrix consisting of the remaining $p-|E'|$ columns.
\begin{lemma}
\label{log:partition:gen}
Under the parameterization $\sqrt{n}\beta^{E'}_n = n^{\delta}\beta^*$ with $\delta \in (0,1/2)$, the approximate log-partition function 
\[
n{\beta^{E'}_n}^T \bQ \beta^{E'}_n/2 + \log \widetilde{\mathbb{P}}(\mathbf{A}_E \sqrt{n}T_n + \mathbf{B}_E\sqrt{n}W_n < b_E\lvert \beta^{E'}_n) = n^{2\delta}\widetilde{C}_n(\bQ\beta^*),
\]
where we define
\[
\widetilde{C}_n(\bQ\beta^*) := {\beta^*}^T \bQ(\bQ+ {P_E^{E'}}^T \mathbf{\Sigma}_{\mathcal{G}}^{-1}P_E^{E'})^{-1}\bQ\beta^*/2 +  {h}^*\left(M_1\bQ\beta^*\right)+ {\beta^*}^T \bQ M_2
\]
with 
$$
M_1= -\begin{bmatrix} P_E^{-E'} & Q_E \end{bmatrix}^T\mathbf{\Sigma}_{\mathcal{G}}^{-1}{P_E^{E'}} (\bQ+ {P_E^{E'}}^T \mathbf{\Sigma}_{\mathcal{G}}^{-1}P_E^{E'})^{-1},$$
$$ M_2=- (\bQ+ {P_E^{E'}}^T \mathbf{\Sigma}_{\mathcal{G}}^{-1}P_E^{E'})^{-1}{P_E^{E'}}^T\mathbf{\Sigma}_{\mathcal{G}}^{-1}r_E/n^{\delta},$$
\noindent and where 
${h}^*(\cdot)$ is the convex conjugate of the function
\begin{equation*}
\begin{aligned}
h(\eta, o) &= \eta^T \mathbf{N}\eta/2+ \mathcal{L}(\eta, o)^{T}(\mathbf{\Sigma}_{\mathcal{G}}^{-1}- \mathbf{\Sigma}_{\mathcal{G}}^{-1}P_E^{E'} (\bQ +  {P_E^{E'}}^T\mathbf{ \Sigma}_{\mathcal{G}}^{-1}P_E^{E'})^{-1}{P_E^{E'}}^T\mathbf{\Sigma}_{\mathcal{G}}^{-1}) \mathcal{L}(\eta, o)/2 \\
&\;\;\;\;\;\;\;\;\;\;\;\;\;\;\;\;\;\;\;\;\;\;\;\;\;\;\;\;\;\;\;\;\;\;\;\;\;\;\;\;\;\;\;\;\;\;\;\;\;\;\;\;\;\;\;\;\;\;\;\;\;\;\;\;\;\;\;\;\;\;\;\;\;\;\;\;\;\;\;\;\;\;\;\;+ n^{-2\delta}\psi_{n^{-\delta}}(o_E, o_{-E})
\end{aligned}
\end{equation*}
with $\mathcal{L}(\eta, o)=P_E^{-E'} \eta + Q_E o + r_E/n^{\delta}$.
\end{lemma}

\begin{lemma}
\label{strong:convexity:gen}
Under the parameterization $\sqrt{n}\beta^{E'}_n = n^{\delta}\beta^*$ with $\delta \in (0,1/2)$ and when $\mathbf{X}_{E'}$ is of full column rank, the approximate log-partition function $n^{2\delta}\widetilde{C}_n(\bQ\beta^*)$,
corresponding to the approximate negative log-likelihood,
\[n(\widehat{\beta}^{E'} -\beta^{E'}_n)^T \bQ(\widehat{\beta}^{E'} -\beta^{E'}_n)/2 + \log \widetilde{\mathbb{P}}(\mathbf{A}_E \sqrt{n}T_n + \mathbf{B}_E\sqrt{n}W_n < b_E\lvert \beta^{E'}_n),\]
is strongly convex.
Furthermore, the index of strong convexity for $\widetilde{C}_n(\bQ\beta^*)$ is bounded below by $\lambda_{\text{min}}$, the smallest eigenvalue of 
\[
(\bQ+ {P_E^{E'}}^T \mathbf{\Sigma}_{\mathcal{G}}^{-1}P_E^{E'})^{-1}.
\]
\end{lemma}

Using Lemma \ref{strong:convexity:gen}, we are now able to prove a consistency result for the approximate selective MLE. 
We note here again that randomization is crucial for the theorem below to hold, as we saw already for the univariate example. 

\begin{theorem}
\label{consistency:gen}
When $\sqrt{n}\beta^{E'}_n = n^{\delta}\beta^*$ for $\delta \in (0,1/2)$, the approximate selective MLE given in \eqref{eq:approx-sel-mle} is $n^{1/2-\delta}$-consistent for $\beta_n^{E'}$ under the selection-adjusted law \eqref{eq:sel:lik}:
\[
\mathbb{P}(n^{1/2-\delta}\|\widehat{\beta}^{E'}_S -\beta^{E'}_n\|>\epsilon \big|\;  \mathbf{A}_E \sqrt{n}T_n + \mathbf{B}_E\sqrt{n}W_n < b_E) \longrightarrow 0
\]
as $n\to \infty$.
\end{theorem}

We move on to show a consistency result for the posterior distribution with respect to a fixed prior. 
We will need the following lemma, where we obtain finite-sample bounds on the log-likelihood ratios at the approximate selective-MLE and an arbitrary value for the parameter. 

\begin{lemma}
\label{likelihood:ratio:carved:gen}
Assume the conditions and parameterization in Lemma \ref{strong:convexity:gen}. 
Denote the logarithm of the approximate truncated likelihood by 
\[
\widetilde{L}_S^{n}(\beta^{E'}_n) = \sqrt{n}\widehat{\beta}^{E'}\bQ\sqrt{n}\beta^{E'}_n - n^{2\delta}\widetilde{C}_n (n^{1/2-\delta}\bQ\beta^{E'}_n )-  n\widehat{\beta}^{E'}\bQ\widehat{\beta}^{E'}/2,
\]
where $\widetilde{C}_n (n^{1/2-\delta}\bQ\beta^{E'}_n )=\widetilde{C}_n (\bQ\beta^*)$ is defined in Lemma \ref{strong:convexity:gen}. 
Then we have 
\begin{equation*}
\label{lik:ratios:carved}
-n\cdot (\widehat{\beta}^{E'}_S-\beta^{E'}_n)^T \bQ (\widehat{\beta}^{E'}_S-\beta^{E'}_n)/2 \leq \widetilde{L}_S^{n}(\beta^{E'}_n)-\widetilde{L}_S^{n}(\widehat{\beta}^{E'}_S) \leq -n\widetilde\lambda_{\text{min}} \cdot(\widehat{\beta}^{E'}_S-\beta^{E'}_n)^T (\widehat{\beta}^{E'}_S-\beta^{E'}_n)/2
\end{equation*}
where $\widetilde\lambda_{\text{min}}$ is the smallest eigenvalue of $\bQ (\bQ+ {P_E^{E'}}^T \mathbf{\Sigma}_{\mathcal{G}}^{-1}P_E^{E'})^{-1}\bQ$, and $\widehat{\beta}^{E'}_S$ is the approximate selective-MLE in \eqref{eq:approx-sel-mle}.
\end{lemma}

\begin{theorem}
\label{posterior:consistency:gen}
Assume the conditions in Theorem \ref{consistency:gen}. 
Consider a ball of radius $\delta$ around the truth $\beta^{E'}_n$, 
\[
\mathcal{B}(\beta^{E'}_n, \delta) \equiv \{b_n : \|b_n - \beta^{E'}_n\| \leq \delta\},
\]
and suppose that $\pi$ is a prior that assigns nonzero probability to $\mathcal{B}(\beta^{E'}_n, \delta)$ for any $\delta>0$. 
Then
\[
\mathbb{P}(\Pi_S(\mathcal{B}^c(\beta^{E'}_n, \delta)\lvert \widehat{\beta}^{E'})>\epsilon \big|\;  \mathbf{A}_E \sqrt{n}T_n + \mathbf{B}_E\sqrt{n}W_n < b_E) \to 0 \text{ as } n \to \infty
\]
for any $\epsilon>0$, where $\Pi_S(\cdot)$ denotes the posterior probability under the approximate truncated likelihood, 
\[
\Pi_S\left(\mathcal{B}^c(\beta^{E'}_n, \delta)\big| \widehat{\beta}^{E'} \right):= \dfrac{\underset{\mathcal{B}^c(\beta^{E'}_n, \delta)}{\int}\pi(b_n)\cdot \exp(\widetilde{L}_S^{n}(b_n))db_n}{\int \pi(b_n)\cdot\exp(\widetilde{L}_S^{n}(b_n))db_n},
\]
for $\widetilde{L}_S^{n}(b_n) = \sqrt{n}\widehat{\beta}^{E'}\bQ \sqrt{n}b_n - n^{2\delta}\widetilde{C}_n (n^{1/2-\delta}\bQ b_n) - n\widehat{\beta}^{E'}\bQ\widehat{\beta}^{E'}/2$. 
\end{theorem}

\medskip
The next result has implications for the computation of the approximate selective MLE. 

\begin{theorem}[Convexity of approximate MAP with general approximation]
\label{lem:map-convex}
Let $\pi(\cdot)$ be a log-concave prior. 
For any barrier function $\psi_s(\cdot)$, minimizing the negative of the approximate log-posterior based on the approximation in \eqref{eq:optimization-general},
\[
\-\log\pi(\beta^{E'}_n) + n(\widehat{\beta}^{E'} -\beta^{E'}_n)^T \bQ(\widehat{\beta}^{E'} -\beta^{E'}_n)/2 + \log \widetilde{\mathbb{P}}(\mathbf{A}_E \sqrt{n}T_n + \mathbf{B}_E\sqrt{n}W_n < b_E\lvert \beta^{E'}_n), 
\]
in $\beta^{E'}_n$, is a convex optimization problem for any $n\in \mathbb{N}$. 
\end{theorem}

\begin{remark}[Uniform convergence on compact sets] \label{unif:convergence}
The approximation $\log \widetilde{\mathbb{P}}(\mathbf{A}_E \sqrt{n}T_n + \mathbf{B}_E\sqrt{n}W_n < b_E\lvert \beta^{E'}_n)$ for the (log- ) selection probability is continuous in $\beta^{E'}_n$ and so is the true selection probability in $\beta^{E'}_n$. 
Hence the difference 
\[
n^{-2\delta}\left
\{\log \widetilde{\mathbb{P}}(\mathbf{A}_E \sqrt{n}T_n + \mathbf{B}_E\sqrt{n}W_n < b_E \lvert \beta^{E'}_n) - \log {\mathbb{P}}(\mathbf{A}_E \sqrt{n}T_n + \mathbf{B}_E\sqrt{n}W_n < b_E\lvert \beta^{E'}_n)\right\}
\]
 converges uniformly on a compact subset $\Theta\subset \mathbb{R}^{E'}$ of the parameter space.
\end{remark}

\smallskip
Finally, it is natural to ask how our approximate MLE compares to the exact MLE,
\begin{equation}
\label{mle:exact}
\begin{aligned}
\breve{\beta}^{E'}_S &= \argmin_{\beta^{E'}_n} \Big\{ 
n(\widehat{\beta}^{E'} -\beta^{E'}_n)' \bQ(\widehat{\beta}^{E'} -\beta^{E'}_n)/2 \\
&\;\;\;\;\;\;\;\;\;\;\;\;\;\;\;+ \log {\mathbb{P}}(\mathbf{A}_E \sqrt{n}T_n + \mathbf{B}_E\sqrt{n}W_n < b_E\lvert \beta^{E'}_n)\Big\}.
\end{aligned}
\end{equation}
The following theorem asserts that the approximate version converges to the exact MLE. We remark that the readers can see \cite{hjort2011asymptotics} to understand the proof of Theorem \ref{mle:convergence}. We provide a proof in the Appendix for completeness.

\begin{theorem}
\label{mle:convergence}
Under the parameterization in Theorem \ref{consistency:gen} for a $\delta\in (0,1/2)$, for any $\epsilon>0$ we have
\[
\mathbb{P}(n^{1/2-\delta} \|\widehat{\beta}^{E'}_S -\breve{\beta}^{E'}_S\|>\epsilon\lvert \;\mathbf{A}_E \sqrt{n}T_n + \mathbf{B}_E\sqrt{n}W_n < b_E) \to 0. 
\]
as $n \to \infty$.
\end{theorem}

\section{HIV drug-resistance data}\label{sec:hiv}
In this section we apply our methods to the HIV dataset analyzed in \cite{rhee2006genotypic}, \cite{bi2020inferactive}. 
With an attempt to understand the genetic basis of drug resistance in HIV, \cite{rhee2006genotypic} used markers of inhibitor mutations to predict susceptibility to 16 antiretroviral drugs. 
We follow \cite{bi2020inferactive} and focus on the protease inhibitor subset of the data, and on one particular drug, Lamivudine (3TC), where the goal is to identify mutations associated with response to 3TC. 
There are $n=633$ cases and $p=91$ different mutations occurring more than 10 times in the sample. 

\begin{figure}[h]
  \centering
    \includegraphics[scale=.1]{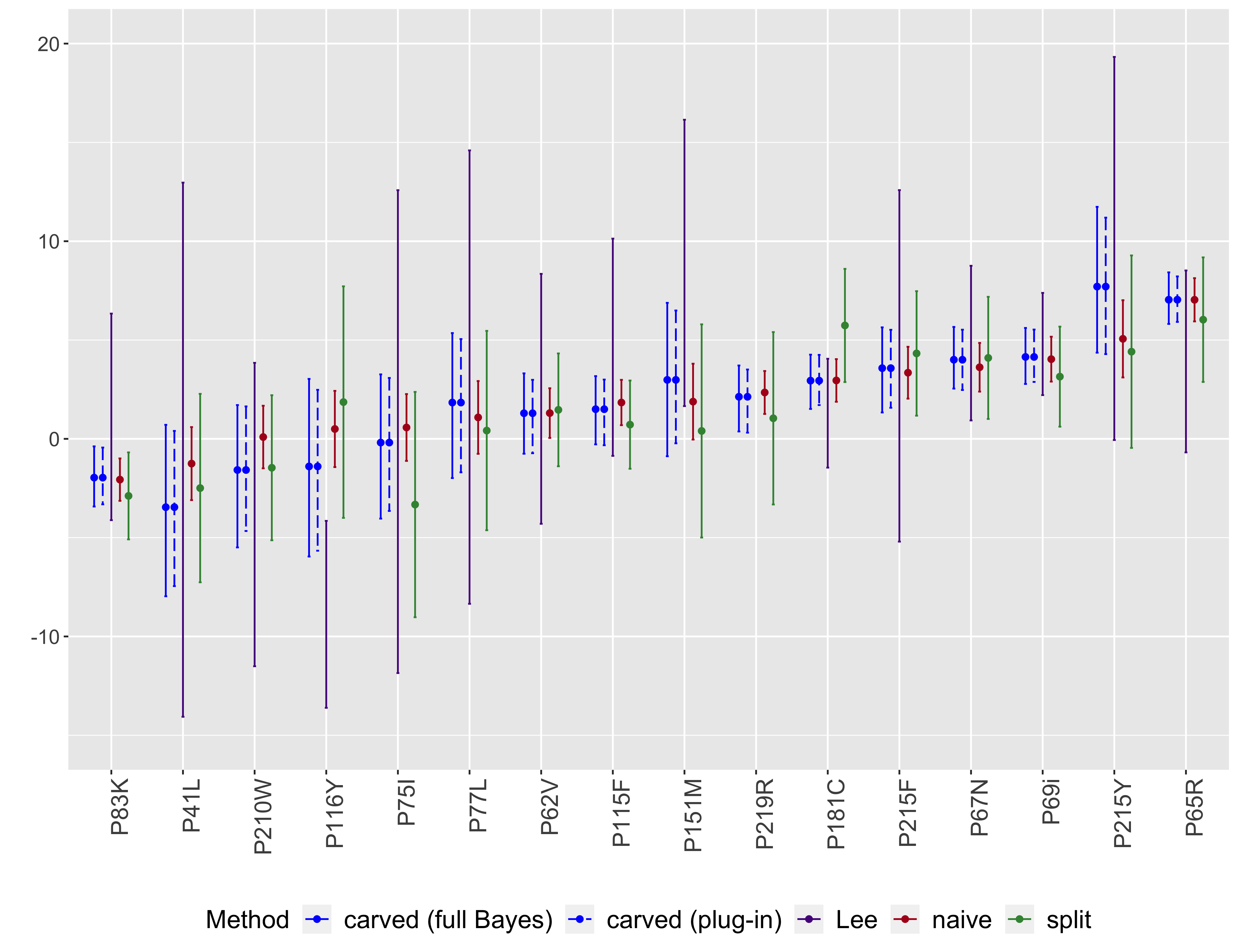}
  \caption{Interval and point estimates for selected features. To allow convenient visualization, the figure is not showing mutation `P184V', which has a different scale from the other variables in the selected set.}
  \label{fig:hiv}
\end{figure}

In the selection stage we applied the Lasso to a 80\% split of the data, with the regularization parameter set to the theoretical value proposed in \citet{negahban2009unified}. 
This resulted in 17 selected mutations, corresponding to the nonzero Lasso estimates. 
Figure \ref{fig:hiv} shows 90\% interval estimates constructed for selected variables (excluding mutation `P184V') according to four different methods: ``naive" is the usual, unadjusted intervals using the entire data; ``split" uses only the 20\% left-out portion of the data to construct the intervals; ``Lee" is the adjusted confidence intervals of \cite{lee2016exact} and based on the original 80\% portion used for selection; finally, ``carved" are the intervals relying on the methods we propose in the current paper. 
Of course, we cannot assume a given $\sigma^2$ in this analysis, and we offer two different implementations to handle an unknown variance. 
The first is to estimate $\sigma^2$ from a least squares fit against the $91$ available predictors, and then simply plug this estimate in (as if it were data independent). 
The second approach is to add a prior on $\sigma^2$ as described in Section \ref{sec:computation}. 
Specifically, we model 
\begin{equation}\label{prior:IG}
\pi(\beta^{E'} \lvert \sigma^2) \propto \sigma^{-|E'|}\exp(-\|\beta^{E'}\|^2 /(2c^2\sigma^2)),\ \ \ \ \ \ \ \pi(\sigma^2) \propto \text{Inv-Gamma}(a;b),
\end{equation}
where we denote $\text{Inv-Gamma}(a;b) := (\sigma^2)^{-a-1}\exp(-b/\sigma^2)$, and where $c^2$ is a large constant. 
In this example we used $a=b=0.1$ for the other hyperparameters. 
The two implementations are referred to as ``plug-in" and ``full Bayes" in Figure \ref{fig:hiv}. 

\smallskip
All four methods are implemented assuming a linear model consisting of the selected variables. 
If the model were correctly specified, ``split" and ``Lee" would both produce valid interval estimates, although the first uses only the held-out data, and the second relies only on the portion of the data used for selection. 
Our ``carved" intervals are approximate, but they utilize information from both portions of the data. 
The ``naive" intervals are invalid. 

\smallskip
In terms of length, it can be seen that the two implementations of carving (``plug-in" and ``full Bayes") produce similar interval estimates, that are shorter than both ``split" and ``Lee". 
Note that we could have used the entire data to construct the intervals of \cite{lee2016exact}, 
but we chose to use only the ``exploratory" portion to ensure that the same variables are selected as with the other methods. 
That ``Lee" interval estimates are considerably longer than ``split" matches our expectations: the leftover Fisher information in this scenario is smaller than the ``marginal" information from independent data, which results in longer intervals. 

The figure also shows point estimates: for ``naive" these are just the Lasso (nonzero) estimates, for ``split" these are least-squares, and for ``carved" these are (approximate) posterior modes. 
In the absence of knowledge about the underlying true means, it is hard to compare the different estimates and directions of shrinkage; ``split" estimates are unbiased under the assumed model, while ``carved" arguably have smaller variance (because they use additional data). 
Figure \ref{fig:hiv:all} in the Appendix depicts the effect-size estimates of all variables in the selected set, including mutation `P184V'.

\section{Computation}\label{sec:computation}
We provide computational details for sampling from the approximate posterior and solving for the approximate MAP (MLE) problem. 
Note that both of these problems require computing the gradient of the approximate log-posterior, which amounts to solving for the optimizing variables in a certain convex optimization problem. Throughout this section, we assume that $\pi$ is a log-concave prior. We also assume that the columns of $\bX$ are scaled by $\sqrt{n}$ (but suppress the subscript $n$). 
The crucial elements in our implementation are:
\begin{itemize}
\item Equation \eqref{grad:sampler:final} gives the exact form of the gradient of the log-posterior, which we compute at each draw of the sampler. Note that it involves $\hat{b}(\bQ \beta^E_{(K)})$, the optimizing variables to the optimization problem 
\[
\sup_{b \in \RR^{|E'|}}\left\{ b^T \bQ \beta^{E'}_{(K)} -( b^T \bQ b/2 +  H_{\psi}(b))\right\},
\]
where $H_{\psi}$ is defined within the section and $\beta^{E'}_{(K)}$ is the $K$-th draw of a sampler.
\item To solve for an approximate selection-adjusted MAP/MLE, we employ gradient descent on the objective of the MAP/MLE problem. The computation at each step of the descent again involves solving for the optimizing variables of the optimization problem stated above.
\end{itemize}
Treating first the case where $\sigma^2$ is known, assume again without loss of generality that $\sigma^2=1$. 
To sample from the log-concave (approximate) posterior incorporating \eqref{eq:optimization-general}, we use a Langevin random walk to obtain a sample of size $n_S$ from the (approximate) selection-adjusted posterior $\widetilde{\pi}_S(\cdot)$. 
As a function of the previous draw, the $(K+1)$-th draw for $\beta^{E'}$ of the Langevin sampler is computed as
\begin{equation}
\label{sampler}
\beta^{E'}_{(K+1)} = \beta^{E'}_{(K)} +\gamma\cdot \grad\log\widetilde{\pi}_S(\beta^{E'}_{(K)}\lvert \widehat{\beta}^{E'}) + \sqrt{2\gamma}\cdot \epsilon_{(K)}
\end{equation}
where $\epsilon_{(K)}$ for $K=1,2,...,n_S$ are independent draws from a centered gaussian with unit variance, and $\gamma$ is a predetermined step size. 
The sampler takes a noisy step along the gradient of the log-posterior, with no accept-reject step---compare to the usual Metropolis Hastings (MH) algorithm.
Hence, at each draw of the sampler, the main computational cost is calculating the gradient of the approximate (log-) selection-adjusted posterior. 

Recalling the approximation based on \eqref{eq:optimization-general}, we note that the approximate log-posterior 
is
$$
 \log \pi(\beta^{E'}_{(K)}) - {\beta^{E'}_{(K)}}^T \bQ \beta^{E'}_{(K)}/2  +  {\beta^{E'}_{(K)}}^T \bQ \widehat{\beta}^{E'} - \log \widetilde{\mathbb{P}}( \widehat{E}= E,\widehat{S}^{\widehat{E}} = s^E \lvert \beta^{E'}_{(K)}) 
$$
\begin{equation}
\label{opt:approx:comp}
\log \widetilde{\mathbb{P}}( \widehat{E}= E, \widehat{S}^{\widehat{E}} = s^E \lvert \beta^{E'}_{(K)}) = -\inf_{b \in \RR^{|E'|}} \left\{(b-\beta^{E'}_{(K)})^T \bQ (b- \beta^{E'}_{(K)})/2 +  H_{\psi}(b)\right\}.
\end{equation}
Above, $H_{\psi}$ is a function of $b = (\eta, o) \in \{(\eta, o): \eta \in \mathbb{R}^{p-|E'|}, o\in \mathbb{R}^{p}\}$, and given by
\begin{equation*}
\begin{aligned}
& H_{\psi}(b) =  \Scale[0.9]{\displaystyle\inf\limits_{(\eta, o) \in \RR^{2p-|E'|}} \Bigg\{\eta^T \mathbf{N} \eta/2+ \left(P_E \begin{pmatrix} b & \eta \end{pmatrix}^T  + Q_E o + r_E\right)^T \mathbf{\Sigma}_{\mathcal{G}}^{-1} \left(P_E \begin{pmatrix} b & \eta \end{pmatrix}^T + Q_E o + r_E\right)\Big/2}\\
&+\displaystyle\sum_{i=1}^{E} \log\left(1+ {1}/{s_{i,E} o_{i,E}}\right) + \sum_{i=1}^{p-|E|} \log\left(1+ {1}/{(\lambda_{i,-E} - o_{i,-E})}\right)+ \log\left(1+ {1}/{
(\lambda_{i,-E} + o_{i,-E})}\right)\Bigg\}.
\end{aligned}
\end{equation*}
Denoting by $\bar{H}^*(\cdot)$ the conjugate of $\bar{H}(b) = b^T \bQ b/2 +  H_{\psi}(b)$, we can write 
\begin{equation*}
\begin{aligned}
\log \widetilde{\mathbb{P}}( \widehat{E}= E, \widehat{S}^{\widehat{E}} = s^E \lvert \beta^{E'}_{(K)}) &= -{\beta^{E'}_{(K)}}^T\bQ \beta^{E'}_{(K)}/2 + \sup_{b \in \RR^{E}}\left\{ b^T \bQ \beta^{E'}_{(K)} -( b^T \bQ b/2 +  H_{\psi}(b))\right\}\\
&=-{\beta^{E'}_{(K)}}^T\bQ \beta^{E'}_{(K)}/2 + \bar{H}^*(\bQ \beta^{E'}_{(K)}).
\end{aligned}
\end{equation*}
Letting $\hat{b}(\bQ \beta^{E'}_{(K)})$ be the maximizer of $\{b^T \bQ \beta^{E'}_{(K)} -( b^T \bQ b/2 +  H_{\psi}(b))\}$, we have
\[
\grad\log \widetilde{\mathbb{P}}( \widehat{E}= E, \widehat{S}^{\widehat{E}} = s^E \lvert \beta^{E'}_{(K)}) = -\bQ \beta^{E'}_{(K)} +  \bQ\grad \bar{H}^{-1}(\bQ \beta^{E'}_{(K)}) = -\bQ \beta^{E'}_{(K)} + \bQ \hat{b}(\bQ \beta^{E'}_{(K)}),
\]
and, finally, the gradient of the log-posterior in \eqref{sampler} equals
\begin{equation}
\label{grad:sampler:final}
\grad  \log \pi(\beta^{E'}_{(K)}) + \bQ \widehat{\beta}^{E'} - \bQ \hat{b}(\bQ \beta^{E'}_{(K)}).
\end{equation}
Alternatively, we could employ a Metropolis-Hastings algorithm that would require the value of the optimization problem in \eqref{opt:approx:comp} to approximate $\log {\mathbb{P}}( \widehat{E}= E, \widehat{S}^{\widehat{E}} = s^E \lvert \beta^{E'}_{(K)})$, which equals 
$$
-{\beta^{E'}_{(K)}}^T\bQ \beta^{E'}_{(K)}/2 + \bar{H}^*(\bQ \beta^{E'}_{(K)})
$$ 
in order to compute the acceptance ratio each time.

\medskip
For the MAP problem, we note that the approximate MAP minimizes the convex objective
\[\text{minimize}_{\beta^{E'} \in \mathbb{R}^{|E'|}} - \log \pi(\beta^{E'})-{\beta^{E'}}^T \bQ \widehat{\beta}^{E'} + \bar{H}^*(\bQ \beta^{E'}).\]
This reduces to the MLE problem when $\pi(\beta^{E'})\propto 1$. Employing a gradient descent algorithm to compute the approximate MAP, we note that the $K$-th update can be written as 
\begin{equation}
\label{MAP:final}
\widehat{\beta}^{E'}_{S; \;(K+1)}= \widehat{\beta}^{E'}_{S; \;(K)} -\eta^T \cdot (\bQ\hat{b}(\bQ\widehat{\beta}^{E'}_{S; \;(K)} ) -\grad \log \widetilde{\pi}(\widehat{\beta}^{E'}_{S; \;(K)} )-\bQ  \widehat{\beta}^{E'}),
\end{equation}
involving again the optimizer $\hat{b}(\bQ{\beta}^{E'})$, obtained from solving $$\text{maximize}_{\beta^{E'}} \left\{b^T \bQ \beta^{E'} -( b^T \bQ b/2 +  H_{\psi}(b))\right\}.$$

Next, suppose that $\sigma^2$ is unknown. 
In that case, we propose to replace \eqref{eq:prior} with a joint prior on $\beta^{E'}$ and $\sigma^2$. 
For example, we can take the conjugate prior \eqref{prior:IG}, a large $c^2$ entailing a diffuse prior for $\beta^{E'}$ conditionally on $\sigma^2$. 
Below, we describe a Gibbs sampler to alternately sample $\beta^{E'}$ and $\sigma^2$.


\smallskip 

Let $\widetilde{L}_S(\beta^{E'}, \sigma^2)$ denote the approximate (log) selection-adjusted likelihood
where we plug in our approximation for the selection probability, given by \eqref{approx:sigma} 
\begin{equation}
\label{opt:approx:comp:sigma}
\begin{aligned}
& \log \widetilde{\mathbb{P}}( \widehat{E}= E, \widehat{S}^{\widehat{E}} = s^E\;  \lvert \; \beta^{E'}_{(K)}, \sigma^2) \\
&\;\;\;\;\;\;\;\;\;\;\;\;\;\;\;\;\;\;\;\;\;\;\;\;\;\;= -(\sigma^2)^{-1}\cdot \inf_{b \in \RR^{|E'|}} \left\{(b-\beta^{E'}_{(K)})^T \bQ (b- \beta^{E'}_{(K)})/2 +  H_{\psi}(b)\right\}.
\end{aligned}
\end{equation}
We employ a Gibbs scheme to draw a new sample $(\beta^{E'}_{(K+1)}, \sigma^2_{(K+1)})^T$ from the resulting posterior, whose logarithm equals
$$ \log \pi(\beta^{E'}, \sigma^2) + \widetilde{L}_S(\beta^{E'}, \sigma^2)$$
up to a constant, under the prior in \eqref{prior:IG}.

Using our previous notation, let us denote the posterior distribution for $\beta^{E'}$ conditional on $\sigma^2$ by $\widetilde{\pi}_S(\beta^{E'} \lvert \widehat{\beta}^{E'}, \sigma^2)$.
Conditional on the $K$-th update for $\sigma^2_{(K)}$, we make a fresh draw $\beta^{E'}_{(K+1)}$ through a noisy update along the gradient of the log-posterior, as described previously in \eqref{sampler}:
$$
(\beta^{E'})\text{-Update:} \ \  \ \ \beta^{E'}_{(K+1)} \leftarrow \beta^{E'}_{(K)} +\gamma\cdot \grad\log\widetilde{\pi}_S(\beta^{E'}_{(K)}\lvert \widehat{\beta}^{E'}, \sigma^2_{(K)}) + \sqrt{2\gamma}\cdot \epsilon_{(K)}.
$$
Observe that the gradient of the log-posterior which leads to our sample update equals
\begin{equation}
\label{grad:sampler:final:sigma}
 -\beta_{(K)}^{E'}/c^2\sigma^2_{(K)} + \bQ \widehat{\beta}^{E'}/\sigma^2_{(K)} - \bQ \hat{b}(\bQ \beta^{E'}_{(K)})/\sigma^2_{(K)}.
\end{equation}
We alternate this step with updates for the variance parameter; see for example \citet{tong2019mala}.
Conditional on $\beta^{E'}_{(K+1)}$, we see that the choice of an inverse-gamma prior for the variance parameter serves as a conjugate prior as it does usually in Bayesian linear regression.  
Specifically, the posterior distribution of $\sigma^2_{(K+1)}$ conditional on $\beta^{E'}_{(K+1)}$ is an inverse-gamma random variable with density proportional to
\begin{equation*}
\begin{aligned}
& (\sigma^2)^{-|E'|/2-p- a-1}\exp\Big(-(b- \widetilde{L}_S(\beta_{(K+1)}^{E'}, 1)+ (\beta_{(K+1)}^{E'})^T\beta_{(K+1)}^{E'}/2c^2))/\sigma^2\Big).
\end{aligned}
\end{equation*}
That is, we now sample
$$(\sigma^2)\text{-Update:} \ \  \ \ \sigma^2_{(K+1)} \leftarrow  \text{Inv-Gamma}(|E'|/2+p +a; \widetilde{b}),$$
where the updated hyperparameter $\widetilde{b}(\beta^{E'}_{(K+1)})$ equals
$$b- \widetilde{L}_S(\beta_{(K+1)}^{E'}, 1)+ (\beta_{(K+1)}^{E'})^T\beta_{(K+1)}^{E'}/2c^2.$$

\section{Discussion}\label{sec:discussion}
To address the problem of inference after variable selection, we adopt the point of view where inference is based on the likelihood when truncated to the event including all possible realizations that lead the researcher to posing the same question. 
The methods we propose are based on an approximation to the adjustment factor---the denominator in the selection-adjusted likelihood---which applies to a large class of selection rules, including such that involve randomization. 
By working directly with the full truncated likelihood, we obviate the need to differentiate between various cases according to choices of the statistician: for example, the approach of \citet{fithian2014optimal} requires computations that are different in essence under the {\it selected model} and under the {\it saturated model}, whereas with the tools we develop, there is no essential difference between the two. 
Similarly, our methods are amenable to data-carving. 
In \citet{panigrahieQTL} the approximation proposed in our paper is employed to obtain tractable pivotal quantities, and more recently \citet{panigrahi2019approximate} used this approximation in a maximum-likelihood approach. 


\smallskip
There is certainly room for further research and extensions of the current work. 
On the methodological side, it would be interesting to investigate if a variational Bayes approach can be taken instead of implementing MCMC sampling schemes for posterior updates. 
From a theoretical point of view, we have shown consistency properties of the posterior that appends a ``carved" likelihood to a prior, that is, we proved that such a posterior concentrates around the true underlying parameter with probability converging to one as the sample size increases. 
Empirical evidence showing that our credible intervals (under a diffuse prior) are similar to the frequentist post-selection intervals, suggests that the frequentist guarantees that we provided can be strengthened, for example by presenting a Bernstein--von Mises-type of result. 


\section{Acknowledgements}
\noindent The idea of providing Bayesian adjusted inference after variable selection was previously proposed by Daniel Yekutieli and Edward George and presented at the 2012 Joint Statistical Meetings in San Diego. 
Our interest re-arose with the recent developments on exact post-selection inference in the linear model. 
A.W. is thankful to Daniel and Ed for helpful conversations and to Ed for pointing out the reference to \citet{bayarri1987bayesian}. A.W. was partially supported by ERC grant 030-8944 and by ISF grant 039-9325. J.T. was supported in part by ARO grant 70940MA. S.P. is thankful to Xuming He, Liza Levina, Rina Foygel Barber and Yi Wang for reading an initial version of the draft and offering valuable comments and insights. 
The authors would also like to sincerely thank and acknowledge Chiara Sabatti for the long discussions and for her suggestions along the way.

\bibliographystyle{plainnat}
\bibliography{References-bayesian-selective.bib}

\begin{thebibliography}{44}
\providecommand{\natexlab}[1]{#1}
\providecommand{\url}[1]{\texttt{#1}}
\expandafter\ifx\csname urlstyle\endcsname\relax
  \providecommand{\doi}[1]{doi: #1}\else
  \providecommand{\doi}{doi: \begingroup \urlstyle{rm}\Url}\fi

\bibitem[Arcones(2002)]{arcones2002moderate}
Miguel~A Arcones.
\newblock Moderate deviations for m-estimators.
\newblock \emph{Test}, 11\penalty0 (2):\penalty0 465--500, 2002.

\bibitem[Bayarri and DeGroot(1987)]{bayarri1987bayesian}
MJ~Bayarri and MH~DeGroot.
\newblock Bayesian analysis of selection models.
\newblock \emph{The Statistician}, pages 137--146, 1987.

\bibitem[Benjamini and Meir(2014)]{benjamini2014selective}
Yoav Benjamini and Amit Meir.
\newblock Selective correlations-the conditional estimators.
\newblock \emph{arXiv preprint arXiv:1412.3242}, 2014.

\bibitem[Berk et~al.(2013)Berk, Brown, Buja, Zhang, Zhao,
  et~al.]{berk2013valid}
Richard Berk, Lawrence Brown, Andreas Buja, Kai Zhang, Linda Zhao, et~al.
\newblock Valid post-selection inference.
\newblock \emph{The Annals of Statistics}, 41\penalty0 (2):\penalty0 802--837,
  2013.

\bibitem[Bi et~al.(2020)Bi, Markovic, Xia, and Taylor]{bi2020inferactive}
Nan Bi, Jelena Markovic, Lucy Xia, and Jonathan Taylor.
\newblock Inferactive data analysis.
\newblock \emph{Scandinavian Journal of Statistics}, 47\penalty0 (1):\penalty0
  212--249, 2020.

\bibitem[Borovkov and Mogul'skii(1978)]{borovkov1978probabilities}
AA~Borovkov and AA~Mogul'skii.
\newblock Probabilities of large deviations in topological spaces. i.
\newblock \emph{Siberian Mathematical Journal}, 19\penalty0 (5):\penalty0
  697--709, 1978.

\bibitem[Buja et~al.(2015)Buja, Berk, Brown, George, Pitkin, Traskin, Zhao, and
  Zhang]{buja2015models}
Andreas Buja, Richard~A Berk, Lawrence~D Brown, Edward~I George, Emil Pitkin,
  Mikhail Traskin, Linda Zhao, and Kai Zhang.
\newblock Models as approximations-a conspiracy of random regressors and model
  deviations against classical inference in regression.
\newblock \emph{Statistical Science}, page~1, 2015.

\bibitem[Consortium et~al.(2017)]{gtex2017genetic}
GTEx Consortium et~al.
\newblock Genetic effects on gene expression across human tissues.
\newblock \emph{Nature}, 550\penalty0 (7675):\penalty0 204--213, 2017.

\bibitem[Dwork et~al.(2014)Dwork, Feldman, Hardt, Pitassi, Reingold, and
  Roth]{dwork2014preserving}
Cynthia Dwork, Vitaly Feldman, Moritz Hardt, Toniann Pitassi, Omer Reingold,
  and Aaron Roth.
\newblock Preserving statistical validity in adaptive data analysis.
\newblock \emph{arXiv preprint arXiv:1411.2664}, 2014.

\bibitem[Dwork et~al.(2015)Dwork, Feldman, Hardt, Pitassi, Reingold, and
  Roth]{dwork2015generalization}
Cynthia Dwork, Vitaly Feldman, Moritz Hardt, Toniann Pitassi, Omer Reingold,
  and Aaron Roth.
\newblock Generalization in adaptive data analysis and holdout reuse.
\newblock \emph{arXiv preprint arXiv:1506.02629}, 2015.

\bibitem[Efron(2011)]{efron2011tweedie}
Bradley Efron.
\newblock TweedieÕs formula and selection bias.
\newblock \emph{Journal of the American Statistical Association}, 106\penalty0
  (496):\penalty0 1602--1614, 2011.

\bibitem[Efron(2012)]{efron2012large}
Bradley Efron.
\newblock \emph{Large-scale inference: empirical Bayes methods for estimation,
  testing, and prediction}, volume~1.
\newblock Cambridge University Press, 2012.

\bibitem[Eichelsbacher and L{\"o}we(2003)]{eichelsbacher2003moderate}
Peter Eichelsbacher and Matthias L{\"o}we.
\newblock Moderate deviations for iid random variables.
\newblock \emph{ESAIM: Probability and Statistics}, 7:\penalty0 209--218, 2003.

\bibitem[Fithian et~al.(2014)Fithian, Sun, and Taylor]{fithian2014optimal}
William Fithian, Dennis Sun, and Jonathan Taylor.
\newblock Optimal inference after model selection.
\newblock \emph{arXiv preprint arXiv:1410.2597}, 2014.

\bibitem[Gao et~al.(2017)Gao, Martini, Zhang, Yuan, Zhang, Simianer, and
  Li]{gao2017incorporating}
Ning Gao, Johannes~WR Martini, Zhe Zhang, Xiaolong Yuan, Hao Zhang, Henner
  Simianer, and Jiaqi Li.
\newblock Incorporating gene annotation into genomic prediction of complex
  phenotypes.
\newblock \emph{Genetics}, 207\penalty0 (2):\penalty0 489--501, 2017.

\bibitem[Gunter et~al.(2011)Gunter, Zhu, and Murphy]{gunter2011variable}
L~Gunter, J~Zhu, and SA~Murphy.
\newblock Variable selection for qualitative interactions.
\newblock \emph{Statistical methodology}, 8\penalty0 (1):\penalty0 42--55,
  2011.

\bibitem[Hjort and Pollard(2011)]{hjort2011asymptotics}
Nils~Lid Hjort and David Pollard.
\newblock Asymptotics for minimisers of convex processes.
\newblock \emph{arXiv preprint arXiv:1107.3806}, 2011.

\bibitem[Kanter and Proppe(1977)]{kanter1977reduction}
Marek Kanter and Harold Proppe.
\newblock Reduction of variance for gaussian densities via restriction to
  convex sets.
\newblock \emph{Journal of Multivariate Analysis}, 7\penalty0 (1):\penalty0
  74--81, 1977.

\bibitem[Kivaranovic and Leeb(2018)]{kivaranovic2018expected}
Danijel Kivaranovic and Hannes Leeb.
\newblock Expected length of post-model-selection confidence intervals
  conditional on polyhedral constraints.
\newblock \emph{arXiv preprint arXiv:1803.01665}, 2018.

\bibitem[Lee and Taylor(2014)]{lee2014exact}
Jason~D Lee and Jonathan~E Taylor.
\newblock Exact post model selection inference for marginal screening.
\newblock In \emph{Advances in Neural Information Processing Systems}, pages
  136--144, 2014.

\bibitem[Lee et~al.(2016)Lee, Sun, Sun, Taylor, et~al.]{lee2016exact}
Jason~D Lee, Dennis~L Sun, Yuekai Sun, Jonathan~E Taylor, et~al.
\newblock Exact post-selection inference, with application to the lasso.
\newblock \emph{The Annals of Statistics}, 44\penalty0 (3):\penalty0 907--927,
  2016.

\bibitem[Markovic and Taylor(2016)]{markovic2016bootstrap}
Jelena Markovic and Jonathan Taylor.
\newblock Bootstrap inference after using multiple queries for model selection.
\newblock \emph{arXiv preprint arXiv:1612.07811}, 2016.

\bibitem[McKeague and Qian(2015)]{mckeague2015adaptive}
Ian~W McKeague and Min Qian.
\newblock An adaptive resampling test for detecting the presence of significant
  predictors.
\newblock \emph{Journal of the American Statistical Association}, 110\penalty0
  (512):\penalty0 1422--1433, 2015.

\bibitem[Negahban et~al.(2009)Negahban, Yu, Wainwright, and
  Ravikumar]{negahban2009unified}
Sahand Negahban, Bin Yu, Martin~J Wainwright, and Pradeep~K Ravikumar.
\newblock A unified framework for high-dimensional analysis of $ m $-estimators
  with decomposable regularizers.
\newblock In \emph{Advances in Neural Information Processing Systems}, pages
  1348--1356, 2009.

\bibitem[Panigrahi(2018)]{panigrahi2018asymptotic}
Snigdha Panigrahi.
\newblock Carving model-free inference.
\newblock \emph{arXiv preprint arXiv:1811.03142}, 2018.

\bibitem[Panigrahi and Taylor(2019)]{panigrahi2019approximate}
Snigdha Panigrahi and Jonathan Taylor.
\newblock Approximate selective inference via maximum likelihood.
\newblock \emph{arXiv preprint arXiv:1902.07884}, 2019.

\bibitem[Panigrahi et~al.(2019)Panigrahi, Zhu, and Sabatti]{panigrahieQTL}
Snigdha Panigrahi, Junjie Zhu, and Chiara Sabatti.
\newblock {Selection-adjusted inference: an application to confidence intervals
  for cis-eQTL effect sizes}.
\newblock \emph{Biostatistics}, 07 2019.
\newblock ISSN 1465-4644.
\newblock \doi{10.1093/biostatistics/kxz024}.
\newblock URL \url{https://doi.org/10.1093/biostatistics/kxz024}.
\newblock kxz024.

\bibitem[Reid et~al.(2014)Reid, Taylor, and Tibshirani]{reid2014post}
Stephen Reid, Jonathan Taylor, and Robert Tibshirani.
\newblock Post-selection point and interval estimation of signal sizes in
  gaussian samples.
\newblock \emph{arXiv preprint arXiv:1405.3340}, 2014.

\bibitem[Rhee et~al.(2006)Rhee, Taylor, Wadhera, Ben-Hur, Brutlag, and
  Shafer]{rhee2006genotypic}
Soo-Yon Rhee, Jonathan Taylor, Gauhar Wadhera, Asa Ben-Hur, Douglas~L Brutlag,
  and Robert~W Shafer.
\newblock Genotypic predictors of human immunodeficiency virus type 1 drug
  resistance.
\newblock \emph{Proceedings of the National Academy of Sciences}, 103\penalty0
  (46):\penalty0 17355--17360, 2006.

\bibitem[Robbins(1988)]{robbins1988uv}
H~Robbins.
\newblock The uv method of estimation.
\newblock \emph{Statistical Decision Theory and Related Topics IV}, 1:\penalty0
  265--270, 1988.

\bibitem[Russo and Zou(2015)]{russo2015controlling}
Daniel Russo and James Zou.
\newblock Controlling bias in adaptive data analysis using information theory.
\newblock \emph{arXiv preprint arXiv:1511.05219}, 2015.

\bibitem[Simon and Simon(2013)]{simon2013estimating}
Noah Simon and Richard Simon.
\newblock On estimating many means, selection bias, and the bootstrap.
\newblock \emph{arXiv preprint arXiv:1311.3709}, 2013.

\bibitem[Simonsohn et~al.(2014)Simonsohn, Nelson, and Simmons]{simonsohn2014p}
Uri Simonsohn, Leif~D Nelson, and Joseph~P Simmons.
\newblock P-curve: A key to the file-drawer.
\newblock \emph{Journal of Experimental Psychology: General}, 143\penalty0
  (2):\penalty0 534, 2014.

\bibitem[Stingo et~al.(2011)Stingo, Chen, Tadesse, and
  Vannucci]{stingo2011incorporating}
Francesco~C Stingo, Yian~A Chen, Mahlet~G Tadesse, and Marina Vannucci.
\newblock Incorporating biological information into linear models: A bayesian
  approach to the selection of pathways and genes.
\newblock \emph{The annals of applied statistics}, 5\penalty0 (3), 2011.

\bibitem[Taylor et~al.(2014)Taylor, Lockhart, Tibshirani, and
  Tibshirani]{taylor2014exact}
J~Taylor, R~Lockhart, RJ~Tibshirani, and R~Tibshirani.
\newblock Exact post-selection inference for sequential regression procedures.
\newblock \emph{Journal of the American Statistical Association}, 111\penalty0
  (514):\penalty0 600--620, 2014.

\bibitem[Taylor and Tibshirani(2018)]{taylor2018post}
Jonathan Taylor and Robert Tibshirani.
\newblock Post-selection inference for-penalized likelihood models.
\newblock \emph{Canadian Journal of Statistics}, 46\penalty0 (1):\penalty0
  41--61, 2018.

\bibitem[Taylor et~al.(2013)Taylor, Loftus, and Tibshirani]{taylor2013tests}
Jonathan Taylor, Joshua Loftus, and Ryan Tibshirani.
\newblock Tests in adaptive regression via the kac-rice formula.
\newblock \emph{arXiv preprint arXiv:1308.3020}, 2013.

\bibitem[Tian et~al.(2018)Tian, Taylor, et~al.]{tian2018selective}
Xiaoying Tian, Jonathan Taylor, et~al.
\newblock Selective inference with a randomized response.
\newblock \emph{The Annals of Statistics}, 46\penalty0 (2):\penalty0 679--710,
  2018.

\bibitem[Tong et~al.(2019)Tong, Morzfeld, and Marzouk]{tong2019mala}
XT~Tong, M~Morzfeld, and YM~Marzouk.
\newblock Mala-within-gibbs samplers for high-dimensional distributions with
  sparse conditional structure.
\newblock \emph{arXiv preprint arXiv:1908.09429}, 2019.

\bibitem[Weinstein et~al.(2013)Weinstein, Fithian, and
  Benjamini]{weinstein2013selection}
Asaf Weinstein, William Fithian, and Yoav Benjamini.
\newblock Selection adjusted confidence intervals with more power to determine
  the sign.
\newblock \emph{Journal of the American Statistical Association}, 108\penalty0
  (501):\penalty0 165--176, 2013.

\bibitem[Yekutieli(2012)]{yekutieli2012adjusted}
Daniel Yekutieli.
\newblock Adjusted bayesian inference for selected parameters.
\newblock \emph{Journal of the Royal Statistical Society: Series B (Statistical
  Methodology)}, 74\penalty0 (3):\penalty0 515--541, 2012.

\bibitem[Yu et~al.(2019)Yu, Bien, and Tibshirani]{yu2019reluctant}
Guo Yu, Jacob Bien, and Ryan Tibshirani.
\newblock Reluctant interaction modeling.
\newblock \emph{arXiv preprint arXiv:1907.08414}, 2019.

\bibitem[Zhong and Prentice(2008)]{zhong2008bias}
Hua Zhong and Ross~L Prentice.
\newblock Bias-reduced estimators and confidence intervals for odds ratios in
  genome-wide association studies.
\newblock \emph{Biostatistics}, 9\penalty0 (4):\penalty0 621--634, 2008.

\bibitem[Z{\"o}llner and Pritchard(2007)]{zollner2007overcoming}
Sebastian Z{\"o}llner and Jonathan~K Pritchard.
\newblock Overcoming the winnerÕs curse: estimating penetrance parameters from
  case-control data.
\newblock \emph{The American Journal of Human Genetics}, 80\penalty0
  (4):\penalty0 605--615, 2007.

\end{thebibliography}

\section{Appendix}
\subsection{Proofs for Section 3}
\label{A:1}

\begin{proof}
\emph{Theorem \ref{carved:gaussian}:}
We begin by noting that $\log \mathbb{P}(n^{1/2-\delta}(\bar{Y}_n + W_n) \in \mathcal{K}\lvert \beta_n)$ can be bounded above by
$$\log \mathbb{E}(\exp(n^{1/2+\delta} \alpha \bar{Y}_n + n^{1/2+\delta}\gamma W_n + u)\lvert \beta_n)$$ for a choice of $\alpha, \gamma \text{ and } u$ such that
\[n^{1/2+\delta}\cdot(\alpha \bar{z} + \gamma \bar{w}) +  u \geq 0\; \text{ whenever  } \;n^{1/2-\delta}(\bar{z} + \bar{w}) \in \mathcal{K}.\]
Setting $$u = -\textstyle\inf_{\bar{z}, \bar{w}: n^{1/2-\delta}(\bar{z} + \bar{w})\in \mathcal{K}}\ (n^{1/2+\delta}\alpha \bar{z} + n^{1/2+\delta}\gamma \bar{w}),$$
it follows that
\begin{equation*}
\begin{aligned}
& \log \mathbb{P}(n^{1/2-\delta}(\bar{Y}_n + W_n) \in \mathcal{K}\lvert \beta_n)\\
& \leq  \sup_{\bar{z}, \bar{w}: n^{1/2-\delta}(\bar{z} + \bar{w})\in \mathcal{K}}\left\{-n^{1/2+\delta}\alpha \bar{z} -n^{1/2+\delta}\gamma \bar{w} + n^{2\delta}\alpha \beta^* + \dfrac{n^{2\delta}}{2}\alpha^2 + \dfrac{n^{2\delta}(1-\rho)}{2\rho}\gamma^2 \right\}\\
&=n^{2\delta}\cdot\sup_{z, w: (z + w)\in \mathcal{K}}\left\{ -\alpha z - \gamma w + \alpha \beta^* + \dfrac{\alpha^2}{2} + \dfrac{(1-\rho)\gamma^2}{2\rho} \right\}.
\end{aligned}
\end{equation*}
Since the above display holds for any arbitrary $\alpha, \gamma$, we can write
\begin{equation*}
\begin{aligned}
&\log \mathbb{P}(n^{1/2-\delta}(\bar{Y}_n + W_n) \in \mathcal{K}\lvert \beta_n)\\
&\leq  -n^{2\delta}\cdot \sup_{\alpha, \gamma}\left\{\inf_{z, w: (z + w)\in \mathcal{K}}\alpha z -\left(\alpha \beta^* + \dfrac{\alpha^2}{2}  \right)+ \gamma w - \dfrac{(1-\rho)\gamma^2}{2\rho}\right\}\\
&= -n^{2\delta}\cdot\inf_{z, w: (z + w)\in \mathcal{K}}\left\{\sup_{\alpha, \gamma}\alpha z -\left(\alpha \beta^* + \dfrac{\alpha^2}{2} \right)+ \gamma w - \dfrac{(1-\rho)\gamma^2}{2\rho}  \right\}\\
&=  -n^{2\delta}\inf_{z, w: (z+w) \in \mathcal{K}} \dfrac{(z-\beta^*)^2}{2} + \dfrac{\rho w^2}{2(1-\rho)}.
\end{aligned}
\end{equation*}
The penultimate equation follows by a minimax argument and the last step calculates the conjugates of the moment generating functions of gaussian random variables with variances $1$ and $\dfrac{1-\rho}{\rho}$ respectively. 

The large deviation limit when $\delta=1/2$ follows from Cramer's Theorem on the real line and the moderate deviations limit follows from Theorem 2.2 and Remark 2.32, \cite{eichelsbacher2003moderate} when $\delta \in (0,1/2)$. These theorems state the following limit
\[\lim_{n\to \infty} \frac{1}{n^{2\delta}}\log \mathbb{P}[\sqrt{n} (\bar{Y}_n + W_n)/n^{\delta}\in \mathcal{K} \lvert \beta_n] = -\inf_{(z,w): z+w \in \mathcal{K}} R(z, w);\]
$R(z, w) = \sup_{x, y}\{ x z + y w - \log\mathbb{E}[ \exp(x Z + y W)\lvert \beta^*]\}$, $(Z,W)\sim \mathcal{N}(\mu, \Sigma), \;\mu =\begin{pmatrix} \beta^* \\0 \end{pmatrix}$ and $\Sigma =\begin{bmatrix} 1 & 0 \\ 0 & {(1-\rho)}/{\rho}\end{bmatrix}$. We have the proof by plugging in the gaussian MGF and finally, observing that the rate function equals
\[R(z, w) = {(z-\beta^*)^2}/{2} + {\rho w^2}/{2(1-\rho)}.\]
\end{proof}

\begin{proof}[Proof of Corollary \ref{carved:LDP:approximate:selective:posterior}]
It follows from Theorem \ref{carved:gaussian} that the sequence of true truncated posteriors, written as
\[\log\pi_S(\beta_n\lvert\bar{y}_n) =\log \pi(\beta_n) -n(
\bar{y}_n - \beta_n)^2/2 -\log \mathbb{P}(\sqrt{n}(\bar{Y}_n + W_n) /n^{\delta} \in \mathcal{K}\lvert \beta_n)\] can be approximated by 
\[\log \pi(\beta_n) -n(\bar{y}_n - \beta_n)^2/2 + n^{2\delta}\cdot\inf_{(z,w):z+w \in \mathcal{K}}\left\{ \dfrac{(z-\beta^*)^2}{2} + \dfrac{\rho w^2}{2(1-\rho)}\right\}.\] 
Note that the limiting sequence of objectives
\[ \dfrac{(z-\beta^*)^2}{2} + \dfrac{\rho w^2}{2(1-\rho)} +\frac{1}{n^{2\delta}}\psi_{n^{-\delta}} (z+ w)\]
are convex in $(z, w)$. 
Furthermore, the above sequence converges to the continuous, convex objective 
\[\dfrac{(z-\beta^*)^2}{2} + \dfrac{\rho w^2}{2(1-\rho)} + I_{\mathcal{K}}(z+w),\; I_{\mathcal{K}}(\bar{z}) =\begin{cases} 
      0  & \text{ if } \bar{z} \in \mathcal{K} \\
      \infty  & \text{ otherwise}\\
         \end{cases}
\]
under the condition $n^{-2\delta}\psi_{n^{-\delta}} (z + w)  \to I_{\mathcal{K}}(z+w)$ for all $(z,w) \in \mathbb{R}^{2}$ as $n \to \infty$. Finally, observing that the limiting objective has a unique minimum, we reach the conclusion of the Corollary.
\end{proof}

\subsection{Proofs for Section 4}
\label{A:2}

\begin{proof}
\emph{Proposition \ref{asymptotic:distribution}: }
Noting that the least squares estimator $\widehat{\beta}^{E'}$ satisfies
\begin{equation}
 \label{ols:restricted}
 \dfrac{1}{\sqrt{n}} \mathbf{X}_{E'}^T \left(\vec{y} - \bX_{E'} \widehat{\beta}^{E'}\right)= 0,
 \end{equation}
we can write 
 \begin{equation*}
\begin{aligned}
0=\dfrac{1}{\sqrt{n}} \mathbf{X}_{E'}^T \left(\vec{y} - \bX_{E'} \widehat{\beta}^{E'}\right) &= \dfrac{1}{\sqrt{n}} \mathbf{X}_{E'}^T \left(\vec{y} - \bX_{E'} {\beta}^{E'}\right) -\left(\dfrac{\bX_{E'}^T \bX_{E'}}{n}\right)\sqrt{n}( \widehat{\beta}^{E'} -{\beta}^{E'})\\
 &=  \dfrac{1}{\sqrt{n}} \bX_{E'}^T \left(\vec{y} - \bX_{E'} {\beta}^{E'}\right) - \mathbf{{Q}}\sqrt{n}( \widehat{\beta}^{E'} -{\beta}^{E'}) \\
 &\;\;- \left(\dfrac{\bX_{E'}^T \bX_{E'}}{n}- \mathbf{{Q}}\right)\sqrt{n}(  \widehat{\beta}^{E'} -{\beta}^{E'}).
 \end{aligned}
 \end{equation*}
Thus, follows from observing ${\bX_{E'}^T \bX_{E'}}/{n}- \mathbf{{Q}}= o_p(1)$ and $\sqrt{n}(  \widehat{\beta}^{E'} -{\beta}^{E'})= O_p(1)$ (see Proposition 7 \cite{buja2015models}) that $\left(\dfrac{\bX_{E'}^T \bX_{E'}}{n}- \mathbf{{Q}}\right)\sqrt{n}( \widehat{\beta}^{E'} -{\beta}^{E'})= o_p(1)$ and hence,
\begin{equation}
\label{suff:stat}
\sqrt{n}( \widehat{\beta}^{E'} -{\beta}^{E'})=  
\mathbf{{Q}}^{-1} \dfrac{ \bX_{E'}^T}{\sqrt{n}} \left(\vec{y} -\bX_{E'} {\beta}^{E'}\right) + o_p(1).
\end{equation}
Further, it is easy to note that the asymptotic variance of $\sqrt{n}(\widehat{\vec{\beta}}^{E'} -\vec{\beta}^{E'})$ equals $\sigma^2\mathbf{Q}^{-1}$ under the above model assumptions.

Defining $\mathbf{{C}}$ as $\mathbb{E}_{P}(\bX_{-E'}^T \bX_{E'}/n)$, we can expand the term $\dfrac{\bX_{-E'}^T}{\sqrt{n}} \left(\vec{y} - \bX_{E'} \widehat{\beta}^{E'}\right)$, which identifies
\begin{equation*}
\sqrt{n}N_{-E'}=  \dfrac{\bX_{-E'}^T}{\sqrt{n}} \left(\vec{y} - \bX_{E'} {\beta}^{E'}\right) -\mathbf{{C}}\sqrt{n}(\widehat{\beta}^{E'} - {\beta}^{E'})- \left(\dfrac{\bX_{-E'}^T \bX_{E'}}{n}- \mathbf{{C}}\right)\sqrt{n}(\widehat{\beta}^{E'} - {\beta}^{E'}).
 \end{equation*}
 Plugging in $\sqrt{n}(\widehat{\beta}^{E'} - {\beta}^{E'})$ from \eqref{suff:stat} and noting that $\left(\dfrac{\bX_{-E'}^T \bX_{E'}}{n}- \mathbf{{C}}\right)\sqrt{n}(\widehat{\beta}^{E'} - {\beta}^{E'})=o_p(1)$, we have
 \begin{equation}
\label{nuisance:stat}
\sqrt{n}N_{-E'}=  \dfrac{\bX_{-E'}^T}{\sqrt{n}} \left(\vec{y} -  \bX_{E'} {\beta}^{E'}\right) -\mathbf{{C}}\mathbf{{Q}}^{-1}\dfrac{ \bX_{E'}^T}{\sqrt{n}} \left(\vec{y} - \bX_{E'} {\beta}^{E'}\right) + o_p(1).
 \end{equation}
Note that the asymptotic variance of $\sqrt{n}N_{-E'}$ equals $\sigma^2 \mathbf{N}^{-1}$,
where $ \mathbf{N}^{-1} = (\mathbf{P}- \mathbf{C}\mathbf{Q}^{-1}\mathbf{C}^T)$, with $\mathbf{P}= \mathbb{E}_{P}(\bX_{-E'}^T \bX_{-E'}/n)$. Further, the asymptotic covariance between $\sqrt{n}(\widehat{\vec{\beta}}^{E'} -\vec{\beta}^{E'})$ and $\sqrt{n} N_{-E'}$ is $0$, which follows from 
\[\Scale[0.90]{\mathbb{E}_{P}\left(\text{Cov}\left(\mathbf{Q}^{-1} \dfrac{ \bX_{E'}^T}{\sqrt{n}} \left(\vec{y} - \bX_{E'}\beta^{E'}\right), \dfrac{\bX_{-E'}^T}{\sqrt{n}} \left(\vec{y} - \bX_{E'}\beta^{E'}\right) -\mathbf{C}\mathbf{Q}^{-1}\dfrac{ \bX_{E'}^T}{\sqrt{n}} \left(\vec{y} - \bX_{E'}\beta^{E'}\right)\Big\lvert \bX\right)\right)=0.}\]

Now we look at the randomization term. Let $\beta_E^*({\beta}^{E'})= \mathbb{E}_P[\widehat{\beta}_E^\lambda]$, where $\widehat{\beta}_E^\lambda \in \mathbb{R}^{|E|}$ is the vector of the non zero coordinates of the Lasso solution in \eqref{carved:lasso}. The randomization term in \eqref{carved:randomization} equals
\begin{equation}
\begin{aligned}
\label{randomization:asymptotic}
\Omega_n &= -\dfrac{\bX^T}{\sqrt{n}}\left(\vec{y}- \bX_E\beta_E^*\right)+\dfrac{{\bX^{\mathcal{S}}}^T}{\rho\sqrt{n}}\left(\vec{y}^{\mathcal{S}}- \bX^{\mathcal{S}}_E\beta_E^*\right) + o_p(1),
\end{aligned}
\end{equation}
where the remainder term is $\left(\dfrac{\bX^T \bX_E}{n}- A\right)\sqrt{n}(\widehat{\beta}_E^\lambda - \beta_E^*) - \left(\dfrac{{\bX^{\mathcal{S}}}^T \bX^{\mathcal{S}}_E}{\rho n}-A\right)\sqrt{n}(\widehat{\beta}_E^\lambda - \beta_E^*),$ and $A= \mathbb{E}_P({\bX^T \bX_E}/{n})= \mathbb{E}_P({{\bX^{\mathcal{S}}}^T \bX^{\mathcal{S}}_E}/{\rho n})$.
Clearly, from equations \eqref{suff:stat}, \eqref{nuisance:stat} and \eqref{randomization:asymptotic}, it follows that 
\[\sqrt{n}(T_n-\mathbb{E}[T_n]) \sim N(0, \mathbf{\Sigma}_{P}); \;\Omega_n \sim N(0, \mathbf{\Sigma}_{\mathcal{G}}); \;\boldsymbol{\Sigma}_{P} = \sigma^2\begin{bmatrix} \mathbf{Q}^{-1} & 0 \\ 0 & \mathbf{N}^{-1} \end{bmatrix}.\]


The proof is finally complete by noting the block diagonal structure of the covariance between $\sqrt{n}T_n, \Omega_n$.  Observe that the covariance 
\[\text{Cov}\left( \dfrac{\bX^T}{\sqrt{n}}\left(\vec{y}- \bX_E\beta_E^*\right)-\dfrac{{\bX^{\mathcal{S}}}^T}{\rho\sqrt{n}}\left(\vec{y}^{\mathcal{S}}- \bX^{\mathcal{S}}_E\beta_E^*\right), \bX^T (\vec{y}-\bX_{E'}\beta^{E'}) \right)=0\]
which proves the asymptotic independence between $\sqrt{n}(T_n-\mathbb{E}[T_n])$ and $\Omega_n$. 
\end{proof}

\begin{proof}
\emph{Proposition \ref{K.K.T.:selection}:} 
We denote $\widehat{\beta}_E^\lambda \in \mathbb{R}^{|E|}$ as the vector of the selected coordinates of the Lasso solution, not shrunk to $0$. From the definition of $\Omega_n $ in \eqref{carved:randomization} and the K.K.T. conditions of LASSO, it follows that
\[\Omega_n = -\dfrac{\bX^T}{\sqrt{n}} \left(\vec{y} - \bX_E \widehat{\beta}_E^\lambda\right) + \begin{pmatrix}\lambda s^E\\ z_{-E}\end{pmatrix} \text{ where the subgradient vector equals at the solution} \]
\[\dfrac{\partial}{\partial\beta}(\lambda \|\beta\|_1)= \begin{pmatrix}\lambda s^E\\ z_{-E}\end{pmatrix}, \text{ and } \|z_{-E}\|_{\infty}<\lambda.\]

\noindent Noting that $$-{\bX^T}\vec{y}/{\sqrt{n}}= -\bX^T \bX_{E'} \widehat{\beta}^{E'}/{\sqrt{n}} -\bX^T (\vec{y} - \bX_{E'} \widehat{\beta}^{E'})/{\sqrt{n}},$$
the K.K.T. map now equals
$$\Omega_n =-\dfrac{\bX^T \bX_{E'}}{n} \sqrt{n}\widehat{\beta}^{E'} -\bX^T (\vec{y} - \bX_{E'} \widehat{\beta}^{E'})/{\sqrt{n}} +\dfrac{\bX^T \bX_E}{n} \sqrt{n}\widehat{\beta}_E^\lambda+ \begin{pmatrix}\lambda s^E\\ z_{-E}\end{pmatrix}.$$

\medskip

Using the facts that $({\bX_E^T \bX_{E'}}/{n}-\mathbf{P}^{E, E'}) = o_p(1)$, $({\bX_{-E}^T \bX_{E'}}/{n}-\mathbf{F}^{E, E'}) = o_p(1)$, $(\bX_{E}^T \bX_{E}/n-\mathbf{Q}^E)=o_p(1) $ and  $(\bX_{-E}^T \bX_{E}/n- \mathbf{C}^E)=o_p(1) $, coupled with the K.K.T. map above, we can write
\begin{equation}
\label{KKT:map}
\Omega_n = -\begin{bmatrix} \mathbf{P}^{E, E'} & \mathcal{I}^{E, E'} \\ \mathbf{F}^{E, E'} & \mathcal{J}^{E, E'}\end{bmatrix} \sqrt{n}T_n + \begin{bmatrix} \mathbf{Q}^{E} & 0 \\ \mathbf{C}^{E} & I\end{bmatrix} \begin{pmatrix}\sqrt{n}\widehat{\beta}_E^\lambda \\ z_{-E}\end{pmatrix} + \begin{pmatrix}\lambda s^E\\ 0\end{pmatrix} + o_p(1).
\end{equation}
The constraints equivalent to selection of 
$\left( \widehat{E},\widehat{S}^E \right) = \left( E, s^E \right)$ are given by
\[\text{diag}\left(\textnormal{sgn}(\widehat{\beta}_E^\lambda)\right) =  s^E, \|z_{-E}\|_{\infty} <\lambda\]
which can be equivalently written using \eqref{KKT:map} as
\begin{equation*}
\begin{aligned}
&-\text{diag}(s^E)(\mathbf{Q}^E)^{-1}\begin{pmatrix} \mathbf{P}^{E, E'} & \mathcal{I}^{E, E'} \end{pmatrix}\sqrt{n}T_n  -\begin{pmatrix} \text{diag}(s^E) (\mathbf{Q}^E)^{-1} & 0 \end{pmatrix}\Omega_n + o_p(1) \\
&\;\;\;\;\;\;\;\;\;\;\;\;\;\;\;\;\;\;\;\;\;\;\;\;\;\;\;\;\;\;\;\;\;\;\;\;\;\;\;\;\;\;\;\;\;\;\;\;\;\;\;\;\;\;\;\;\;\;\;\;\;\;\;\;\;\;\;\;\;\;\;\;\;\;\;\;\;\;\;\;\;\;\;\;\;\;\;\;\;\;\;\;\;\;< -\lambda\cdot \text{diag}(s^E) (\bQ^E)^{-1} s^E;
\end{aligned}
\end{equation*}
\begin{equation*}
\begin{aligned}
&\left(\begin{pmatrix}  \mathbf{F}^{E, E'} & \mathcal{J}^{E, E'}\end{pmatrix} - \mathbf{C}^E(\mathbf{Q}^E)^{-1} \begin{pmatrix} \mathbf{P}^{E, E'} & \mathcal{I}^{E, E'} \end{pmatrix}\right)\sqrt{n}T_n + \begin{pmatrix} -\mathbf{C}^E(
\mathbf{Q}^E)^{-1} & I \end{pmatrix} \Omega_n + o_p(1)  \\
&\;\;\;\;\;\;\;\;\;\;\;\;\;\;\;\;\;\;\;\;\;\;\;\;\;\;\;\;\;\;\;\;\;\;\;\;\;\;\;\;\;\;\;\;\;\;\;\;\;\;\;\;\;\;\;\;\;\;\;\;\;\;\;\;\;\;\;\;\;\;\;\;\;\;\;\;\;\;\;\;\;\;\;\;\;\;\;\;\;\;\;\;\;\;< \lambda \cdot\left(1 -\mathbf{C}^E(\mathbf{Q}^E)^{-1} s^E\right);
\end{aligned}
\end{equation*}
and
\begin{equation*}
\begin{aligned}
&\left(-\begin{pmatrix}  \mathbf{F}^{E, E'} & \mathcal{J}^{E, E'}\end{pmatrix} + \mathbf{C}^E(\mathbf{Q}^E)^{-1} \begin{pmatrix} \mathbf{P}^{E, E'} & \mathcal{I}^{E, E'} \end{pmatrix}\right)\sqrt{n}T_n + \begin{pmatrix} \mathbf{C}^E(\mathbf{Q}^E)^{-1} & -I\end{pmatrix} \Omega_n + o_p(1)  \\
&\;\;\;\;\;\;\;\;\;\;\;\;\;\;\;\;\;\;\;\;\;\;\;\;\;\;\;\;\;\;\;\;\;\;\;\;\;\;\;\;\;\;\;\;\;\;\;\;\;\;\;\;\;\;\;\;\;\;\;\;\;\;\;\;\;\;\;\;\;\;\;\;\;\;\;\;\;\;\;\;\;\;\;\;\;\;\;\;\;\;\;\;\;\;< \lambda\cdot \left(1 + \mathbf{C}^E(\mathbf{Q}^E)^{-1} s^E\right).
\end{aligned}
\end{equation*}
\end{proof}

\subsection{Proofs for Section 5}
\begin{proof}
\emph{Theorem \ref{thm:approx}:}
\noindent Under the parameterization $\sqrt{n}\vec{\beta}^{E'}_n = n^{\delta}\beta^*$, observe that the selection probability can be written as
\[\mathbb{P}\left( \begin{bmatrix} \mathbf{A}_E &  \mathbf{B}_E \end{bmatrix}(\sqrt{n}  \bar{Z}_n + E_n) +o_p(1)< b_E  -n^{\delta}\mathbf{A}_E\begin{pmatrix}\beta^* \\ 0 \end{pmatrix} \Big\lvert \beta^{E'}_n\right).\]

To prove the theorem, observe from Proposition \ref{asymptotic:distribution} that the asymptotic distribution of $\sqrt{n}\bar{Z}_n$ is a multivariate normal with a mean $0$ and covariance matrix given by $\boldsymbol{\Sigma}=\begin{bmatrix} \boldsymbol{\Sigma}_{P} & 0 \\ 0 & \boldsymbol{\Sigma}_{\mathcal{G}} \end{bmatrix}$ with $\boldsymbol{\Sigma}_{P} = \begin{bmatrix} \mathbf{Q}^{-1} & 0 \\ 0 & \mathbf{N}^{-1} \end{bmatrix}$ under $\sigma^2=1$. Further, $\sqrt{n}\bar{Z}_n$ satisfies a moderate deviations principle \citep{borovkov1978probabilities} under the existence of an exponential moment, assumed in \eqref{moment:condition}. Hence,
\begin{equation}
\label{mdp:Z}
\lim_{n\to \infty} \dfrac{1}{n^{2\delta}} \log \mathbb{P}(n^{-\delta}\begin{bmatrix} \mathbf{A}_E &  \mathbf{B}_E \end{bmatrix}\sqrt{n}  \bar{Z}_n <  - \mathbf{A}_E\begin{pmatrix}\beta^* & 0\end{pmatrix}^T\lvert \beta^{E'}_n) =- \displaystyle\inf_{(b'+ \beta^*, \eta', w') \in \mathcal{H}_0} \mathcal{R}(b', \eta', w').
\end{equation}
In the above display, the half-space is
$$\mathcal{H}_0 = \{ (b, \eta, w): \mathbf{A}_E(b, \eta)^T+ \mathbf{B}_E w< 0 \}$$ and the rate function equals
$$\mathcal{R}(b', \eta', w') = \sup_{(x, y, z) \in \mathbb{R}^{2p}} \left\{x^T b' + y^T\eta' + z^T w' - \log \mathbb{E}[\exp(x^T \widetilde{\beta}^{E'} + y^T \tilde{N}_{-E'} + z^T \tilde{W}_n)]\right\},$$
where the random variable $(\widetilde{\beta}^{E'}, \tilde{N}_{-E'}, \tilde{W}_n)$ is distributed as a Gaussian random variable with mean $0$ and covariance $\boldsymbol{\Sigma}$.
A simplification of the rate function by computing the conjugate of the gaussian log-MGF at $(b, \eta, w)$ yields the right-hand side 
as 
\[\displaystyle\inf_{(b'+ \beta^*, \eta', w') \in \mathcal{H}_0} \dfrac{b'^T \bQ b'}{2} + \dfrac{\eta'^T \mathbf{N} \eta'}{2} +\dfrac{w'^T \boldsymbol{\Sigma}_{\mathcal{G}}^{-1} w'}{2}\]
\[=\displaystyle\inf_{(b, \eta, w) \in \mathcal{H}_0} \dfrac{(b- \beta^*)^T \bQ (b- \beta^*)}{2} + \dfrac{\eta^T \mathbf{N} \eta}{2} +\dfrac{w^T \boldsymbol{\Sigma}_{\mathcal{G}}^{-1} w}{2}. \]

Now, observe that the probability we set out to approximate satisfies
\begin{equation*}
\begin{aligned}
& \dfrac{1}{n^{2\delta}}\Big( \log \mathbb{P}(n^{-\delta}\begin{bmatrix} \mathbf{A}_E &  \mathbf{B}_E \end{bmatrix}(\sqrt{n}  \bar{Z}_n + E_n) +o_p(1)< n^{-\delta} b_E  - \mathbf{A}_E\begin{pmatrix}\beta^* & 0\end{pmatrix}^T\lvert \beta^{E'}_n) \\
&\;\;\;\;\;\;\;- \log \mathbb{P}(n^{-\delta}\begin{bmatrix} \mathbf{A}_E &  \mathbf{B}_E \end{bmatrix}(\sqrt{n}  \bar{Z}_n + E_n) <  - \mathbf{A}_E\begin{pmatrix}\beta^* & 0\end{pmatrix}^T\lvert \beta^{E'}_n)\Big)
\to 0
\end{aligned}
\end{equation*}
as $n\to \infty$ under \eqref{b:condition}. Further, the assumption in \eqref{error:condition} coupled with the moderate deviations limit in \eqref{mdp:Z} leads to the conclusion:
\begin{equation*}
\begin{aligned}
& \lim_{n\to \infty} \dfrac{1}{n^{2\delta}} \log \mathbb{P}(n^{-\delta}\begin{bmatrix} \mathbf{A}_E &  \mathbf{B}_E \end{bmatrix}(\sqrt{n}  \bar{Z}_n + E_n) < - \mathbf{A}_E\begin{pmatrix}\beta^* & 0\end{pmatrix}^T\lvert \beta^{E'}_n) \\
&=\displaystyle\inf_{(b, \eta, w) \in \mathcal{H}_0} \dfrac{(b- \beta^*)^T \bQ (b- \beta^*)}{2} + \dfrac{\eta^T \mathbf{N} \eta}{2} +\dfrac{w^T \boldsymbol{\Sigma}_{\mathcal{G}}^{-1} w}{2}.
\end{aligned}
\end{equation*}


To conclude the proof of the Theorem, note that 
\[\Scale[0.88]{\displaystyle\lim_{n\to \infty} \displaystyle\inf_{(b, \eta, w) \in \mathcal{H}_n} \dfrac{(b- \beta^*)^T \bQ (b- \beta^*)}{2} + \dfrac{\eta^T \mathbf{N} \eta}{2} +\dfrac{w^T \boldsymbol{\Sigma}_{\mathcal{G}}^{-1} w}{2} = \displaystyle\inf_{(b, \eta, w) \in \mathcal{H}_0} \dfrac{(b- \beta^*)^T \bQ (b- \beta^*)}{2} + \dfrac{\eta^T \mathbf{N} \eta}{2} +\dfrac{w^T \boldsymbol{\Sigma}_{\mathcal{G}}^{-1} w}{2}}.\]
\end{proof}

\begin{proof}
\emph{Proposition \ref{eq:change-variables-optimization}:}
Substituting $w$ in the objective trivially yields the objective function in the transformed variables. We are left to verify the equivalence of constraints.
To complete the proof, consider the map in \eqref{KKT:map} in the proof of Proposition \ref{K.K.T.:selection} ignoring the $o_p(1)$ term 
\[n^{\delta} w = P_E n^{\delta}  \begin{pmatrix} b \\ \eta \end{pmatrix} + Q_E n^{\delta} o + r_E\]
which is based on solving the carved Lasso objective in \eqref{carved:lasso} with $\sqrt{n} T_n =n^{\delta} \begin{pmatrix} b \\ \eta \end{pmatrix}$, the implicit randomization 
$\Omega_n = n^{\delta} w$ and $\begin{pmatrix}\sqrt{n}\widehat{\beta}_E^\lambda \\ z_{-E} \end{pmatrix}= n^{\delta} o$. The K.K.T. conditions characterizing the solution of the carved lasso objective in this case are given by
\[\{o\in \mathbb{R}^p: \textnormal{sgn}(n^{\delta}o_{E}) = s^E,\;\|n^{\delta}o_{-E}\|_{\infty} \leq \lambda\}.\]
It follows from Proposition \ref{K.K.T.:selection} that the above constraints are equivalent to polyhedral constraints on $(b, \eta, w)$
\[\mathcal{H}_n=\left\{(b, \eta, w)\in \mathbb{R}^{2p}: A_E \begin{pmatrix} b\\ \eta \end{pmatrix} + B_E w \leq  n^{-\delta} b_E\right\}.\]
Thus, the constraints on optimizing variables $\{(b, \eta, w) \in \mathcal{H}_n\}$ under the change of variables map \eqref{change:variables} are equivalent to the constraints $\{(b, \eta, o) \in \RR^{2p}: o \in \mathcal{O}_n\}$
\end{proof}

\subsection{Proofs for Section 6}

\begin{proof}[Proof of Lemma \ref{strong:convexity}]
To see a proof, we first note that the optimization in the approximation in \eqref{univar-barrier} can be equivalently written as
\[-\rho\cdot n^{2\delta} \cdot \inf_{\bar{z} \in \RR} \left\{(\bar{z} -\beta^*)^2/2 + \frac{1}{\rho\cdot n^{2\delta}}\psi_{n^{-\delta}} (\bar{z})\right\}\]
through a variable substitution $w = \bar{z}-z$, followed by optimizing over $z$. Observe that with the equivalent approximating optimization, $\widetilde{L}_S^{n}(\beta_n)$ equals
\[\sqrt{n}\bar{y}_n n^{\delta}\beta^* - n^{2\delta}{\beta^*}^2/2 -n\bar{y}_n^2/2 + \rho n^{2\delta}\cdot{\beta^*}^2/2 - n^{2\delta}\cdot \sup_{\bar{z}\in \mathbb{R}} \bar{z} \rho \beta^* - \left\{\rho\bar{z}^2/2 +{n^{-2\delta}}\cdot\psi_{n^{-\delta}} (\bar{z}) \right\}\]
\[=\sqrt{n}\bar{y}_n n^{\delta}\beta^*-n\bar{y}_n^2/2 - n^{2\delta}\widetilde{C}_n\left(\beta^*\right),\;\;\;\;\;\;\;\;\;\;\;\;\;\;\;\;\;\;\;\;\;\;\;\;\;\;\;\;\;\;\;\;\;\;\;\;\;\;\;\;\;\;\;\;\;\;\;\;\;\;\;\;\;\;\;\;\;\;\;\;\;\;\;\;\;\;\;\;\;\;\;\;\;\;\;\;\;\;\;\;\;\;\;\;\;\;\;\;\]
where $\widetilde{C}_n\left(\beta^*\right) =  (1-\rho)\cdot {\beta^*}^2/2 + \bar{H}_n^*\left(\rho\beta^*\right).$
Observe that $\widetilde{C}_n(.)$ is strongly convex as $ (1-\rho)\cdot{\beta^*}^2/2$ is strongly convex with index $(1-\rho)$ and $\bar{H}_n^*(\rho\beta^*)$ is a convex function in $\beta^*$. It is straight forward from here to see that the indices of convexity are bounded below by $(1-\rho).$ 
\end{proof}

\begin{proof}[Proof of Lemma \ref{identity:convexity}]
Denoting $\hat{\beta}^* = n^{1/2-\delta}\widehat{\beta}_S$, the randomized selective MLE $\widehat{\beta}_S$ satisfies 
\[\sqrt{n}\bar{y}_n = n^{\delta} \grad \widetilde{C}_n(n^{1/2-\delta}\widehat{\beta}_S) = n^{\delta} \grad \widetilde{C}_n\left(\hat{\beta}^*\right),\; \text{ that is } \hat\beta^* = \grad\widetilde{C}_n^{-1}(n^{1/2-\delta}\bar{y}_n).\] Thus, we have
\[n^{1-2\delta}(\widehat{\beta}_S-\beta_n)^2=(\hat{\beta}^*-\beta^*)^2 = (\grad\widetilde{C}_n^{-1}(n^{1/2-\delta}\bar{y}_n) - \beta^*)^2 =(\grad\widetilde{C}_n^{*}(n^{1/2-\delta}\bar{y}_n) - \grad\widetilde{C}_n^{*}(\grad \widetilde{C}_n(\beta^*)))^2.\]
Lemma \ref{strong:convexity} shows that $\widetilde{C}_n(\beta^*)$ is strongly convex with indices of convexity $m_n\geq M = (1-\rho)$ and the proof is complete by using the fact that the convex conjugate of a strongly convex function with index $m_n$ is Lipschitz smooth with Lipschitz index $1/m_n$.
Hence, we have
\[n^{1-2\delta}(\widehat{\beta}_S-\beta_n)^2 \leq \cfrac{1}{m_n^2}(n^{1/2-\delta}\bar{y}_n- \grad \widetilde{C}_n(\beta^*))^2\leq \cfrac{1}{(1-\rho)^2}(n^{1/2-\delta}\bar{y}_n - \grad \widetilde{C}_n(\beta^*))^2.\]
\end{proof}

\begin{proof}[Proof of Theorem \ref{consistency:approximate:MLE}]
For a fixed $\epsilon >0$, Markov's inequality and Lemma \ref{identity:convexity} yields
\begin{equation*}
\begin{aligned}
\mathbb{P}(n^{1/2-\delta}|\widehat{\beta}_S-\beta_n|>\epsilon \lvert\; \sqrt{n}(\bar{Y}_n + W_n)/n^{\delta} \in \mathcal{K} ) 
&\leq \dfrac{\mathbb{E}[(\sqrt{n}\bar{Y}_n -n^{\delta}\grad \widetilde{C}_n(\beta^*))^2\lvert\; \sqrt{n}(\bar{Y}_n + W_n)/n^{\delta} \in \mathcal{K}]}{n^{2\delta}(1-\rho)^2\epsilon^2}.
\end{aligned}
\end{equation*}

Denote ${C}_n(\beta^*)$ as the true counterpart of the approximate sequence $\widetilde{C}_n(\beta^*)$ with the (log-) exact selection probability plugged in. That is, letting 
$$n^{2\delta}{C}_n(\beta^*) = n^{2\delta} {\beta^*}^2/2 + \log\bar{\Phi}\left(-\sqrt{\rho}\cdot n^{\delta}\beta^*\right),$$
we have
\begin{equation*}
\begin{aligned}
& \dfrac{\mathbb{E}\left[(\sqrt{n}\bar{Y}_n -n^{\delta}\grad \widetilde{C}_n(\beta^*))^2\;\Big\lvert\;\sqrt{n}(\bar{Y}_n + W_n)/n^{\delta} \in \mathcal{K}\right]}{n^{2\delta}(1-\rho)^2\cdot\epsilon^2}\\  
&= \dfrac{\mathbb{E}\left[(\sqrt{n}\bar{Y}_n -n^{\delta}\grad {C}_n(\beta^*))^2\;\Big\lvert\;\sqrt{n}(\bar{Y}_n + W_n)/n^{\delta} \in \mathcal{K}\right]}{n^{2\delta}(1-\rho)^2\cdot\epsilon^2} +   \dfrac{(\grad {C}_n(\beta^*)-\grad\widetilde{C}_n(\beta^*))^2}{(1-\rho)^2\cdot\epsilon^2}\nonumber\\
&= \dfrac{\Var\left[\sqrt{n}\bar{Y}_n\;\lvert\;\sqrt{n}(\bar{Y}_n+ W_n)/n^{\delta} \in \mathcal{K}\right]}{n^{2\delta}(1-\rho)^2\cdot\epsilon^2}+   \dfrac{(\grad {C}_n(\beta^*)-\grad\widetilde{C}_n(\beta^*))^2}{(1-\rho)^2\cdot\epsilon^2}\nonumber \\
&\leq \dfrac{\Var\left[\sqrt{n}\bar{Y}_n\right]}{n^{2\delta}(1-\rho)^2\cdot\epsilon^2} +  \dfrac{(\grad {C}_n(\beta^*)-\grad\widetilde{C}_n(\beta^*))^2}{(1-\rho)^2\cdot\epsilon^2}
\end{aligned}
\end{equation*}
The last step uses the fact that variance of a gaussian random variable reduces when restricted to a convex set (see \cite{kanter1977reduction} for a proof) and thus, the first term converges to $0$ as $n \to \infty$. Convergence of the second term to $0$ follows from a combination of corollary \ref{carved:LDP:approximate:selective:posterior} and the properties of convexity and differentiability $\widetilde{C}_n(\beta^*)$. 
\end{proof}

\begin{proof}[Proof of Theorem \ref{thm:inconsistency}]
The maximizer of \eqref{soft:max:MLE} is 
\[
\widehat{\beta}_S = \bar{Y}_n - \frac{1}{\sqrt{n}(\sqrt{n}\bar{Y}_n){(\sqrt{n}\bar{Y}_n + 1)}}. 
\]
Denoting $Z_n = (\sqrt{n}|\beta^*|)\sqrt{n}\bar{Y}_n$ and $b(z)= \log\left(1+\dfrac{1}{z}\right)$ we have
$$
\begin{aligned}
\widehat{\beta}_S - \beta_n &= \bar{Y}_n - \beta^* - n^{-1/2} \frac{1}{\sqrt{n}\bar{Y}_n{(\sqrt{n}\bar{Y}_n + 1)}}\\
& = \bar{Y}_n - \beta^* +  n^{-1/2} \nabla b(n^{1/2} \bar{Y}_n)\\
& = \frac{Z_n}{n |\beta^*|}- \beta^*  +  n^{-1/2}\nabla b\left(\frac{Z_n}{n^{1/2}|\beta^*|} \right).
\end{aligned}
$$
For $\beta^*<0$, $Z_n = (n^{1/2}|\beta^*|) n^{1/2} \bar{Y}_n$ converges in law to an exponential 
random variable with mean 1. Further note that, 
\[z\nabla b(z) \to -K \text{ as } z\downarrow 0 \]
for $K=1$.
Using these two facts, we have $\dfrac{Z_n}{n |\beta^*|}=o_p(1)$ and  
\[n^{-1/2}\nabla b\left(\frac{Z_n}{n^{1/2}|\beta^*|} \right) = \left(\frac{|\beta^*|}{Z_n}\right)\frac{Z_n}{n^{1/2}|\beta^*|}\nabla b\left(\frac{Z_n}{n^{1/2}|\beta^*|} \right) = \left(\frac{|\beta^*|}{Z_n}\right) W_n \]
where $W_n \to -K \text{ as } n \to \infty.$ Hence, we can approximate the sequence of random variables $\widehat{\beta}_n - \beta^*$ in distribution by the random variable \[-\frac{K|\beta^*|}{Z} - \beta^* \text{ where } Z \sim \text{Exp}(1).\]
Letting $\delta = -\beta^*>0$, we conclude the proof by noting that
$$
\begin{aligned}
\mathbb{P}\left(|\widehat{\beta}_S - \beta_n|>\frac{\delta}{2}\Big\lvert \sqrt{n} \bar{Y}_n >0\right)  &\approx \mathbb{P}\left(\Big\lvert \frac{K|\beta^*|}{Z} + \beta^*\Big \lvert>\frac{\delta}{2} \;\Big\lvert \sqrt{n} \bar{Y}_n >0\right)\\
& \geq \mathbb{P}\left(-\frac{K|\beta^*|}{Z} - \beta^*>\frac{\delta}{2} \;\Big\lvert \sqrt{n} \bar{Y}_n>0\right) \\
& = \mathbb{P}\left(Z > -\frac{K|\beta^*|}{\beta^* + \delta/2}\;\Big\lvert \sqrt{n} \bar{Y}_n>0\right)\\
&= \exp\left(-\frac{K|\beta^*|}{\beta^* + \delta/2}\right) = \exp(2K)>0.
\end{aligned}
$$
\end{proof}

\begin{proof}[Proof of Theorem \ref{posterior:consistency:uni}]
Denoting $\widehat{\beta}_S$ to be the selection-adjusted MLE, fix $\epsilon>0$. Now, observe that
\begin{equation*}
\begin{aligned}
\Pi_S\left(\mathcal{B}^c( \beta_n, \delta)\lvert \bar{y}_n\right) &= \dfrac{\underset{\mathcal{B}^c( \beta_n, \delta)}{\int}\pi(b_n)\cdot \exp( \widetilde{L}_S^{n}(b_n))db_n}{\int \pi(b_n)\cdot\exp(\widetilde{L}_S^{n}(b_n))db_n}\\
&= \dfrac{\underset{\mathcal{B}^c( \beta_n, \delta)}{\int}\pi(b_n)\cdot \exp\{{\widetilde{L}_S^{n}(b_n)}-{\widetilde{L}_S^{n}(\widehat{\beta}_S)}\}db_n}{\int \pi(b_n)\cdot \exp\{{\widetilde{L}_S^{n}(b_n)}-{\widetilde{L}_S^{n}(\widehat{\beta}_S)}\}db_n}\\
&\leq \dfrac{\underset{\mathcal{B}^c(\beta_n, \delta)}{\int}\pi(b_n)\cdot \exp(-n (1-\rho) \cdot(\widehat{\beta}_S-b_n)^2/2)db_n}{\underset{\mathcal{B}(\beta_n, \delta)}{\int} \pi(b_n)\cdot \exp(-n\cdot (\widehat{\beta}_S-b_n)^2/2)db_n}.
\end{aligned}
\end{equation*}
Under the considered parameterization, let $\hat{\beta}^*= n^{1/2-\delta}\widehat{\beta}_S$.
Now, the last inequality follows by noting that
\[{\widetilde{L}_S^{n}(b_n)}-{\widetilde{L}_S^{n}(\widehat{\beta}_S)} =  \sqrt{n}\bar{y}_n( n^{\delta}\beta^* -n^{\delta}\hat\beta^*)- n^{2\delta}\widetilde{C}_n\left(\beta^*\right) + n^{2\delta}\widetilde{C}_n\left(\hat{\beta}^*\right)\]
and a Taylor expansion around of $\widetilde{C}_n\left(\beta^*\right)$ around $\hat{\beta}^*$ yields 
\[{\widetilde{L}_S^{n}(b_n)}-{\widetilde{L}_S^{n}(\widehat{\beta}_S)} = -n \cdot (\widehat{\beta}_S-b_n)^2 \grad^2 \widetilde{C}_n\left( \mathcal{R}(\hat{\beta}^*, \beta^*)\right) /2,\]
as the selection-adjusted MLE satisfies $\sqrt{n}\bar{y}_n = n^{\delta}\grad \widetilde{C}_n(\hat{\beta}^*).$
Finally, for any $\beta'$ we note that $(1-\rho) \leq \grad^2 \widetilde{C}_n(\beta') \leq 1$ and thus we have the last inequality. 
\medskip

Fix $\delta>0$, let $r \in (0,1)$ and $s<r \in (0,1)$ and observe that
\begin{equation*}
\begin{aligned}
&\mathbb{P}(\|\widehat{\beta}_S-\beta_n\| \leq r\delta \lvert\; \sqrt{n}(\bar{Y}_n + W_n)/n^{\delta} \in \mathcal{K}) \\
&\leq  \mathbb{P}(\|\widehat{\beta}_S-b_n\| \geq (1-r)\delta \text{ for all } b_n \in \mathcal{B}^c(\beta_n, \delta) \text{ and }\\
& \;\;\;\;\;\;\;\;\;\;\;\;\; \|\widehat{\beta}_S-b_n\| \leq (s+r)\delta \text{ for all } b_n \in \mathcal{B}(\beta_n, s\delta);  \lvert\;  \sqrt{n}(\bar{Y}_n + W_n)/n^{\delta} \in \mathcal{K})\\
&\leq  \mathbb{P}\left(\Pi_S(\mathcal{B}^c(\beta_n, \delta)\lvert  \bar{Y}_n) \leq \dfrac{\exp(-n(1-\rho)(1-r)^2\delta^2/2)\pi(\mathcal{B}^c(\beta_n, \delta))}{\exp(- n(r+s)^2\delta^2/2)\pi(\mathcal{B}(\beta_n, s\delta))}\Big\lvert\; \sqrt{n}(\bar{Y}_n + W_n)/n^{\delta} \in \mathcal{K}\right)\\
&= \mathbb{P}\Big(\Pi_S(\mathcal{B}^c(\beta_n, \delta)\lvert \bar{Y}_n) \leq \dfrac{\exp(-n\cdot((1-\rho)(1-r)^2-(r+s)^2)\delta^2/2)\pi(\mathcal{B}^c(\beta_n, \delta))}{\pi(\mathcal{B}(\beta_n, s\delta))}\\
&\;\;\;\;\;\;\;\;\;\;\;\;\;\;\;\;\;\;\;\;\;\;\;\;\;\;\;\;\;\;\;\;\;\;\;\;\;\;\;\;\;\;\;\;\;\;\;\;\;\;\;\;\;\;\;\;\;\;\;\;\;\;\;\;\;\;\;\;\;\;\;\;\;\;\;\;\;\;\;\;\;\;\;\;\;\;\;\;\;\;\;\;\;\;\;\;\;\;\;\;\;\;\;\;\;\Big\lvert\; \sqrt{n}(\bar{Y}_n + W_n)/n^{\delta} \in \mathcal{K}\Big)\\
& \leq \mathbb{P}\left(\Pi_S(\mathcal{B}^c(\beta_n, \delta)\lvert \bar{Y}_n) \leq \epsilon \Big\lvert\; \sqrt{n}(\bar{Y}_n + W_n)/n^{\delta} \in \mathcal{K}\right) \text{ for sufficiently large } n.
\end{aligned}
\end{equation*}
We can choose $r, s<r \in (0,1)$ in the penultimate step such that
\[(1-\rho)(1-r)^2-(r+s)^2> (1-\rho)(1-r)^2-4r^2>0.\]
In fact, choosing $r>0$ and smaller than the positive root $\sqrt{(1-\rho)} (2- \sqrt{(1-\rho)})/(4-(1-\rho))$, follows the last step as $\exp(-n\cdot((1-\rho)(1-r)^2-(r+s)^2)\delta^2/2)$ can be made smaller than $\epsilon>0$ for sufficiently large $n$. The above argument implies that 
\[\mathbb{P}(\|\widehat{\beta}_S-\beta_n\| \leq r\delta \lvert\; \sqrt{n}(\bar{Y}_n + W_n)/n^{\delta} \in \mathcal{K}) \to 1\] as $n\to \infty$ which, in turn leads to consistency of the selective posterior under the selective law at $\beta_n$.
\end{proof}

\begin{proof}
\emph{Lemma \ref{log:partition:gen}:}
To prove this, we start with the optimization objective involved in the approximation $\log \widetilde{\mathbb{P}}(\mathbf{A}_E \sqrt{n}T_n + \mathbf{B}_E\sqrt{n}W_n < b_E\lvert \beta^{E'}_n)$. Noting that the optimizing variables $b$ are constraint-free and the optimization problem in $b$ is a quadratic, we begin by optimizing over $b\in \RR^{E'}$ in the approximating optimization
\begin{equation*}
\begin{aligned}
& \inf_{(b, \eta, o) \in \RR^{2p}} \Big\{(b-\beta^*)^T \bQ (b- \beta^*)/2 + \eta^T \mathbf{N} \eta/2 \\
&\;\;\;\;\;\;\;\;\;\;\;\;\;\;\;+ \left(P_E^{E'} b + P_E^{-E'} \eta + Q_E o + r_E/n^{\delta}\right)^T \mathbf{\Sigma}_{\mathcal{G}}^{-1} \left(P_E^{E'} b + P_E^{-E'} \eta  + Q_E o + r_E/n^{\delta}\right)\Big/2\\
&\;\;\;\;\;\;\;\;\;\;\;\;\;\;\;+ \psi_{n^{-\delta}}(o_E, o_{-E})\Big\}.
\end{aligned}
\end{equation*}
Denoting $ \mathcal{L}(\eta, o)= P_E^{-E'} \eta + Q_E o + r_E/n^{\delta}$, optimizing over $b$, we have the above problem equivalent to an optimization in $(\eta, o)$:
\begin{equation*}
\begin{aligned}
& {\beta^*}^T\bQ \beta^*/2 -{\beta^*}^T (\bQ(\bQ+ {P_E^{E'}}^T \mathbf{\Sigma}_{\mathcal{G}}^{-1}P_E^{E'})^{-1}\bQ)\beta^*/2  + {\beta^*}^T \bQ (\bQ+ {P_E^{E'}}^T \mathbf{\Sigma}_{\mathcal{G}}^{-1}P_E^{E'})^{-1}{P_E^{E'}}^T\mathbf{\Sigma}_{\mathcal{G}}^{-1}r_E/n^{\delta}\\
&+ \inf_{(\eta, o)}\Big\{ {\beta^*}^T \bQ (\bQ+ {P_E^{E'}}^T \mathbf{\Sigma}_{\mathcal{G}}^{-1}P_E^{E'})^{-1}{P_E^{E'}}^T\mathbf{\Sigma}_{\mathcal{G}}^{-1}\begin{bmatrix} P_E^{-E'} & Q_E \end{bmatrix} \begin{pmatrix} \eta \\ o\end{pmatrix} + \eta^T \mathbf{N} \eta/2 \\
& \;\;\;\;\;\;\;+\mathcal{L}(\eta, o)^{T}(\mathbf{\Sigma}_{\mathcal{G}}^{-1}- \mathbf{\Sigma}_{\mathcal{G}}^{-1}P_E^{E'} (\bQ +  {P_E^{E'}}^T \mathbf{\Sigma}_{\mathcal{G}}^{-1}P_E^{E'})^{-1}{P_E^{E'}}^T\mathbf{\Sigma}_{\mathcal{G}}^{-1}) \mathcal{L}(\eta, o)/2 + \psi_{n^{-\delta}}(o_E, o_{-E})\Big\}.
\end{aligned}
\end{equation*}
Clearly, then the sequence of approximate log-partition functions equals
\begin{align*} 
& n^{2\delta}\Big({\beta^*}^T (\bQ(\bQ+ {P_E^{E'}}^T \mathbf{\Sigma}_{\mathcal{G}}^{-1}P_E^{E'})^{-1}\bQ)\beta^*/2 -{\beta^*}^T \bQ (\bQ+ {P_E^{E'}}^T \mathbf{\Sigma}_{\mathcal{G}}^{-1}P_E^{E'})^{-1}{P_E^{E'}}^T\mathbf{\Sigma}_{\mathcal{G}}^{-1}r_E/n^{\delta}\\
& +\sup_{(\eta, o)} \Big\{{\beta^*}^T \bQ M_1^T (\eta, o)^T - \mathcal{L}(\eta, o)^{T}(\mathbf{\Sigma}_{\mathcal{G}}^{-1}- \mathbf{\Sigma}_{\mathcal{G}}^{-1}P_E^{E'} (\bQ +  {P_E^{E'}}^T \mathbf{\Sigma}_{\mathcal{G}}^{-1}P_E^{E'})^{-1}{P_E^{E'}}^T\mathbf{\Sigma}_{\mathcal{G}}^{-1}) \mathcal{L}(\eta, o)/2\\
&\;\;\;\;\;\;\;\;\;\;\;\;\; -\eta^T \mathbf{N} \eta/2- \psi_{n^{-\delta}}(o_E, o_{-E})\Big\}.
\end{align*}
The above expression can be written as $n^{2\delta}\widetilde{C}_n(\bQ\beta^*)$ where $\widetilde{C}_n(\bQ\beta^*)$ equals
\[{\beta^*}^T (\bQ(\bQ+ {P_E^{E'}}^T \mathbf{\Sigma}_{\mathcal{G}}^{-1}P_E^{E'})^{-1}\bQ)\beta^*/2 +  {h}^*\left(M_1\bQ\beta^*\right)+ {\beta^*}^T \bQ M_2\]
with $M_1, M_2$ and ${h}^*(\cdot)$ is defined in the Lemma.
\end{proof}

\begin{proof}
\emph{Lemma \ref{strong:convexity:gen}:}
Based on the representation formula for $\widetilde{C}_n(\bQ\beta^*)$ derived in Lemma \ref{log:partition:gen} that gives 
\[\widetilde{C}_n(\bQ\beta^*) = {\beta^*}^T \bQ(\bQ+ {P_E^{E'}}^T \mathbf{\Sigma}_{\mathcal{G}}^{-1}P_E^{E'})^{-1}\bQ\beta^*/2 +  {h}^*\left(M_1\bQ\beta^*\right)+ {\beta^*}^T \bQ M_2,\]
 we are able to represent sequence $\widetilde{C}_n(\bQ\beta^*)$ as the sum of a positive definite quadratic form and a convex function, under the condition that $\mathbf{X}_{E'}$ is of full column rank. 
Thus, follows the strong convexity of the sequence 
$\widetilde{C}_n(\bQ\beta^*)$. Finally, the indices of strong convexity are bounded below by $\lambda_{\text{min}}$ which follows from observing
\[{\beta^*}^T \bQ(\bQ+ {P_E^{E'}}^T \mathbf{\Sigma}_{\mathcal{G}}^{-1}P_E^{E'})^{-1}\bQ\beta^*/2 \succ \lambda_{\text{min}} \cdot {\beta^*}^T \bQ\bQ\beta^*. \]
\end{proof}

\begin{proof}[Proof of Theorem \ref{consistency:gen}]
Denoting $\bar{\alpha}= \bQ\beta^*$, a natural parameterization in the approximate log-likelihood
\[-n(\widehat{\beta}^{E'} -\beta^{E'}_n)^T \bQ(\widehat{\beta}^{E'} -\beta^{E'}_n)/2 - \log \widetilde{\mathbb{P}}(\mathbf{A}_E \sqrt{n}T_n + \mathbf{B}_E\sqrt{n}W_n < b_E\lvert \beta^{E'}_n)\]
\[\;\;\;\;\;\;\;\;\;\;\;\;\;\;= n^{\delta}{\beta^*}^{T} \bQ \sqrt{n}\widehat{\beta}^{E'} - n^{2\delta}\widetilde{C}_n(\bQ\beta^*) - n(\widehat\beta^{E'})^T\bQ \widehat{\beta}^{E'}= n^{\delta}\bar{\alpha}^T \sqrt{n}\widehat{\beta}^{E'} - n^{2\delta}\widetilde{C}_n(\bar{\alpha}) - n(\widehat\beta^{E'})^T\bQ \widehat{\beta}^{E'},\]
let the MLE of the parameters $\bar{\alpha}$ be $\widehat{\bar{\alpha}}$. An estimating equation for the MLE for $\bar{\alpha}$ can be now written as
\[n^{1/2-\delta}\widehat{\beta}^{E'} = \grad\widetilde{C}_n(\widehat{\bar{\alpha}}).\] 
Now, using the strong convexity of $\widetilde{C}_n(\cdot)$ in Lemma \ref{strong:convexity:gen}, we make 
 the crucial observation that the MLE sequence satisfies the following contraction inequality
\[\|\widehat{\bar{\alpha}}-\bar{\alpha}\|^2 = \|\grad\widetilde{C}_n^{-1}(n^{1/2-\delta}\widehat{\beta}^{E'}) - \bar\alpha\|^2 =\|\grad\widetilde{C}_n^{*}(n^{1/2-\delta}\widehat{\beta}^{E'}) - \grad\widetilde{C}_n^{*}(\grad \widetilde{C}_n(\bar{\alpha}))\|^2.\]
which leads to the contraction inequality
\begin{equation}
\label{imp:contraction}
\|\widehat{\bar{\alpha}}-\bar{\alpha}\|^2 \leq \dfrac{1}{\lambda_{\text{min}}^2} \|n^{1/2-\delta}\widehat{\beta}^{E'} - \grad \widetilde{C}_n(\bar{\alpha})\|^2
\end{equation}
with $\lambda_{\text{min}}$ defined in Lemma \ref{strong:convexity:gen}. The fact we use in deriving the above contraction is that the convex conjugate of a strongly convex function with index $M$ is Lipschitz smooth with Lipschitz index $M^{-1}$.\\

Denoting $\lambda_{\text{min}}^{\bQ}$ as the smallest eigen value of $\bQ\bQ$, we observe that 
\[n^{1-2\delta}\|\widehat{\beta}^{E'}_S-\beta^{E'}_n\|^2  \leq (\lambda_{\text{min}}^{\bQ})^{-1}\|\widehat{\bar\alpha}-\bar\alpha\|^2.\]

Now, fix an $\epsilon_0 >0$. Let $n^{2\delta}{C}_n(\bQ\beta^*)= n{\beta^{E'}_n}^T \bQ \beta^{E'}_n/2 +\log\mathbb{P}(\mathbf{A}_E \sqrt{n}T_n + \mathbf{B}_E\sqrt{n}W_n < b_E\lvert \beta^{E'}_n)$ represent the exact log-partition functions where ${C}_n(\bQ\beta^*)$ is the true counterpart of the approximate sequence $\widetilde{C}_n(\bQ\beta^*)$ with the (log-) exact selection probability plugged in. Applying the contraction in \eqref{imp:contraction}, coupled with the Markov's inequality, we obtain
\begin{equation*}
\begin{aligned}
&\mathbb{P}(\|\widehat{\bar\alpha}-\bar\alpha\|>\epsilon_0 \lvert\;\mathbf{A}_E \sqrt{n}T_n + \mathbf{B}_E\sqrt{n}W_n < b_E)\\
&\leq  \dfrac{\mathbb{E}\left[\|\sqrt{n}\widehat{\beta}^{E'} -n^{\delta}\grad \widetilde{C}_n(\bar\alpha)\|^2\;\Big\lvert\;\mathbf{A}_E \sqrt{n}T_n + \mathbf{B}_E\sqrt{n}W_n < b_E\right]}{n^{2\delta}\cdot\lambda_{\text{min}}^2\cdot\epsilon_0^2}\\  
&=   \dfrac{\mathbb{E}\left[\|\sqrt{n}\widehat{\beta}^{E'} -n^{\delta}\grad {C}_n(\bar\alpha)\|^2\;\Big\lvert\;\mathbf{A}_E \sqrt{n}T_n + \mathbf{B}_E\sqrt{n}W_n < b_E\right]}{n^{2\delta}\cdot\lambda_{\text{min}}^2\cdot\epsilon_0^2} +   \dfrac{\|\grad {C}_n(\bar\alpha)-\grad\widetilde{C}_n(\bar\alpha)\|^2}{\lambda_{\text{min}}^2\cdot\epsilon_0^2}\nonumber\\
&= \dfrac{O(1)}{n^{2\delta}\lambda_{\text{min}}^2\cdot\epsilon_0^2}+   \dfrac{\|\grad {C}_n(\beta^*)-\grad\widetilde{C}_n(\beta^*)\|^2}{\lambda_{\text{min}}^2\cdot\epsilon_0^2}\nonumber.
\end{aligned}
\end{equation*}
The last step uses a selective Central Limit Theorem proved in \cite{panigrahi2018asymptotic} to conclude that
\[\mathbb{E}\left[(\sqrt{n}\widehat{\beta}^{E'} -n^{\delta}\grad {C}_n(\bQ\beta^*))^2\;\Big\lvert\;\mathbf{A}_E \sqrt{n}T_n + \mathbf{B}_E\sqrt{n}W_n < b_E\right] = O(1)\]
 and the second term converges to $0$ due to the properties of convexity and differentiability $\widetilde{C}_n(\cdot)$. Thus
 \[\mathbb{P}(\|\widehat{\bar\alpha}-\bar\alpha\|>\epsilon_0 \lvert\; \mathbf{A}_E \sqrt{n}T_n + \mathbf{B}_E\sqrt{n}W_n < b_E)\to 0 \text{ as } n\to \infty.\]
\noindent The proof of consistency is now complete by noting that
\begin{equation*}
\begin{aligned}
&\mathbb{P}(n^{1/2-\delta}\|\widehat{\beta}^{E'}_S -\beta^{E'}_n\|>\epsilon \lvert\;  \mathbf{A}_E \sqrt{n}T_n + \mathbf{B}_E\sqrt{n}W_n < b_E) \\
&\leq\mathbb{P}(\|\hat{\bar\alpha}_n-\bar\alpha\|>(\lambda_{\text{min}}^{\bQ})^{1/2}\epsilon \lvert\;  \mathbf{A}_E \sqrt{n}T_n + \mathbf{B}_E\sqrt{n}W_n < b_E).
\end{aligned}
\end{equation*}
Hence, setting $\epsilon_0 = (\lambda_{\text{min}}^{\bQ})^{1/2}\epsilon$, we have
$n^{1/2-\delta}\|\widehat{\beta}^{E'}_S -\beta^{E'}_n\|$ converges in probability to $0$ as $n\to \infty$ under the selective law, which proves consistency of the selective MLE at a rate of $n^{1/2-\delta}$.
\end{proof}

\begin{proof}[Proof of Lemma \ref{likelihood:ratio:carved:gen}]
The difference of the logarithms of the carved likelihood at $\beta^{E'}_n\equiv n^{\delta-1/2}\beta^*$ and the selective MLE $\widehat{\beta}^{E'}_S\equiv n^{\delta-1/2}\hat{\beta}^*$ is given by
\begin{equation*}
\widetilde{L}_S^{n}(\beta^{E'}_n)-\widetilde{L}_S^{n}(\widehat{\beta}^{E'}_S) = \sqrt{n}(\widehat{\beta}^{E'})^T \cdot n^{\delta}\bQ(\beta^* - \hat{\beta}^*) - n^{2\delta} \cdot\left(\widetilde{C}_n (\bQ\beta^*) - \widetilde{C}_n (\bQ\hat{\beta}^*)\right).
\end{equation*}
Using the estimating equation for the selective MLE for $\beta^*$: $n^{1/2-\delta}\widehat{\beta}^{E'} = \grad\widetilde{C}_n(\bQ\hat{\beta}^*)$ and 
a first order Taylor series expansion of $\widetilde{C}_n (\bQ\beta^*)$ around $\bQ\hat{\beta}^*$ yields the difference of log-likelihoods as
\[-n^{2\delta} (\hat{\beta}^*-\beta^*)^T \bQ\grad^2\widetilde{C}_n (\mathcal{R}(\bQ\hat{\beta}^*, \bQ\beta^*)) \bQ (\hat{\beta}^*-\beta^*)/2,\]
where $ (\hat{\beta}^*-\beta^*)^T \bQ\grad^2\widetilde{C}_n (\mathcal{R}(\bQ\hat{\beta}^*, \bQ\beta^*)) \bQ (\hat{\beta}^*-\beta^*)$ is the remainder term of the expansion.
Note that, it follows from the representation formula for $\widetilde{C}_n (\bQ\beta^*)$ in Lemma \ref{log:partition:gen} for a $\beta^*$ that $\bQ\grad^2\widetilde{C}_n (\bQ\beta^*)\bQ \succ \widetilde\lambda_{\text{min}} \cdot I$,
where $\widetilde\lambda_{\text{min}}$ is the smallest eigen value of $\bQ (\bQ+ {P_E^{E'}}^T \mathbf{\Sigma}_{\mathcal{G}}^{-1}P_E^{E'})^{-1}\bQ$.
Further, $\bQ\grad^2\widetilde{C}_n (\bQ\beta^*)\bQ \prec \bQ$, from which follows the conclusion of the Lemma by noting that $n^{\delta} (\hat{\beta}^*-\beta^*)=\sqrt{n}(\widehat{\beta}^{E'}_S-\beta^{E'}_n)$.
\end{proof}

\begin{proof}[Proof of Theorem \ref{posterior:consistency:gen}]
Fixing $\epsilon>0$, we let $\widehat{\beta}^{E'}_S$ denote the selection-adjusted MLE obtained by maximizing the truncated likelihood in \eqref{eq:approx-sel-mle} and recall that $\widetilde\lambda_{\text{min}}$ is the smallest eigen value of $\bQ(\bQ+ {P_E^{E'}}^T \mathbf{\Sigma}_{\mathcal{G}}^{-1}P_E^{E'})^{-1}\bQ$ defined in Lemma \ref{likelihood:ratio:carved:gen}. Now, we compute the posterior probability of $\mathcal{B}^c(\beta^{E'}_n, \delta)$ under the approximate carved posterior:
\begin{equation*}
\begin{aligned}
\Pi_S\left(\mathcal{B}^c(\beta^{E'}_n, \delta)\lvert \widehat{\beta}^{E'}\right) &= \dfrac{\underset{\mathcal{B}^c(\beta^{E'}_n, \delta)}{\int}\pi(b_n)\cdot \exp( \widetilde{L}_S^{n}(b_n))db_n}{\int \pi(b_n)\cdot\exp(\widetilde{L}_S^{n}(b_n))db_n}\\
&= \dfrac{\underset{\mathcal{B}^c(\beta^{E'}_n, \delta)}{\int}\pi(b_n)\cdot \exp\{{\widetilde{L}_S^{n}(b_n)}-{\widetilde{L}_S^{n}(\widehat{\beta}^{E'}_S)}\}db_n}{\int \pi(b_n)\cdot \exp\{{\widetilde{L}_S^{n}(b_n)}-{\widetilde{L}_S^{n}(\widehat{\beta}^{E'}_S)}\}db_n}\\
&\leq \dfrac{\underset{\mathcal{B}^c(\beta^{E'}_n, \delta)}{\int}\pi(b_n)\cdot \exp(-n \widetilde\lambda_{\text{min}} \cdot(\widehat{{\beta}}^{E'}_S-b_n)^T (\widehat{{\beta}}^{E'}_S-b_n)/2) db_n}{\underset{\mathcal{B}(\beta^{E'}_n, \delta)}{\int} \pi(b_n)\cdot \exp(-n\cdot (\widehat{{\beta}}^{E'}_S-b_n)^T \bQ (\widehat{{\beta}}^{E'}_S-b_n)/2)db_n}.
\end{aligned}
\end{equation*}
The last inequality follows from the conclusion of  Lemma \ref{likelihood:ratio:carved:gen} that bounds the likelihood ratios at $b_n$ and the selective MLE $\widehat{\beta}^{E'}_S$ from both above and below. Fix $\delta>0$, let $r \in (0,1)$ and $s<r \in (0,1)$ and let $\lambda^{\bQ}_{\text{max}}$ denote the largest eigen value of $\bQ\succ 0$. Finally, observe that
\begin{equation*}
\begin{aligned}
&\mathbb{P}(\|\widehat{\beta}^{E'}_S-\beta^{E'}_n\| \leq r\delta \lvert\; \mathbf{A}_E \sqrt{n}T_n + \mathbf{B}_E\Omega_n < b_E) \\
&\leq  \mathbb{P}(\|\widehat{\beta}^{E'}_S-b_n\| \geq (1-r)\delta \text{ for all } b_n \in \mathcal{B}^c(\beta^{E'}_n, \delta) \text{ and }\\
& \;\;\;\;\;\;\;\;\;\;\;\;\; \|\widehat{\beta}^{E'}_S-b_n\| \leq (s+r)\delta \text{ for all } b_n \in \mathcal{B}(\beta^{E'}_n, s\delta);  \lvert\;  \mathbf{A}_E \sqrt{n}T_n + \mathbf{B}_E\Omega_n < b_E)\\
&\leq  \mathbb{P}\Big(\Pi_S(\mathcal{B}^c(\beta^{E'}_n, \delta)\lvert  \widehat{\beta}^{E'}) \leq \dfrac{\exp(-n\widetilde\lambda_{\text{min}}(1-r)^2\delta^2/2)\pi(\mathcal{B}^c(\beta^{E'}_n, \delta))}{\exp(- n\lambda^{\bQ}_{\text{max}}(r+s)^2\delta^2/2)\pi(\mathcal{B}(\beta^{E'}_n, s\delta))}\\
&\;\;\;\;\;\;\;\;\;\;\;\;\;\;\;\;\;\;\;\;\;\;\;\;\;\;\;\;\;\;\;\;\;\;\;\;\;\;\;\;\;\;\;\;\;\;\;\;\;\;\;\;\;\;\;\;\;\;\;\;\;\;\;\;\;\;\;\;\;\;\;\;\;\;\;\;\;\;\;\;\;\;\;\;\;\;\;\;\;\;\;\;\;\;\;\;\;\;\;\;\;\;\;\;\;\Big\lvert\; \mathbf{A}_E \sqrt{n}T_n + \mathbf{B}_E\Omega_n < b_E\Big)\\
&= \mathbb{P}\Big(\Pi_S(\mathcal{B}^c(\beta^{E'}_n, \delta)\lvert \widehat{\beta}^{E'}) \leq \dfrac{\exp(-n\cdot(\widetilde\lambda_{\text{min}}(1-r)^2-\lambda^{\bQ}_{\text{max}}(r+s)^2)\delta^2/2)\pi(\mathcal{B}^c(\beta^{E'}_n, \delta))}{\pi(\mathcal{B}(\beta^{E'}_n, s\delta))}\\
&\;\;\;\;\;\;\;\;\;\;\;\;\;\;\;\;\;\;\;\;\;\;\;\;\;\;\;\;\;\;\;\;\;\;\;\;\;\;\;\;\;\;\;\;\;\;\;\;\;\;\;\;\;\;\;\;\;\;\;\;\;\;\;\;\;\;\;\;\;\;\;\;\;\;\;\;\;\;\;\;\;\;\;\;\;\;\;\;\;\;\;\;\;\;\;\;\;\;\;\;\;\;\;\;\;\Big\lvert\; \mathbf{A}_E \sqrt{n}T_n + \mathbf{B}_E\Omega_n < b_E\Big)\\
& \leq \mathbb{P}\left(\Pi_S(\mathcal{B}^c(\beta^{E'}_n, \delta)\lvert \widehat{\beta}^{E'}) \leq \epsilon \Big\lvert\; \mathbf{A}_E \sqrt{n}T_n + \mathbf{B}_E\Omega_n < b_E\right) \text{ for sufficiently large } n.
\end{aligned}
\end{equation*}
Observe that the last step follows by choosing $r, s<r \in (0,1)$ in the penultimate step such that
\[\widetilde\lambda_{\text{min}}(1-r)^2-\lambda^{\bQ}_{\text{max}}(r+s)^2> \widetilde\lambda_{\text{min}}(1-r)^2-\lambda^{\bQ}_{\text{max}}4r^2>0.\]
Such a choice is possible by noting that roots of the above quadratic in $r$, $(\widetilde\lambda_{\text{min}} \pm 2(\widetilde\lambda_{\text{min}}{\lambda^{\bQ}_{\text{max}}})^{1/2})/(\widetilde\lambda_{\text{min}}-4\lambda^{\bQ}_{\text{max}})$ are opposite to each other in signs and that $\widetilde\lambda_{\text{min}} - 4\lambda^{\bQ}_{\text{max}}<0$. Thus, follows the last step as $\exp(-n\cdot(\widetilde\lambda_{\text{min}}(1-r)^2-\lambda^{\bQ}_{\text{max}}(r+s)^2)\delta^2/2)$ can be made smaller than $\epsilon>0$ for sufficiently large $n$.

Finally, the conclusion of Theorem \ref{consistency:gen} implies that 
\[\mathbb{P}(\|\widehat{\beta}^{E'}_n-\beta^{E'}_n\| \leq r\delta \lvert\; \mathbf{A}_E \sqrt{n}T_n + \mathbf{B}_E\Omega_n < b_E) \to 1\] as $n\to \infty$ which, in turn leads to consistency of the selective posterior under the selective law at $\beta^{E'}_n$.
\end{proof}

\begin{proof}[Proof of Theorem \ref{lem:map-convex}]
Minimizing the above objective in $\beta_n^{E'}$ is equivalent to minimizing
$$
 - \log \pi(\beta^{E'}_n) + n{\beta^{E'}_n}^T \bQ \beta^{E'}_n/2  - n {\beta^{E'}_n}^T \bQ \widehat{\beta}^{E'} + \log \widetilde{\mathbb{P}}(\mathbf{A}_E \sqrt{n}T_n + \mathbf{B}_E\sqrt{n}W_n < b_E\lvert \beta^{E'}_n). 
$$
First we note that:
\begin{equation*}
\begin{aligned}
& n^{2\delta} \cdot \inf_{b,\eta, o}\Big\{ (b-\beta^*)^T \bQ (b- \beta^*)/2+ \eta^T \mathbf{N} \eta/2\\
&\;\;\;\;\;\;\;\;+ \left(P_E \begin{pmatrix} b & \eta \end{pmatrix}^T  + Q_E o + r_E/n^{\delta}\right)^T \mathbf{\Sigma}_{\mathcal{G}}^{-1} \left(P_E \begin{pmatrix} b & \eta \end{pmatrix}^T + Q_E o + r_E/n^{\delta}\right)\Big/2 \\
&\;\;\;\;\;\;\;\;+ \psi_{n^{-\delta}}(o_E, o_{-E})\Big\}
\end{aligned}
\end{equation*}
is equivalent to
\begin{equation*}
\begin{aligned}
& n \cdot \inf_{b',\eta', o'}\Big\{ (b'-{\beta}^{E'}_n)^T \bQ (b'- {\beta}^{E'}_n)/2+ \eta'^T \mathbf{N} \eta'/2\\
&\;\;\;\;\;\;\;\;+ \left(P_E \begin{pmatrix} b' & \eta' \end{pmatrix}^T  + Q_E o' + r_E/\sqrt{n}\right)^T \mathbf{\Sigma}_{\mathcal{G}}^{-1} \left(P_E \begin{pmatrix} b' & \eta' \end{pmatrix}^T + Q_E o' + r_E/\sqrt{n}\right)\Big/2 \\
&\;\;\;\;\;\;\;\;+ \psi_{n^{-1/2}}(o'_E, o'_{-E})\Big\}
\end{aligned}
\end{equation*}
where $\sqrt{n}b' =  n^{\delta} b , \sqrt{n}\eta' =  n^{\delta} \eta, \sqrt{n}o' =  n^{\delta} o.$ Ignoring the prior for now and denoting 
\begin{equation*}
\begin{aligned}
& \Scale[0.87]{H_{\psi}(b) =\inf_{(\eta, o) \in \RR^{2p-|E'|}} \Big\{\eta^T \mathbf{N} \eta/2+ \left(P_E \begin{pmatrix} b & \eta \end{pmatrix}^T  + Q_E o + r_E/\sqrt{n}\right)^T \mathbf{\Sigma}_{\mathcal{G}}^{-1} \left(P_E \begin{pmatrix} b & \eta \end{pmatrix}^T + Q_E o + r_E/\sqrt{n}\right)\Big/2}\\
&\;\;\;\;\;\;\;\;\;\;\;\;\;\;\;\;\;\;\;\;\;\;\;\;\;\;\;\;\;\;+ \psi_{n^{-1/2}}(o_E, o_{-E})\Big\}
\end{aligned}
\end{equation*}
observe that
$${n\beta^{E'}_n}^T \bQ \beta^{E'}_n/2  - n {\beta^{E'}_n}^T \bQ \widehat{\beta}^{E'} + \log {\widetilde{\mathbb{P}}}_{\boldsymbol{\Sigma}}(\mathbf{A}_E \sqrt{n}T_n + \mathbf{B}_E\sqrt{n}W_n < b_E\lvert \beta_n^{E'})$$ equals
$$
\begin{aligned}
& {n\beta^{E'}_n}^T \bQ \beta^{E'}_n/2  - n {\beta^{E'}_n}^T \bQ \widehat{\beta}^{E'}- n\cdot\inf_{b' \in \mathbb{R}^{|E'|}} \left\{(b'-{\beta}^{E'}_n)^T \bQ (b'- {\beta}^{E'}_n)/2 +  H_{\psi}(b')\right\}\\
&= {n\beta^{E'}_n}^T \bQ \beta^{E'}_n/2  - n {\beta^{E'}_n}^T \bQ \widehat{\beta}^{E'} -{n\beta^{E'}_n}^T \bQ \beta^{E'}_n/2 \\
&\;\;\;+ n\cdot\sup_{b' \in \mathbb{R}^{|E'|}} \left\{b{'}^T \bQ \beta^{E'}_n - b{'}^T \bQ b'/2 -H_{\psi}(b')\right\}\\
&= n\cdot  \bar{H}^*(\bQ \beta^{E'}_n)   - n {\beta^{E'}_n}^T \bQ \widehat{\beta}^{E'}.
\end{aligned}
$$
In the above equation, $\bar{H}^*$ is the convex conjugate of the function of 
$
\bar{H}(z) = b^T \bQ b/2 +  H_{\psi}(b).
$
Hence a MAP estimate minimizes
$n\cdot  \bar{H}^*(\bQ \beta^{E'}_n)   - n {\beta^{E'}_n}^T \bQ \widehat{\beta}^{E'} - \log \pi(\beta^{E'}_n)$
which is convex, whenever $\pi(\cdot)$ is a log-concave prior. 
\end{proof}

\begin{proof}[Proof of Theorem \ref{mle:convergence}]
Under the parameterization $\sqrt{n}\beta^{E'}_n = n^{\delta}\beta^*$, denote by
\[n^{2\delta}\cdot \widetilde{\ell}_S^n(\beta^*) = -n(\widehat{\beta}^{E'} -\beta^{E'}_n)^T\bQ(\widehat{\beta}^{E'} -\beta^{E'}_n)/2 - \log \widetilde{\mathbb{P}}(\mathbf{A}_E \sqrt{n}T_n + \mathbf{B}_E\sqrt{n}W_n < b_E\lvert \beta^{E'}_n)\]
the approximate (log-) selection-adjusted likelihood sequence and the corresponding true likelihood sequence by
\[ n^{2\delta} \cdot{\ell}_S^n(\beta^*) = -n(\widehat{\beta}^{E'} -\beta^{E'}_n)^T \bQ(\widehat{\beta}^{E'} -\beta^{E'}_n)/2 - \log \mathbb{P}(\mathbf{A}_E \sqrt{n}T_n + \mathbf{B}_E\sqrt{n}W_n < b_E\lvert \beta^{E'}_n).\]
Denote the MLE for $\beta^*$ as $\hat{\beta}^*$ where $\hat{\beta}^* = n^{1/2-\delta}\widehat{\beta}^{E'}_S$.
We will show that
\begin{equation*}
\begin{aligned}
&\mathbb{P}\left( R_n(\epsilon) \leq \frac{1}{2}\inf_{s:\|s-\hat{\beta}^*\| =\epsilon}\{\widetilde{\ell}_S^n(\hat{\beta}^*)-\widetilde{\ell}_S^n(s)\}\Big\lvert \mathbf{A}_E \sqrt{n}T_n + \mathbf{B}_E\sqrt{n}W_n < b_E\right) \\
&\leq \mathbb{P}(n^{1/2-\delta}\|\widehat{\beta}^{E'}_S -\breve{\beta}^{E'}_S\|\leq \epsilon \lvert \mathbf{A}_E \sqrt{n}T_n + \mathbf{B}_E\sqrt{n}W_n < b_E)
\end{aligned}
\end{equation*}
where $R_n(\epsilon) = \sup_{\|s-\hat{\beta}^*\|\leq \epsilon}|\ell_S^n (s) -\widetilde{\ell}_S^n(s)|.$
 Let 
$p_n =\hat{\beta}^* +\alpha u $ for a unit vector $u$ and $\alpha>\epsilon$. Then
\[\ell_S^n(\hat{\beta}^* +\epsilon u) = \ell_S^n \left(\left(1-\frac{\epsilon}{\alpha}\right)\hat{\beta}^* +\frac{\epsilon}{\alpha} p_n\right) \geq \left(1-\frac{\epsilon}{\alpha}\right) \ell_S^n(\hat{\beta}^*) +\frac{\epsilon}{\alpha} \ell_S^n(p_n).\]
\begin{equation*}
\begin{aligned}
\frac{\epsilon}{\alpha} \{\ell_S^n(\hat{\beta}^*)-\ell_S^n(p_n)\} &\geq   \ell_S^n(\hat{\beta}^*)- \ell_S^n (\hat{\beta}^*+ \epsilon u)\\
&=(\ell_S^n(\hat{\beta}^*) - \widetilde{\ell}_S^n(\hat{\beta}^*)) - \{ \ell_S^n (\hat{\beta}^*+ \epsilon u)  - \widetilde{\ell}_S^n(\hat{\beta}^* + \epsilon  u)\} +( \widetilde{\ell}_S^n(\hat{\beta}^*)-\widetilde{\ell}_S^n(\hat{\beta}^*+\epsilon u) ) \\
&\geq \inf_{s:\|s-\hat{\beta}^*\| =\epsilon}\{ \widetilde{\ell}_S^n(\hat{\beta}^*)-\widetilde{\ell}_S^n(s)\} -2\cdot\sup_{\|s-\hat{\beta}^*\|\leq \epsilon}|\ell_S^n (s) -\widetilde{\ell}_S^n(s)|
\end{aligned}
\end{equation*}
We note that the event $R_n(\epsilon) \leq \frac{1}{2}\inf_{s:\|s-\hat{\beta}^*\| =\epsilon}\{ \widetilde{\ell}_S^n(\hat{\beta}^*)-\widetilde{\ell}_S^n(s)\},$
implies $\ell_S^n(\hat{\beta}^*)-\ell_S^n(p_n)>0 $ for all $p_n =\hat{\beta}^* +\alpha u $ and for any $\alpha>\epsilon$, which means that the maximizer of $\ell_S^n(.)$ given by $\breve{\beta^*}\equiv n^{1/2-\delta}\breve{\beta}^{E'}_S$ lies inside a $\epsilon$-ball around $\hat{\beta}^*\equiv n^{1/2-\delta}\widehat{\beta}^{E'}_S$, the maximizer of the pseudo selective posterior sequence.
Thus, we have
\begin{equation*}
\begin{aligned}
&\mathbb{P}(n^{1/2-\delta}\|\widehat{\beta}^{E'}_S- \breve{\beta}_S^{E'} \| >\epsilon\lvert \mathbf{A}_E \sqrt{n}T_n + \mathbf{B}_E\sqrt{n}W_n < b_E) \\
&\leq \mathbb{P}\left(R_n(\epsilon) \geq \frac{1}{2}\inf_{s:\|s-\hat{\beta}^*\| =\epsilon}\{\widetilde{\ell}_S^n(\hat{\beta}^*)-\widetilde{\ell}_S^n(s)\} \Big\lvert \mathbf{A}_E \sqrt{n}T_n + \mathbf{B}_E\sqrt{n}W_n < b_E\right).
\end{aligned}
\end{equation*}
To complete the proof, we note from Lemma \ref{likelihood:ratio:carved:gen} that 
$$
 \widetilde{\ell}_S^n(\hat{\beta}^*)-\widetilde{\ell}_S^n(s) \geq 
\frac{\widetilde\lambda_{\text{min}}}{2}\cdot\|s-\hat{\beta}^*\|^2, $$
which implies 
\[\inf_{s:\|s-\hat{\beta}^*\| =\epsilon}\{ \widetilde{\ell}_S^n( \widehat{\beta}^{E'}_S)-\widetilde{\ell}_S^n(s)\} \geq  \frac{\widetilde\lambda_{\text{min}}}{2}\cdot\epsilon^2. \]
Using the fact that the randomized selective MLE $\widehat{\beta}^{E'}_S$ is stochastically bounded under the selective law (follows from Theorem \ref{consistency:gen}) and uniform convergence of $R_n(\epsilon)$ on compact sets (from Remark \ref{unif:convergence})
\begin{equation*}
\begin{aligned}
&\mathbb{P}(n^{1/2-\delta}\|\widehat{\beta}^{E'}_S- \breve{\beta}_S^{E'} \|>\epsilon \lvert  \mathbf{A}_E \sqrt{n}T_n + \mathbf{B}_E\sqrt{n}W_n < b_E)\\
& \leq\mathbb{P}\left(R_n(\epsilon) \geq \frac{1}{2}\inf\limits_{s:\|s-\hat{\beta}^*\| =\epsilon}\{ \widetilde{\ell}_S^n( \hat{\beta}^*)-\widetilde{\ell}_S^n(s)\}\Big\lvert  \mathbf{A}_E \sqrt{n}T_n + \mathbf{B}_E\sqrt{n}W_n < b_E\right)\\
&\leq \mathbb{P}\left(R_n(\epsilon) \geq \frac{\lambda_{\text{min}}}{4}\cdot\epsilon^2\;\Big\lvert  \;\mathbf{A}_E \sqrt{n}T_n + \mathbf{B}_E\sqrt{n}W_n < b_E\right) \to 0.
\end{aligned}
\end{equation*}
\end{proof}

\subsection{Supplement to HIV drug-resistance analysis}
\label{A:5}

\begin{figure}[H]
 \centering
    \includegraphics[scale=.110]{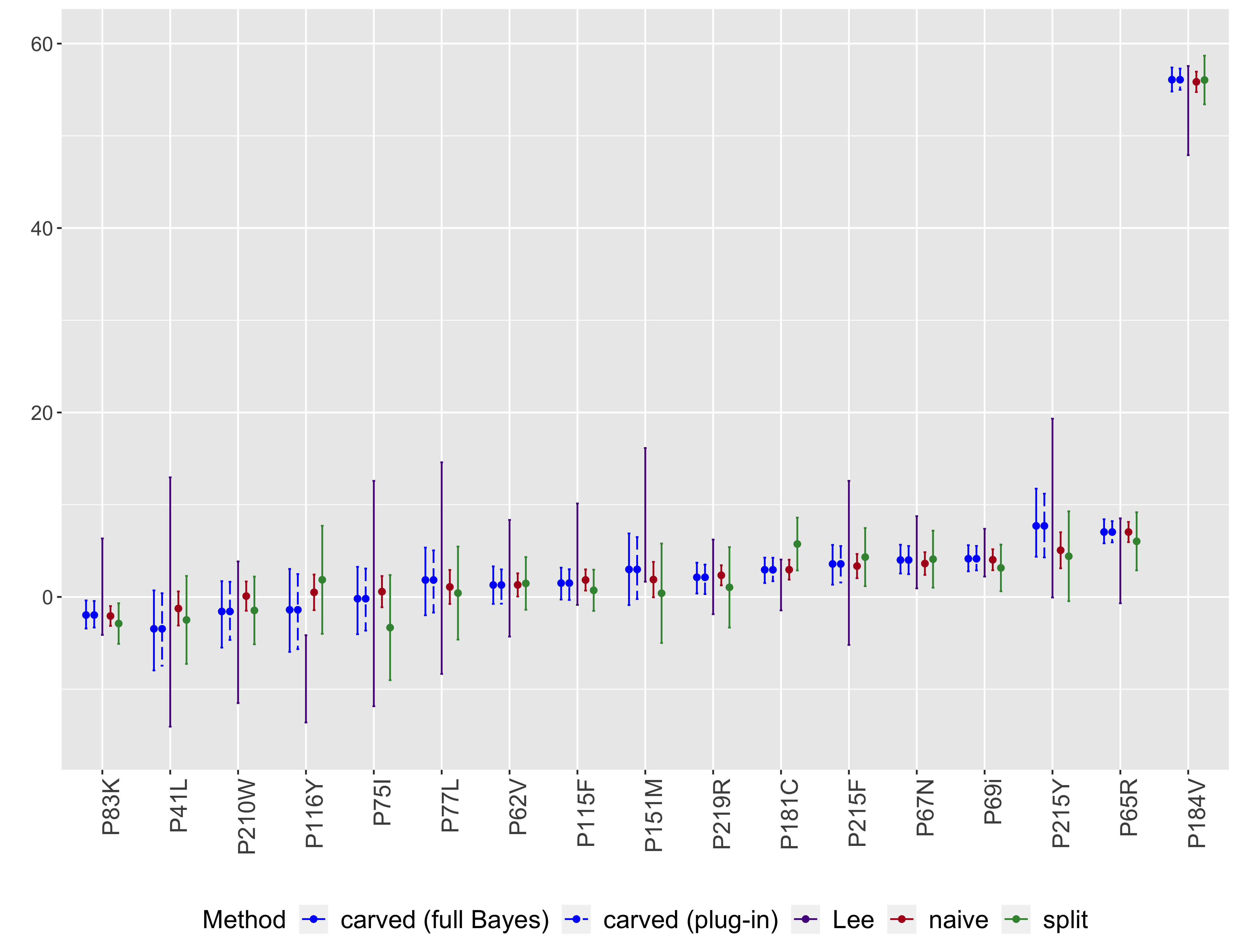}
  \caption{Confidence intervals and point estimates for selected features. }
  \label{fig:hiv:all}
\end{figure}

Figure \ref{fig:hiv:all} depicts the estimates constructed for mutation `P184V', excluded in Figure \ref{fig:hiv} in the main file.

\subsection{Polyhedral selection rules}
\label{A:6}
With the article focusing on Lasso selection in a linear model, here we give examples for other situations that fit our framework. 
In the first example variables are selected based on marginal correlations with the response, a procedure that is sometimes referred to as {\it marginal screening} \citep[e.g., by][]{lee2016exact} and frequently used in statistical genomics \citep[e.g.,][]{mckeague2015adaptive, gtex2017genetic}. 
Our second example, motivated by \citet{taylor2018post}, demonstrates how the methods developed in this work apply more broadly in generalized linear models. 


%
%

\begin{example}
\rm{ 
Consider solving the optimization problem given by 
\begin{equation}
\label{marginal:screening}
\underset{\gamma \in \mathbb{R}^p}{\text{minimize}} \ \ \dfrac{1}{2}\Big\|\gamma - (\bX^{\mathcal{S}})^T \vec{y}^{\mathcal{S}}/\sqrt{n}\rho \Big\|_2^2 + \mathcal{P}^{\;\lambda}_{\ell_\infty}(\gamma),
\end{equation}
for $\rho= n_1/n$, where 
\[ \mathcal{P}^{\;\lambda}_{\ell_\infty}(\gamma) = \begin{cases} 
      0 & \text{ if } \ \|\gamma\|_{\infty} \leq \lambda \\
      \infty  & \text{ otherwise} ;
      \end{cases}
\]
and where $\lambda \in \mathbb{R}^p$ is a vector that determines the cut-off for significance. 
Denoting the solution to \eqref{marginal:screening} by $\widehat{\gamma}^{\;\lambda}$, let
$$\widehat{E} =\{j: \ |\widehat{\gamma}_j^{\;\lambda}| = \lambda\}.$$
If we set
$$\Omega_n = \sqrt{n}W_n = \dfrac{\partial}{\partial\gamma}\left\{ -\dfrac{1}{2}\Big\|\gamma - (\bX^{\mathcal{S}})^T \vec{y}^{\mathcal{S}}/\sqrt{n}\rho \Big\|_2^2 + 
\dfrac{1}{2}\Big\|\gamma - \bX^T \vec{y}/\sqrt{n} \Big\|_2^2\right\}_{\widehat{\gamma}},$$
then, recalling the notation in Proposition \ref{asymptotic:distribution}, we have that
\begin{equation*}
\begin{pmatrix}
\sqrt{n}\left({T}_n - {\mu} \right)\\
\Omega_n
\end{pmatrix}
\in \RR^{2p}
\end{equation*}
is asymptotically normal under the modeling assumption \eqref{E':model}. 
Now we observe that for the program \eqref{marginal:screening}, the event
$$
\left( \widehat{E},\widehat{S}^E \right) = \left( E, s^E \right)
$$
can be rewritten in a polyhedral form,
$$\mathbf{A}_E \sqrt{n}T_n + \mathbf{B}_E\sqrt{n} W_n < b_E.$$
for 
\begin{equation*}
\mathbf{A}_E= \begin{bmatrix} -\text{diag}(s^E)\mathbf{P}^{E, E'} & -\text{diag}(s^E)\mathcal{I}^{E, E'} \\  \mathbf{F}^{E, E'}  & \mathcal{J}^{E, E'} \\  
-\mathbf{F}^{E, E'}  & -\mathcal{J}^{E, E'} \end{bmatrix}, 
\mathbf{B}_E= \begin{bmatrix} -\text{diag}(s^E) & \mathbf{0} \\ \mathbf{0} & \mathbf{I} \\ \mathbf{0} & -\mathbf{I}\end{bmatrix}, \;\; b_E = \lambda \begin{pmatrix}-1 \\ 1 \\ 1 \end{pmatrix}, 
\end{equation*} 
if we ignore the $o_p(1)$ remainder term. 
}

\end{example}

\begin{example}
\rm{
Suppose that $Y_i\in \{0,1\}$, and we are to select variables with the $\ell_1$-penalized logistic likelihood. 
For a matrix $\boldsymbol{D}\in \RR^{k\times p}$ with rows $d_i^T$, $i=1,...,k$, we denote by $\pi(\boldsymbol{D}; \beta)$ the vector in $\mathbb{R}^{k}$ with the $i$-th entry equal to 
$$
(1+ \exp(d_i^T \beta))^{-1} \exp(d_i^T \beta).
$$
Then we select 
\begin{equation}\label{eq:lasso-sel}
\widehat{E} = \{j: \widehat{\beta}_j^{\lambda}\neq 0\},
\end{equation}
where $\widehat{\beta}^{\lambda}$ is the solution to
\begin{equation}
\label{logistic:lasso}
\underset{\beta \in \mathbb{R}^p}{\text{minimize}} \ \ -\dfrac{1}{\sqrt{n}\rho}(\log \pi(\bX^{\mathcal{S}}; \beta))^T \vec{y}^{\mathcal{S}} - \dfrac{1}{\sqrt{n}\rho}\{\log (1- \pi(\bX^{\mathcal{S}}; \beta))\}^T (1- \vec{y}^{\mathcal{S}}) +  \lambda \|\beta\|_1.
\end{equation}
We remark that methods for inference appealing to the Polyhedral lemma in a non-randomized framework have been treated in \cite{taylor2018post}. 
For the query in \eqref{logistic:lasso}, we define the randomization variable as
$$
\Omega_n = \dfrac{1}{\sqrt{n} \rho} (\bX^{\mathcal{S}})^T (\vec{y}^{\mathcal{S}} - \pi(\bX^{\mathcal{S}}; \widehat{\beta}^{\lambda})) - \dfrac{1}{\sqrt{n}}\bX^{T} (\vec{y} - \pi(\bX; \widehat{\beta}^{\lambda})).
$$
Further, we define
\begin{equation}\label{T_n:logistic}
\sqrt{n}\vec{T}_n := 
\begin{pmatrix}
\sqrt{n}\widehat{\beta}^{E'}\\\sqrt{n} \vec{N}_{-E'}
\end{pmatrix} := 
\begin{pmatrix}
\sqrt{n}\widehat{\beta}^{E'} \\ \dfrac{1}{\sqrt{n}} X_{-E'}^T\left(\vec{y} -\pi(\bX_{E'}; \widehat{\beta}^{E'})  \right)  
\end{pmatrix} 
\end{equation}
where $\widehat{\beta}^{E'}$ satisfies the estimating equation given by
$$\dfrac{1}{\sqrt{n}}X_{E'}^T\left(\vec{y} -\pi(\bX_{E'}; \widehat{\beta}^{E'})\right)= 0.$$

Under the model \eqref{E':model}, the event $\{\widehat{E}=E, \widehat{S}^{\widehat{E}} = s^E\}$ can be represented in a polyhedral form 
$$
\mathbf{A}_E \sqrt{n}T_n + \mathbf{B}_E\sqrt{n} W_n < b_E,
$$
on ignoring the $o_p(1)$ remainder term. 
Specifically, the matrices characterizing the convex selection region in Proposition \ref{K.K.T.:selection} are now given by
$$\mathbf{P}^{E, E'} = \mathbb{E}_{P}(\bX_{E}^T \mathcal{D} \bX_{E'}/n), \ \ \ \mathbf{F}^{E, E'}= \mathbb{E}_{P}(\bX_{-E}^T \mathcal{D} \bX_{E'}/n),$$ 
$$\mathbf{Q}^E = \mathbb{E}_{P}(\bX_{E}^T \mathcal{D} \bX_{E}/n), \ \ \ \ \mathbf{C}^E = \mathbb{E}_{P}(\bX_{-E}^T \mathcal{D} \bX_{E}/n),$$ 
where $\mathcal{D}$ is a diagonal matrix with $n$ rows and columns, and with the $j$-th diagonal entry equal to 
$$
\pi_j(\bX_{E'}; \beta^{E'})\cdot (1- \pi_j(\bX_{E'}; \beta^{E'})). 
$$
We emphasize that randomization here is implicit and due to the fact that selection uses only part of the data. 
Hence, it is different from the randomization scheme in \citet{tian2018selective}, where the objective \eqref{logistic:lasso} is perturbed with heavy-tailed noise.
}

\end{example}

\end{document}